\documentclass{egpubl}
\usepackage{eg2026}
 
\ConferencePaper        
\CGFccby

\usepackage[T1]{fontenc}
\usepackage{dfadobe}  

\usepackage{cite}  
\BibtexOrBiblatex
\electronicVersion
\PrintedOrElectronic
\ifpdf \usepackage[pdftex]{graphicx} \pdfcompresslevel=9
\else \usepackage[dvips]{graphicx} \fi

\usepackage{egweblnk} 

\makeatletter
\let\EG@thebibliography\thebibliography
\let\EG@endthebibliography\endthebibliography
\makeatother

\usepackage{float}
\usepackage{enumitem}
\usepackage{amsmath}
\usepackage[nameinlink]{cleveref}
\usepackage{overpic} 
\usepackage{tcolorbox}
\usepackage{hyperref}
\usepackage{soul}
\usepackage{amsfonts}
\usepackage{booktabs}
\usepackage[table]{xcolor}
\usepackage[export]{adjustbox}
\usepackage{cite}

\makeatletter
\let\thebibliography\EG@thebibliography
\let\endthebibliography\EG@endthebibliography
\makeatother

\crefname{equation}{eq.}{eqs.}
\Crefname{equation}{Eq.}{Eqs.}
\crefname{table}{table}{tables}
\Crefname{table}{Table}{Tables}
\crefname{figure}{fig.}{figs.}
\Crefname{figure}{Fig.}{Figs.}
\crefname{section}{sec.}{secs.}
\Crefname{Section}{Sec.}{Secs.}
\Crefrangeformat{equation}{Eqs.~#3#1#4--#5#2#6}
\crefrangeformat{figure}{Figs.~#3#1#4--#5#2#6}
\crefrangeformat{theorem}{Thms.~#3#1#4--#5#2#6}

\newcommand{\ThreePPose}{\textcolor{black}{\mathbf{p}}}
\newcommand{\ThreePHist}{\textcolor{black}{\ThreePPose_{t-h : t}}}
\newcommand{\ThreePChunk}{\textcolor{black}{\ThreePPose_{t : t+T}}}
\newcommand{\ThreePChunkPred}{\textcolor{black}{\mathbf{\hat p}_{t : t+T}}}
\newcommand{\ThreePLogits}{\textcolor{black}{\mathbf{z}_t^{\text{3p}}}}
\newcommand{\GameLogits}{\textcolor{black}{\mathbf{z}_t^{\text{game}}}}
\newcommand{\ColoredNote}{\textcolor{black}{\mathbf{c}}}
\newcommand{\ColoredNotes}{\textcolor{black}{\ColoredNote_t^{: n}}}
\newcommand{\BombNote}{\textcolor{black}{\mathbf{b}}}
\newcommand{\BombNotes}{\textcolor{black}{\BombNote_t^{: n}}}
\newcommand{\Obstacle}{\textcolor{black}{\mathbf{o}}}
\newcommand{\Obstacles}{\textcolor{black}{\Obstacle_t^{: n}}}

\newcommand{\ReconLoss}{\mathcal{L}_{\text{Recon}}}
\newcommand{\MatchingLoss}{\mathcal{L}_{\text{Match}}}

\newcommand{\textline}[1]{%
  \noindent
  \makebox[\linewidth]{%
    \leavevmode\leaders\hrule height \dimexpr0.5ex+0.2pt\relax depth \dimexpr-0.5ex+0.2pt\relax\hfill\kern0pt
    \quad #1 \quad 
    \leavevmode\leaders\hrule height \dimexpr0.5ex+0.2pt\relax depth \dimexpr-0.5ex+0.2pt\relax\hfill\kern0pt
  \par\vspace{0.5em} 
  }%
}

\newcommand{\NRefs}{N_\text{ref}}

\newcommand{\RefSegments}{\mathbf{x}^{:\NRefs}}
\newcommand{\GameSegment}{\mathbf{x}^{\text{game}}_t}
\newcommand{\StyleEncoder}{\mathcal{E}^\text{style}}
\newcommand{\GameEncoder}{\mathcal{E}^\text{game}}
\newcommand{\ThreePEncoder}{\mathcal{E}^\text{3p}}
\newcommand{\ThreePDecoder}{\mathcal{D}^\text{3p}}
\newcommand{\StyleEmbeddings}{\mathbf{z}^{:\NRefs}}

\DeclareMathOperator*{\argmax}{arg\,max}

\newcommand{\RefColorNotes}{\mathbf{c}_{t_i^*}^{:n}}
\newcommand{\RefBombNotes}{\mathbf{b}_{t_i^*}^{:n}}
\newcommand{\RefObstacles}{\mathbf{o}_{t_i^*}^{:n}}
\newcommand{\RefThreePHist}{\mathbf{p}_{t_i^*-h:t_i^*}}
\newcommand{\RefThreePChunk}{\mathbf{p}_{t_i^*:t_i^* + T}}

\title{\textsc{Robo-Saber}: Generating and Simulating Virtual Reality Players}

\author[N. H. Kim et al.]
{\parbox{\textwidth}{
\centering 
N. H. Kim$^{1}$\orcid{0000-0002-9172-624X},
J. M. Liu$^{2}$\orcid{0009-0009-6936-4017},
J. Lehtinen$^{1, 3}$\orcid{0000-0001-9418-4944},
P. H\"am\"al\"ainen$^{1}$\orcid{0000-0001-7764-3459},
J. F. O'Brien$^{2}$\orcid{0000-0001-9513-0542},
and X. B. Peng$^{3, 4}$\orcid{0000-0002-3677-5655}
}
        \\
{\parbox{\textwidth}{\centering 
$^1$Aalto University, Finland$\quad$
$^2$University of California, Berkeley, United States$\quad$
$^3$NVIDIA$\quad$
$^4$Simon Fraser University, Canada
       }
}
}

\begin{document}

\teaser{
  \centering 
    \begin{overpic}[width=0.995\linewidth,trim={0 0 0 0},clip]{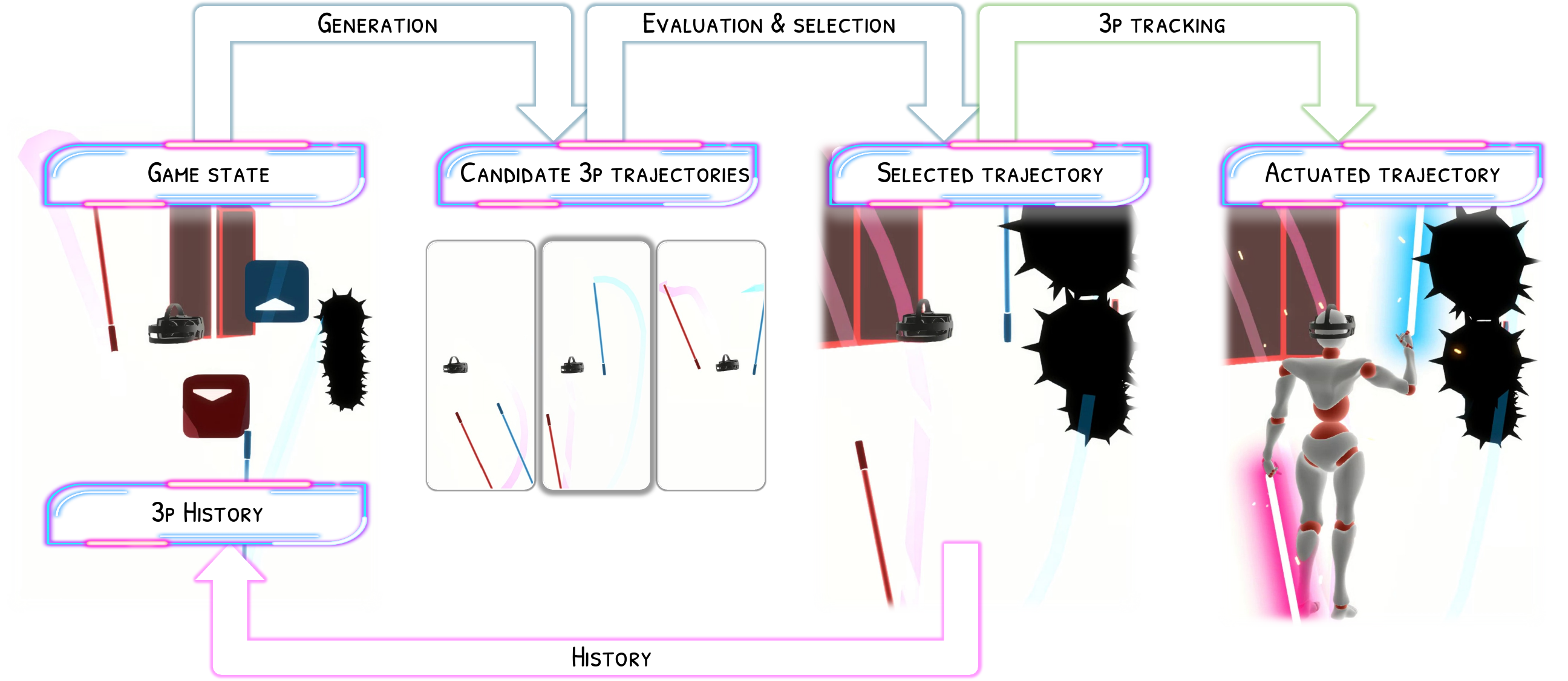}
  \put(22,39){(\Cref{sec:method})}
  \put(46,39){(\Cref{sec:selection})}
  \put(67.5,39){(\Cref{sec:robotic_agent}, optional)}
  \end{overpic}
  \vfill
  \includegraphics[width=0.245\textwidth,trim={700 300 400 200},clip]{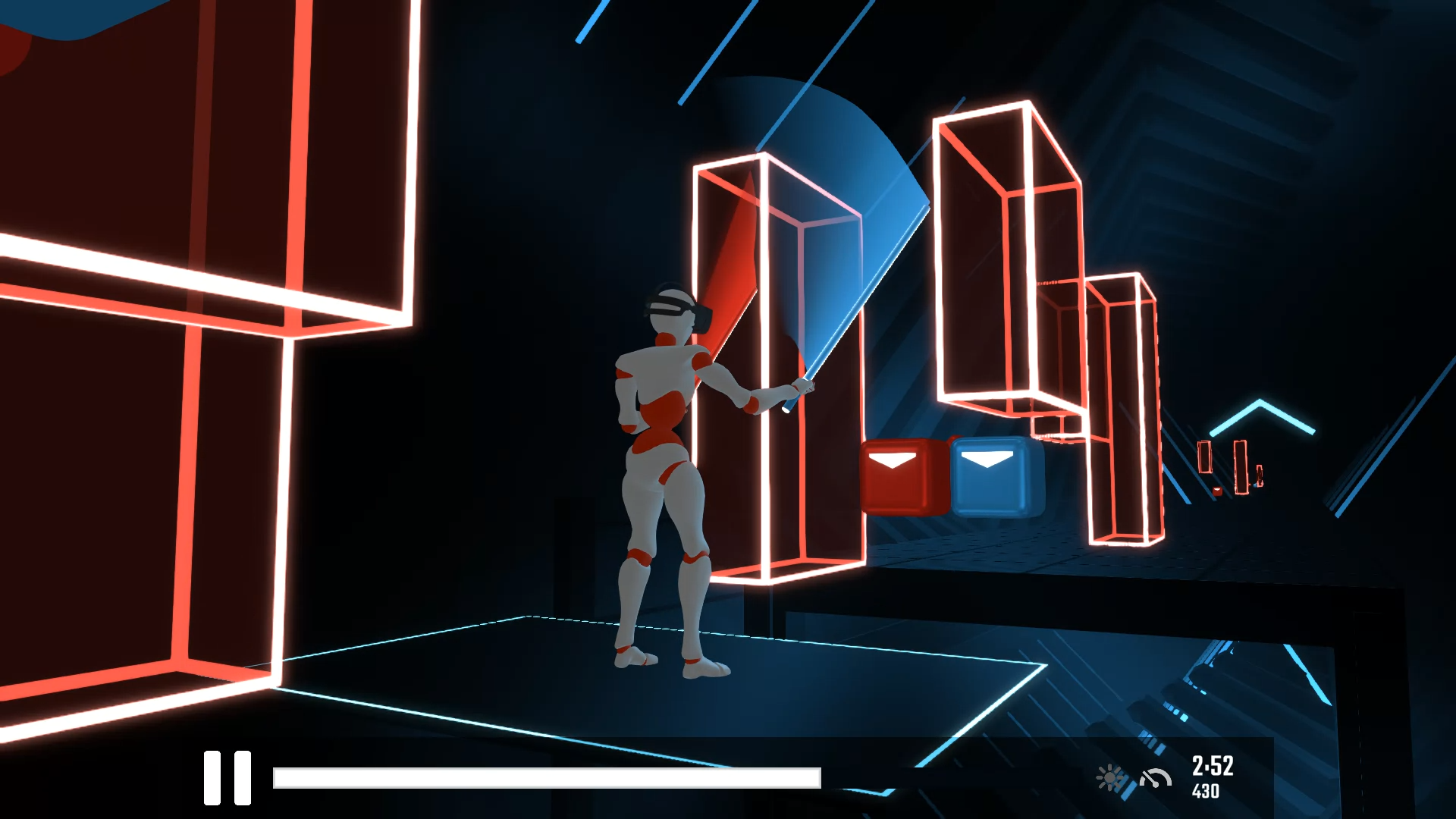}
  \includegraphics[width=0.245\textwidth,trim={700 300 400 200},clip]{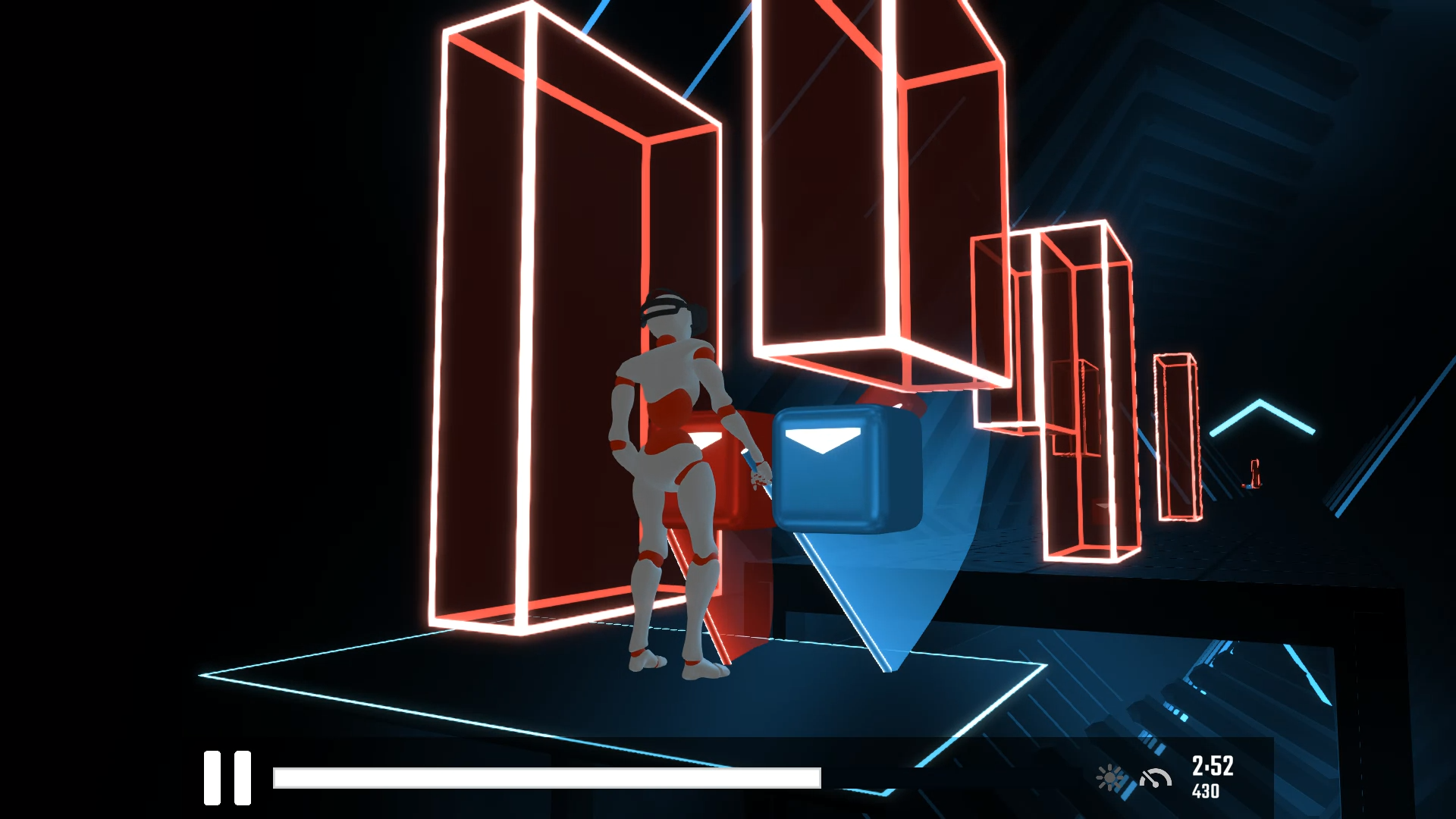}
  \includegraphics[width=0.245\textwidth,trim={700 300 400 200},clip]{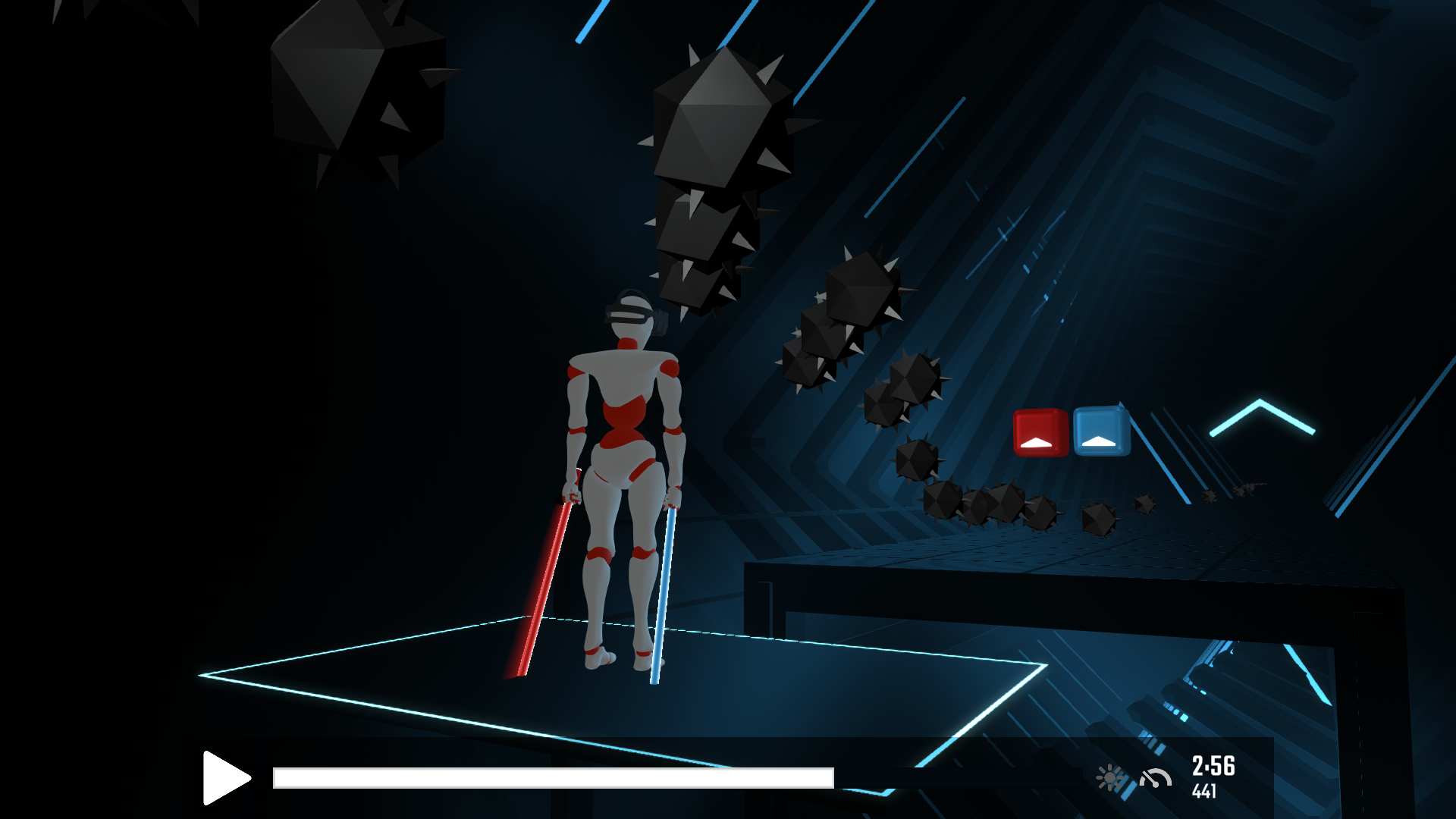}
  \includegraphics[width=0.245\textwidth,trim={700 300 400 200},clip]{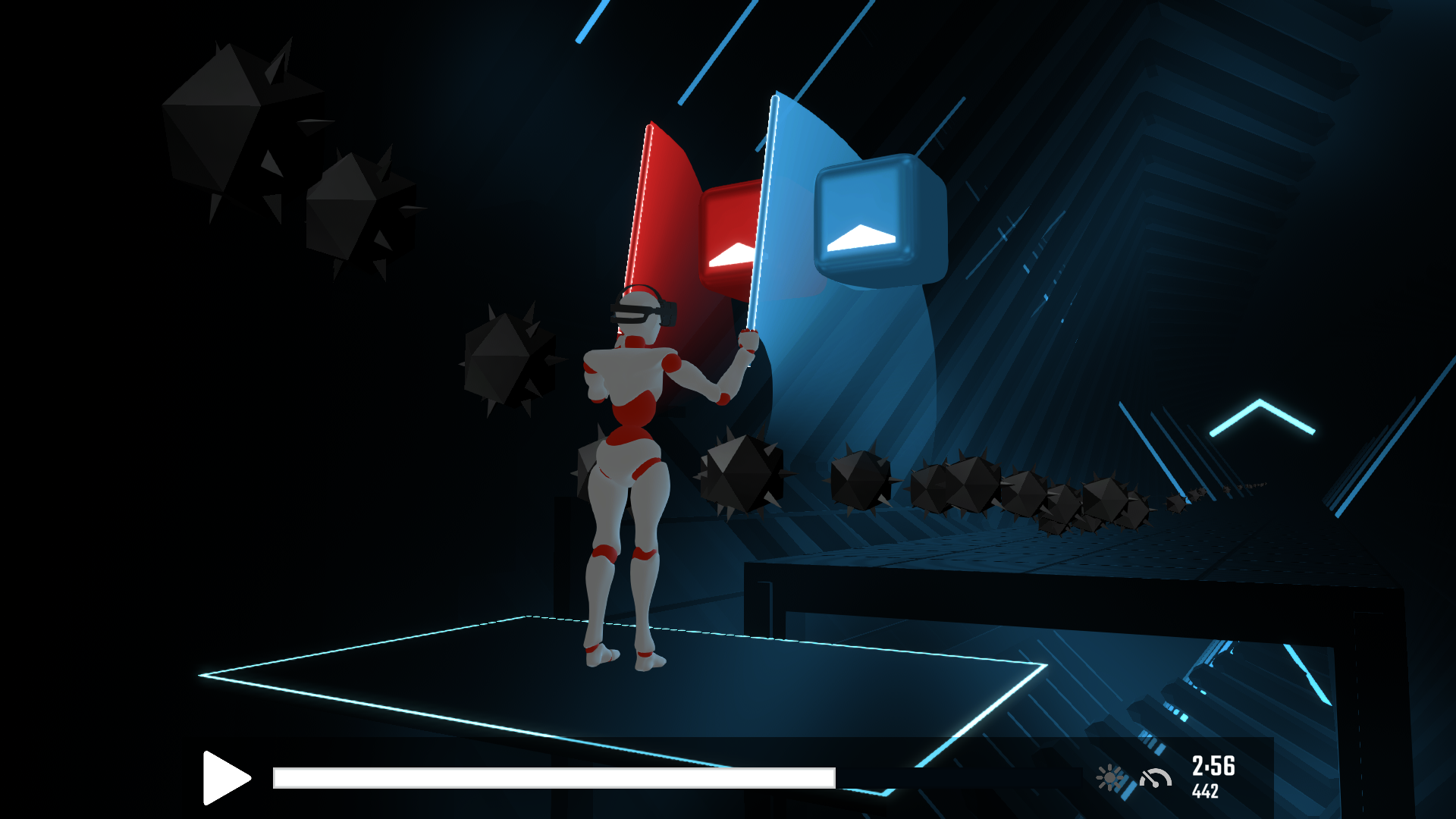}
  \caption{\textbf{Robo-Saber} is a player model for VR games, demonstrated in simulated \emph{Beat Saber} playtesting. \textbf{Top:} System overview. An autoregressive generative model samples candidate three-point ($3p$) trajectories for the headset and handhelds, conditioned on the game state.
  The best $3p$ trajectory is then selected based on simulating the game forward.
  Optionally, a physics-based tracking policy performs whole-body movements based on the $3p$ trajectory.
  \textbf{Bottom:} Examples of Robo-Saber's whole-body gameplay.}
  \label{fig:teaser}
}

\maketitle

\begin{abstract}
We present the first motion generation system for playtesting virtual reality (VR) games.
Our player model generates VR headset and handheld controller movements from in-game object arrangements, guided by style exemplars and aligned to maximize simulated gameplay score.
We train on the large BOXRR-23 dataset and apply our framework on the popular VR game \emph{Beat Saber}.
The resulting model \textbf{Robo-Saber} produces skilled gameplay and captures diverse player behaviors, mirroring the skill levels and movement patterns specified by input style exemplars.
Robo-Saber demonstrates promise in synthesizing rich gameplay data for predictive applications and enabling a physics-based whole-body VR playtesting agent.

\end{abstract}

\section{Introduction}

Traditional virtual reality (VR) development can be cumbersome.
For example, developers are often required to leave their workstations to test the full-body movements involved in their application.
Addressing this problem, simulating human perception and motor control through deep reinforcement learning (DRL) has emerged as a promising approach for automatic testing of VR interaction \cite{ikkala2022breathing,fischer2024sim2vr}. 
However, traditional DRL agents are ill-suited for representing how different users will move and interact, given the variability in body morphology, skill levels, and abilities within the user population.

\emph{Could we learn a player model from real-world data instead?}
The increasing availability of open-source VR gameplay data (\textit{e.g.}, BOXRR-23, \citen{nair2023berkeley}) opens up opportunities for supervised learning as an alternative to \emph{tabula rasa} DRL.
In this work, we seek to produce a model that learns generalizable gameplay capabilities from such a dataset.
Crucially, an ideal model would be calibrated to the diversity in skill levels and behavioral patterns in the training data.
Such a player model could then generate comprehensive synthetic gameplay data for novel content, supporting \emph{in silico} user modeling and automated playtesting use cases.    

For producing this model, the following VR-specific computational challenges remain:
First, many VR scenarios require interaction with 3D objects in richly configurable environments, making it challenging to effectively support the vast space of object configurations.
Second, VR gameplay frequently involves extended-length motions, lasting several minutes or sometimes hours, requiring models to generate long, robust motion sequences that remain aligned with gameplay objectives throughout.
Third, capturing behavioral diversity and output multimodality requires specific design choices that enable the representation of individual styles or behavior modes.
Altogether, these challenges converge on a central problem: \textbf{generative motion planning}--learning to generate robust, well-aligned, and diverse motion plans for a vast space of input scenarios from real-world movement data.

In pursuing the above challenges, we develop and evaluate \textbf{Robo-Saber}, a novel gameplay agent for the VR game \emph{Beat Saber} built on a conditional motion generation model.
\emph{Beat Saber} is especially well-suited to evaluating our framework:
The VR game is an established benchmark across computational and non-computational research domains, widely popular, and offers extensive open-source data (see \Cref{sec:beatsaber} in the Appendix).
We train Robo-Saber on \emph{Beat Saber} gameplay recordings from the large BOXRR-23 dataset \cite{nair2023berkeley}.
Conditioned on in-game object states, Robo-Saber generates diverse three-point (positions and orientations of the headset and two handheld controllers, denoted as $3p$) motion plans aligned with gameplay objectives.
The model utilizes \emph{contextual gameplay exemplars} (\textit{i.e.}, player-specific examples of gameplay movements paired with in-game object states) to distinguish between players based on individual characteristics such as skill level and movement patterns.
Deployed autoregressively, Robo-Saber produces minutes-long gameplay trajectories, enabling automated testing of game content (\emph{maps} in \emph{Beat Saber}).

Robo-Saber is capable of skilled \emph{Beat Saber} gameplay, closely competing with elite-level human players via generated kinematic trajectories.
We demonstrate Robo-Saber's ability to simulate gameplay consistent with the human reference players' skill levels and movement patterns represented by their exemplars.
We build on this result and leverage Robo-Saber for a player modeling use case: the \emph{personalized score prediction} (PSP) scenario, \textit{i.e.}, predicting scores for player-map combinations absent from the training data.
Our results suggest that personalized gameplay can be simulated, enabling one to predict player performance on brand-new game content from a few gameplay examples.
Furthermore, we find that Robo-Saber can synthetically augment training data for a downstream score-prediction model, thereby achieving impressive accuracy.
Finally, we demonstrate that Robo-Saber readily interfaces with physics-based tracking. We examine the effect of tracking on player modeling performance and identify areas of opportunity, contributing towards building a physics-based whole-body VR player model.

Our technical contributions are summarized as follows.
First, we introduce improvements to Categorical Codebook Matching (CCM, \citen{starke2024categorical}):
Our Transformer-based \cite{vaswani2017attention} encoder models support the use of \emph{contextual exemplars} as conditioning signals.
We also replace the mean squared error (MSE) matching loss in CCM with a logit-space Jensen-Shannon divergence (JSD) loss (\Cref{sec:method}).
Furthermore, we develop and evaluate a rejection sampling scheme for motion plans sampled from our model (\Cref{sec:selection}).
We use kinematic gameplay performance evaluated via a custom GPU-accelerated game simulation, which greatly improves generalization compared to deterministic rollout (\Cref{sec:Q2} and \Cref{fig:selection_vs_argmax}).
Applying these techniques to the popular benchmark VR game \emph{Beat Saber}, we produce \emph{the first style-conditioned generative VR player model} that lends itself to predictive applications and physics-based whole-body gameplay synthesis.

\noindent\textit{Project page.} Our source code, lay-friendly summary, and supplementary videos can be found on our project page: \url{https://robo-saber.github.io}

\section{Related Work}

\noindent \emph{Generative models for motion and VR.}
The availability of annotated human motion data has enabled impressive progress in kinematic motion generation approaches.
We build on generative techniques for solving controllable and interactive motion synthesis (\textit{e.g.}, \citen{holden2017phase, zhang2018mode, ling2020character, tevet2023human, shi2024interactive}), as well as more recent research successfully combining generative modeling with physics simulation for expressive physics-based character control (\textit{e.g.}, \citen{xie2023hierarchical,tevet2024closd,huang2025diffuse,xu2025parc}).
A key application domain of motion generation techniques is virtual reality (VR).
For VR, motion generation often incorporates three-point ($3p$) pose data, as most consumer-grade VR equipment allows tracking the positions and orientations of the headset and two handheld controllers.
Much of the prior work targets the task of synthesizing full-body pose to be displayed in VR (\textit{e.g.}, \citen{ye2022neural3points, du2023avatars, starke2024categorical, barquero2025sparse}), \textit{i.e.}, learning a $3p$-to-full-body correspondence from various motion datasets.
However, as movements in VR games are responses to specific tasks (\textit{e.g.}, swinging arms to combat an enemy), modeling $3p$ movements as intentional behaviors requires conditioning on game input.
Towards this end, the BOXRR-23 dataset \cite{nair2023berkeley} provides $3p$ trajectories for well-known VR games, including \emph{Beat Saber}.
In this work, we uniquely leverage the vast amount of \emph{Beat Saber}-specific $3p$ pose data from BOXRR-23 and the open-source custom map database BeatSaver \cite{BeatSaver}, aligning game and pose data to train a conditional motion generation model for producing $3p$ movements directly from game states.
Our incorporation of spatio-temporal goals follows the recent success of generative models for virtual instrument performance (\textit{e.g.}, \citen{wang2024furelise, jiao2025bach, chen2025midi}), extending the idea beyond symbolic music to VR gameplay.
By integrating with a physics-based humanoid tracking controller (\Cref{sec:robotic_agent}), our work also lays the foundation for simulating whole-body gameplay movements. \\ \\
\noindent \emph{Stylized motion generation.}
Generating motions according to specific styles has long been of interest to character animation and generative modeling research.
Early work focused on transferring style between two or more motions using nearest neighbor queries \cite{xia2015realtime}, convolutional autoencoder-based motion manifolds \cite{holden2016deep,holden2017fast}, or spectral methods \cite{yumer2016spectral}.
Subsequently, generative architectures for stylized motion synthesis have advanced, ranging from variational autoencoders \cite{du2019stylistic} to flow models \cite{wen2021autoregressive} and denoising diffusion models \cite{alexanderson2023listen,ao2023gesturediffuclip,zhong2024smoodi,raab2024monkey,sawdayee2025dance,kim2025personabooth}.
A key challenge of the stylization problem is representing the elusive notion of style, which is sometimes manually annotated (\textit{e.g.}, \citen{smith2019efficient,aberman2020unpaired}) or often learned as latent codes from exemplar motions.
These representations have been used as conditioning signals in synthesizing stylized motions over long horizons (\textit{e.g.}, \citen{wen2021autoregressive,ji2025sport,xiao2025motionstreamer}).
Similar to prior work \cite{wen2021autoregressive,tao2022style}, our work uses exemplar motions for autoregressive motion generation while learning a set of latent codes with a Transformer-based style encoder.
Orthogonal to most recent stylized conditional generation work that aligns output motions with input text prompts (\textit{e.g.}, \citen{zhong2024smoodi,ji2025sport}), our work instead aims to align output motions with gameplay objectives.
Moreover, our framework operates with \emph{contextual} exemplars, which provide movement signals as well as task information. \\ \\
\noindent \emph{Computational user modeling.}
Constructing faithful models of user behavior is a long-pursued goal in human-computer interaction that promises to enhance our understanding of human behavior and expedite interactive system design.
Modern approaches frame interaction as Markov decision processes, using deep reinforcement learning (DRL) to discover simulated rational behavior \cite{oulasvirta2022computational}.
DRL-based user models have been successfully applied across a wide range of complexity levels: from puzzle games \cite{roohi2020predicting} to embodied scenarios including touchscreen typing \cite{jokinen2021touchscreen}, gestural interaction \cite{cheema2020predicting}, muscle-and-tendon arm models \cite{ikkala2022breathing}, and VR interaction \cite{fischer2024sim2vr}.
The increasing availability of real behavior data has enabled imitation learning approaches that calibrate to behavioral diversity (\textit{e.g.}, \citen{de2022automated}).
Although DRL remains the dominant paradigm, we believe there is untapped potential in recent advances in data-driven approaches that could leverage massive real-world data.
Thus, we seek to produce user models via supervised learning (\textit{e.g.}, \citen{gudmundsson2018human}) and extend promising results in generative models for human motion synthesis more broadly.
A key promise in applying generative techniques to control and planning is \emph{reward alignment} (\textit{e.g.}, \citen{starke2024categorical, huang2025diffuse, truong2025beyondmimic}), which mirrors a human user's ability to evaluate and refine plans; our use of score-based selection applies this aspect to simulate gameplay.

\section{Method} \label{sec:method}

\begin{figure*}
  \begin{overpic}[width=0.99\linewidth,trim={0, 0, 0, 0},clip]{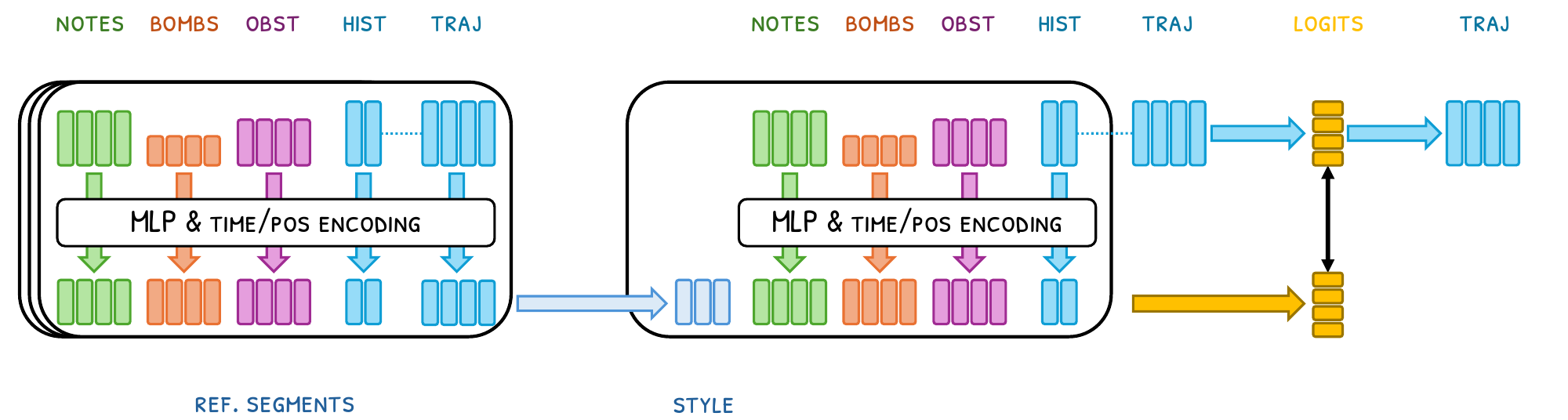}
  \put(5, 22.5){$\RefColorNotes$}
  \put(11, 22.5){$\RefBombNotes$}
  \put(17, 22.5){$\RefObstacles$}
  \put(21, 22.5){$\RefThreePHist$}
  \put(27, 22.5){$\RefThreePChunk$}
  \put(50, 22.5){$\ColoredNotes$}
  \put(55, 22.5){$\BombNotes$}
  \put(61, 22.5){$\Obstacles$}
  \put(66, 22.5){$\ThreePHist$}
  \put(72.5, 22.5){$\ThreePChunk$}
  \put(83, 22.5){$\ThreePLogits$}
  \put(92.5, 22.5){$\ThreePChunkPred$}
  \put(16, 3){$\RefSegments$}
  \put(43.5, 3){$\StyleEmbeddings$}
  \put(55, 3){$\GameSegment$}
  \put(83, 3){$\GameLogits$}
  \put(88, 12.5){\makebox(0,0){\parbox{3cm}{\centering$\MatchingLoss$\\[0.125ex]\hypersetup{linkcolor=magenta}\Cref{eq:jsd}}}}
  \put(36, 7.5){\makebox(0,0){\parbox{3cm}{\centering $\StyleEncoder$\\[1em]\hypersetup{linkcolor=magenta}\Cref{eq:style_encoder}}}}
  \put(77.5, 7.5){\makebox(0,0){\parbox{3cm}{\centering$\GameEncoder$\\[1em]\hypersetup{linkcolor=magenta}\hypersetup{linkcolor=magenta}\Cref{eq:game_encoder}}}}
  \put(80, 18.5){\makebox(0,0){\parbox{3cm}{\centering$\ThreePEncoder$\\[1em]\hypersetup{linkcolor=magenta}\Cref{eq:3p_encoder}}}}
  \put(89, 18.5){\makebox(0,0){\parbox{3cm}{\centering$\ThreePDecoder$\\[1em]\hypersetup{linkcolor=magenta}\Cref{eq:3p_decoder}}}}
  \end{overpic}
  \caption{Overview of our generative model architecture, extending Categorical Codebook Matching (CCM, \citen{starke2024categorical}, \Cref{sec:method}) with Transformer encoder models $\StyleEncoder$ and $\GameEncoder$, as well as a modified loss function based on Jensen-Shannon divergence. 
  }
  \label{fig:generation-overview}
\end{figure*}

Our approach uses a generative motion planning model to produce $3p$ gameplay trajectories.
\textbf{\Cref{fig:teaser} overviews the complete pipeline; \Cref{fig:generation-overview} details the generative model architecture and notation.}

\subsection{Overview and Problem Formulation}

Our objective is to develop a generative model that produces realistic VR gameplay motions conditioned on in-game object configurations.
Specifically, we model the three-point ($3p$) poses of the VR headset and handheld controllers as they respond to dynamic game states.
The key challenge is not only to generate motions that align with gameplay objectives, but also to capture the diversity of player behaviors and styles present in real human gameplay.

To address this, we formulate the task as learning the following conditional distribution:
\begin{align}
    \ThreePChunkPred \sim p\left( \ThreePChunk \mid \ThreePHist, \GameSegment, \RefSegments \right) \label{eq:our_dist_overview}
\end{align}
where $\ThreePChunkPred$ denotes the generated future poses, $\ThreePHist$ represents the length-$h$ pose history, and $\GameSegment$ captures the current in-game object state (\textit{e.g.}, incoming notes, bombs, obstacles).
To capture player-specific characteristics, we condition on $\RefSegments$--short gameplay segments from the same player that serve as style exemplars.
Detailed notation for these components is introduced in \Cref{sec:training_setup}.

Our system follows a generate-simulate-select pipeline (\Cref{fig:teaser}):
First, we use a reference-conditioned Categorical Codebook Matching model (CCM, \Cref{sec:ccm}) to generate multiple candidate $3p$ motion trajectories.
Second, we evaluate these candidates using TorchSaber, a custom GPU-accelerated game simulator (\Cref{sec:selection}).
Finally, we select the trajectory that achieves the highest simulated gameplay performance.
This model can be deployed autoregressively to produce gameplay sequences of arbitrary length.

\noindent \textit{Contextual exemplars.}
While using exemplar motion as a style signal is well-explored in prior work (\textit{e.g.}, \citen{xia2015realtime, holden2016deep,wen2021autoregressive,guo2024generative,raab2024monkey,sawdayee2025dance,kim2025personabooth}), we extend the idea in two ways.
First, each style reference in $\RefSegments$ includes not only the player's motion but also the game state $\GameSegment$ during that motion—allowing the model to learn how players respond to different gameplay situations rather than just their unconditional movement patterns.
Second, we condition on multiple reference examples ($N_\text{ref}$) simultaneously to capture consistent behavioral patterns across varied scenarios.
Our experiments (\Cref{sec:experiments}) further validate the use of contextual exemplars.

\subsection{Reference-Conditioned Categorical Codebook Matching} \label{sec:ccm}

Please refer to \Cref{fig:generation-overview} for the visualization of our codebook matching process.
To model the conditional distribution in \Cref{eq:our_dist_overview}, we implement Categorical Codebook Matching (CCM, \citen{starke2024categorical}).
We use a Gumbel-Softmax variational auto-encoder (GS-VAE) to encode each individual $3p$ trajectory, such that its embedding corresponds to the logits of a categorical distribution.
To support conditioning by reference, we extend \citen{starke2024categorical}'s CCM encoder block, introducing a style encoder $\StyleEncoder$ that embeds style reference examples; the resulting encoder block consisting of two Transformer-based models ($\StyleEncoder$ and $\GameEncoder$) is trained to predict the logits from the corresponding input conditions, \textit{i.e.}, $\ThreePHist$, $\GameSegment$, and $\RefSegments$.

More formally, the length-$T$ pose sequence $\ThreePChunk$ is mapped to the logits (\Cref{sec:training_setup} overviews feature notation and representations in more detail):
\begin{align}
\ThreePLogits = \mathcal{E}^{\text{3p}}\left( \ThreePChunk \right). \label{eq:3p_encoder}
\end{align}
The logits are reshaped into those of a joint categorical distribution consisting of $C$ channels, each with $D$ categories.
With Gumbel-Softmax sampling, we produce from this distribution one-hot sequences of length $C$, each row corresponding to an integer in $\{1, 2, \cdots, D\}$.
With straight-through estimation, the one-hot samples retain their gradients through the reparameterized Gumbel sampling procedure \cite{jang2017categorical}.
The resulting one-hot samples are then decoded into 
\begin{align}
\ThreePChunkPred = \mathcal{D}^\text{3p}\left( \text{GumbelSoftmax}\left(\ThreePLogits\right) \right). \label{eq:3p_decoder}
\end{align}
Hence, the GS-VAE's auto-encoding loss is simply the mean squared error (MSE) loss between the original and reconstructed input, \textit{i.e.}:
\begin{align}
    \ReconLoss &= \frac{1}{T} \big| \big| \ThreePChunk - \ThreePChunkPred \big| \big|^2
\end{align}
Our GS-VAE departs from the multi-layer perceptron (MLP) architecture of \citen{starke2024categorical} and instead utilizes a Transformer architecture \cite{vaswani2017attention} to better leverage the sequential nature of the $3p$ pose sequence.
The details of our Transformer-based GS-VAE are discussed in \Cref{appendix:architecture} in the Appendix.

To match our conditioning signals with a corresponding $\ThreePChunkPred$, we train an encoder block consisting of the style encoder $\StyleEncoder$ and the game segment encoder $\GameEncoder$.
$\StyleEncoder$ first embeds the reference examples into
\begin{align}
\StyleEmbeddings = \StyleEncoder\left(\RefSegments \right). \label{eq:style_encoder}
\end{align}
$\GameEncoder$ predicts the logits 
\begin{align}
\GameLogits = \mathcal{E}^\text{game}\left( \ThreePHist, \GameSegment, \StyleEmbeddings \right). \label{eq:game_encoder}
\end{align}
Taking into account the varying number of objects captured in purview, as well as their continuous timing values (when each object reaches the player), we also employ a Transformer architecture for both $\StyleEncoder$ and $\GameEncoder$.
That is, all of the objects and poses in $\ThreePHist$, $\GameSegment$, and $\RefSegments$ are projected to the $d$-dimensional input space via an MLP prepared for each domain (among notes, bombs, obstacles, poses).

The resulting embeddings then receive position encoding; importantly, position encoding for game objects is based on their timing, which informs their temporal arrangements (as opposed to indices, which only inform their order of appearance).
Pose sequences, however, receive position encoding based on index per usual, as the time between poses is constant.
We register an all-zero sentinel token \cite{devlin2019bert, dosovitskiy2020image} concatenated to the end of the projected sequences and feedforward its corresponding final embedding through an MLP head to produce $\StyleEmbeddings$ and $\GameLogits$.

\noindent \textit{Codebook Matching via Jensen-Shannon divergence.} Per the usual CCM formulation \cite{starke2024categorical}, we encourage the similarity of the two categorical distributions $\ThreePLogits$ and $\GameLogits$. A sufficient match between the two enables the conditional generation in \Cref{eq:our_dist_overview} by connecting $\mathcal{E}^{\text{game}}$ to $\mathcal{D}^{\text{3p}}$ (\textit{i.e.}, omitting $\mathcal{E}^\text{3p}$ at inference).  

While \citen{starke2024categorical} minimize L2-distances between the $C \times D$ one-hot samples coming from the two distributions, we instead minimize the Jensen-Shannon divergence (JSD) between the distributions.
Although this alternative loss introduces a loss weight hyperparameter to training, we note that this approach is more grounded in principle for matching two categorical distributions.
The JSD-based matching loss is computed according to:
\begin{align}
\MatchingLoss &= D_{\text{KL}}\left(\ThreePLogits \big| \big| P \right) + D_{\text{KL}}\left( \GameLogits \big| \big| P \right) \label{eq:jsd}
\end{align}
where $D_{\text{KL}}$ is Kullback-Leibler divergence between two distributions and $P = \frac12 \left( \ThreePLogits + \GameLogits \right)$.
Then, the final loss function is simply the weighted sum of the two loss terms:
\begin{align}
    \mathcal{L} &= \ReconLoss + \lambda_{\text{Match}} \cdot \MatchingLoss
\end{align}
We find $\lambda_{\text{Match}}=$1e-4 to be effective for our experiments.

\subsection{Inference: Candidate Trajectory Selection} \label{sec:selection}

Incorporating rejection sampling in generative motion planning has been explored more recently, \textit{e.g.}, in \citen{starke2024categorical,ioannidis2025diffusion}.
Leveraging the variational capabilities, trained generative models can typically produce multiple viable candidate outputs, among which a suitable one is selected.
While evaluating generated motion plans can be generally non-trivial, often calling for learned proxies (\textit{e.g.}, \citen{ioannidis2025diffusion}), gameplay is fortunately well-suited for simulation.
Consequently, we take the approach of \emph{combining game simulation with GS-VAE's sampling procedure} for generative motion planning.
Given the predicted logits $\GameLogits$, we sample $N_\text{traj}$ \emph{candidate} $3p$ trajectories:
\begin{align}
    \mathbf{\hat p}_{t:t+T}^{:N_\text{traj}} &\sim \mathcal{D}^\text{3p}\left(\text{GumbelSoftmax}\left( \GameLogits \right)\right)
\end{align}
The candidate trajectories are evaluated based on the input game state $\mathbf{x}^\text{game}_t$, which contains sufficient information for simulating the game forward for $T$ frames. The highest-scoring trajectory $\mathbf{\hat p}_{t:t+T}^{j^*}$ is then chosen as the final output, where: 
\begin{align}
    j^*
    &= \argmax_{j=1,2,\cdots,N_\text{traj}}\text{Evaluate}\left(
        \mathbf{\hat p}_{t:t+T}^{j},
        \GameSegment
        \right) \label{eq:evaluation}
\end{align}
Unlike \citen{starke2024categorical}, we use the entire $T$-frame prediction without intermediate re-planning; we find the motion quality for the $3p$ trajectories to be better overall when the output chunks are coherent and continuous.

\noindent \textit{TorchSaber: A GPU-accelerated Beat Saber simulator.}
As \emph{Beat Saber} remains a proprietary, closed-source game, evaluating scores for generated trajectories inside the real game would require interacting with the compiled game or building an additional integration layer.
We instead develop and use TorchSaber, a GPU-accelerated \emph{Beat Saber} simulator, as an alternative (details in \Cref{appendix:torchsaber} in the Appendix).
TS produces simplified, normalized proxy scores that omit parts of the official scoring (\textit{e.g.}, combos and some cut-angle terms), while still correlating strongly with real scores; evaluating the held-out human play data, we obtain Pearson's $r$=0.856 between TS and recoded \emph{Beat Saber} scores.

\noindent \textit{Reward function.}
Based on TorchSaber's simulation results, we assign a reward value $r_j$ to each candidate output $\mathbf{\hat p}_{t:t+T}^{j}$, as defined below (\Cref{appendix:reward_def} of the Appendix provides a detailed discussion of the reward terms):
\begin{align}
    r = r_{\text{TS}}
            - \lambda_{\text{Bomb}}\cdot r_{\text{Bomb}}
            + \lambda_{\text{Obstacle}}\cdot r_{\text{Obstacle}}
\end{align}
\begin{itemize}[leftmargin=*]
    \item $r_{\text{TS}} \in [0, 1]$ is the TS score, calculated for the $3p$ poses and colored notes.
    \item $r_\text{Bomb} \in [0, 1]$ is the bomb penalty, computed as the ratio between the number of bomb hits and the number of appearing bombs.
    \item $r_\text{Obstacle} \in [0, 1]$ is the obstacle distance bonus, computed as the minimum distance between any appearing obstacles' $yz$-bounds and the head $yz$-positions. When the head collides with an obstacle, the value becomes negative.
    \item The $\lambda_*$ values are weights corresponding to each term; see \Cref{table:hyperparameters} of the Appendix for the values used in our experiments.
\end{itemize}

\subsection{Training Setup and Features}
\label{sec:training_setup}

Our system requires a diverse and extensive dataset to generalize properly. For \emph{Beat Saber}, BOXRR-23 \cite{nair2023berkeley} and BeatSaver \cite{BeatSaver} datasets provide this scale and diversity, featuring millions of replay sequences from hundreds of thousands of players and maps.
Given the dataset's characteristics and quality issues, we apply quality control (QC) measures to ensure that our model is trained on correctly aligned, high-quality data (detailed in \Cref{appendix:QC} in the Appendix).
Our post-QC dataset then consists of 2,357,627 replay sequences from the pool of 85,497 maps and 71,070 players.
By aligning the gameplay data of BOXRR-23 with the map data of BeatSaver \cite{BeatSaver}, we produce supervised training samples.

\noindent \textit{Pose representation.}
The 27-dimensional $3p$ pose vector $\mathbf{p}$ (at each timestep) is constructed by concatenating the global $xyz$-coordinates and 6-dimensional orientation \cite{zhou2019continuity} for each of the three parts: the headset and two handhelds.

\noindent \textit{In-game object representation.}
The in-game object configurations follow the Beat Saber Modding Group format \cite{BSMGWiki}.
Each colored note $\ColoredNote$ is represented by grid position indices, note color, and cut direction (dim=4).
Each bomb $\BombNote$ and obstacle $\Obstacle$ is represented by grid position indices; obstacles additionally include width, height, and depth (dim=2 and 5, respectively).
Each object is paired with its absolute timestamp, indicating when it reaches the player.

We use a \emph{lookahead} horizon: only up to $n$ objects within $s$ seconds are considered at timestep $t$.
For each object captured in lookahead, we compute its relative timing, \textit{i.e.}, the difference between the current game timestamp and the object's absolute timestamp.
Using subscript $t$ to denote objects in purview at $t$ and superscript ":n" to denote a sequence containing up to $n$ objects, \textit{e.g.}, $\ColoredNotes = \left( \mathbf{c}_t^1, \mathbf{c}_t^2, \cdots, \mathbf{c}_t^{k_{\mathbf{c}_t}} \right), k_{\mathbf{c}_t} \leq n$, we construct the game state observation:
\begin{align}
\GameSegment = \left( \ColoredNotes \ \BombNotes \ \Obstacles \right).
\end{align}
We pair $\mathbf{x}^{\text{game}}_t$ with $\ThreePHist$ and $\ThreePChunk$.

\noindent \textit{Style reference representation.}
A gameplay reference example is similarly constructed as $\mathbf{x}^{\text{ref}}_{t_i^*} = \left( \RefColorNotes \ \RefBombNotes \ \RefObstacles \ \RefThreePHist \ \RefThreePChunk \right)$ for some timestep $t_i^*$, sampled from a random replay sequence $i$ of the same player.
Using the superscript ":$N_\text{ref}$" to represent a collection of size $N_\text{ref}$, we denote a set of gameplay reference examples as:
\begin{align}
\RefSegments = \left( \mathbf{x}^{\text{ref}}_{t_1^*}, \mathbf{x}^{\text{ref}}_{t_2^*}, \cdots, \mathbf{x}^{\text{ref}}_{t_{N_\text{ref}}^*} \right)
\end{align}
with a random time $t^*_i$ for each player-specific reference $i \in \left\{1, 2, \cdots, N_\text{ref}\right\}$.

\section{Experiments} \label{sec:experiments}
Below, we validate Robo-Saber’s behavior on held-out validation data, using the TS scores of both Robo-Saber and human players as the main quantitative measure.
Importantly, the human $3p$ trajectories are fed into TS to re-compute their scores, so that the scores can be directly compared.

Our primary research question is: \emph{Can Robo-Saber enable automated and personalized VR playtesting?} More specifically:
\begin{itemize}[leftmargin=*,label={}]
    \item \ref{sec:Q1}.~\nameref{sec:Q1}
    \item \ref{sec:Q2}.~\nameref{sec:Q2}
    \item \ref{sec:Q3}.~\nameref{sec:Q3}
    \item \ref{sec:Q4}.~\nameref{sec:Q4}
\end{itemize}
To briefly summarize the results, we find that Robo-Saber is capable of elite-level \emph{Beat Saber} gameplay and the game simulation improves generalization.
The model's generated gameplay, emulating held-out players, achieves a strong correlation of $r=0.789$ with the ground-truth scores on held-out maps. 
Note that this section focuses on the $3p$ (the headset and handhelds) motion generation, which is suitable for VR player modeling where the player's full-body movement does not have to be inferred or visualized. Extending Robo-Saber with physics-based full-body movement generation is explored in \Cref{sec:robotic_agent}.

\noindent \textit {Comparing model variants.} Note that our experiments compare three Robo-Saber variants: (1) a \emph{reference-agnostic} model that does not utilize $\RefSegments$ ($\NRefs=0$), (2) a \emph{reference-aware} model with $\NRefs=1$, and (3) a reference-aware model with $\NRefs=5$.
For evaluating statistical significance, we report the mean $\pm$ standard error in figure legends, wherever appropriate, and apply Wilcoxon's signed-rank test \cite{wilcoxon1992individual} to compute the $p$-value.

\noindent \textit {Holdout data preparation.}
We prepare holdout data as follows.
First, from the post-QC dataset prepared from BOXRR-23, we exclude the most popular 1\% of the maps, as measured by the number of unique players who have played them. This prevents the model from overfitting to most typical patterns and enables a stringent test for generalization. Our visual results are rendered on a selected subset of these 847 most popular maps.

From the remaining maps and the player population, we hold out 10\% from each, selecting uniformly at random.
The holdout data consists of 66,397 player-map pairs between 7,107 players and 8,465 maps.
For our quantitative results, we report the performance on this set.
Importantly, neither any player nor any map in the holdout appears during training.

\makeatletter
\renewcommand{\thesubsection}{Q1}
\makeatother

\subsection{\ul{Is Robo-Saber capable of general \emph{Beat Saber} gameplay?}}
\label{sec:Q1}

\begin{figure}
    \centering
    \includegraphics[height=0.6\linewidth,valign=B]{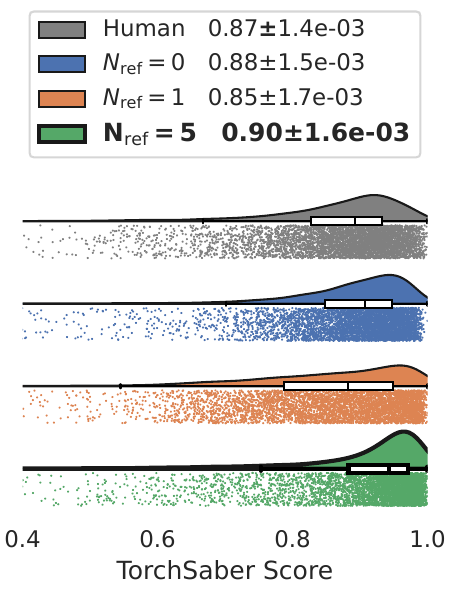}
    \hfill
    \includegraphics[height=0.6\linewidth,valign=B]{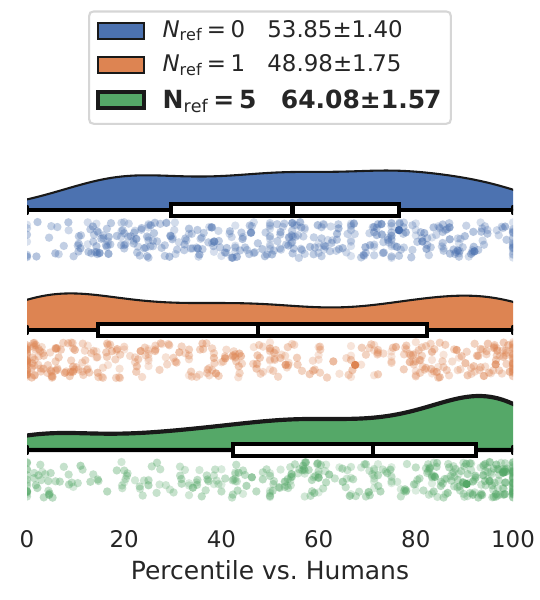}
    \caption{
    Comparing Robo-Saber's performance to that of human players. Robo-Saber trajectories are produced by using $\NRefs=5$ segments from elite (top 5\%) players.
    \textbf{Left:} Human and Robo-Saber TS score distributions across all held-out maps are shown as raincloud plots.
    \textbf{Right:} Robo-Saber's performance relative to human players is quantified as score percentiles. To compute the percentiles, for each difficulty level, we select the 100 most-played maps (400 total) and compare Robo-Saber's score against human scores in each map. The number of human scores available in each map is visualized by opaqueness (max=388, min=3)
    The performance statistics are summarized as mean $\pm$ standard error.
    The reference-aware model with 5 reference segments from top players ($\NRefs=5$) outperforms humans on average.
    }
    \label{fig:robo-vs-humans-scatter}
\end{figure}

We benchmark Robo-Saber's performance against human players on the held-out test set.
To ensure we elicit the best performance from the reference-aware model variant, we sample player-specific $\RefSegments$ from the top 5\% of held-out players.
We use the same $\RefSegments$ for simulating the same player across different maps.
While it is possible for a player to have varying skill levels across gameplay records, we simplify by excluding skill level variation from consideration, as players tend to report their best-effort attempt on leaderboards.
\textbf{Please refer to the supplementary video for qualitative results featuring $3p$ \emph{and} whole-body gameplay animations.}

\noindent \textit{Robo-Saber vs. humans.}
As shown in \Cref{fig:robo-vs-humans-scatter}, Robo-Saber's TS scores closely compete with those of human players; the $\NRefs=5$ variant outperforms human players on average at slightly above the 60th percentile.
Interestingly, $\NRefs=1$ results in lower overall performance than $\NRefs=0$, perhaps because a single randomly chosen reference segment may fail to adequately represent gameplay skills.
For example, the single reference segment may be an idle standing motion, which is insufficient to signal to the model that high-performing samples should be produced.
Providing a stronger style-conditioning signal, \textit{i.e.}, via more exemplars, remedies this and produces evidently better performance. \\ \\
\noindent \textit{Performance across difficulty levels.}
The top part of \Cref{fig:physics_addendum} shows the breakdown of percentile distribution across difficulty levels, revealing that Robo-Saber performs better than the 50th percentile against human scores up to the Expert difficulty.
However, interpreting performance on Expert+ difficulty should take into account the heavy skew towards elite-level players.

\makeatletter
\renewcommand{\thesubsection}{Q2}
\makeatother

\subsection{\ul{Does incorporating TorchSaber impact generalization?}} \label{sec:Q2}

\begin{figure}
    \centering
    \includegraphics[width=0.99\linewidth, clip,trim={0 0 0 0}]{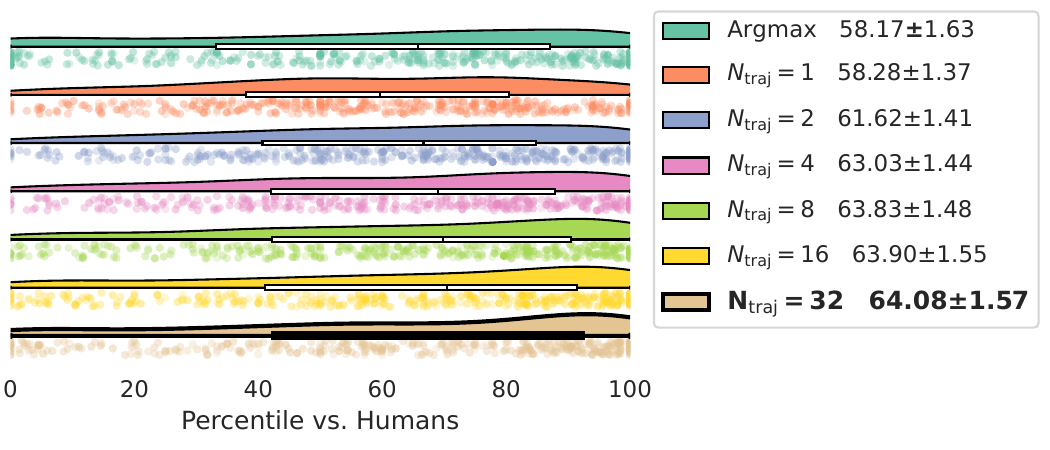}
    \caption{
    Sampling-based candidate trajectory selection improves performance compared to using deterministic Argmax selection for GS-VAE during inference.    
    The percentiles are evaluated on the same held-out maps as \Cref{fig:robo-vs-humans-scatter} with $\NRefs=5$ and $\RefSegments$ sampled from elite (top 5\%) players.
    The TS score percentiles are summarized as mean $\pm$ standard error for each configuration. 
    }
    \label{fig:selection_vs_argmax}
\end{figure}

Our candidate selection strategy uses rejection sampling, with TS as a simulator to evaluate each candidate. This alignment step impacts how well our model generalizes. Choosing only the trajectory corresponding to the peak of the output logit (using argmax actions) can lead to worse performance, because the model may not always produce correct logits at each stage, especially in previously unseen game states.
To address this, our system generates $N_{\text{traj}}$ sample trajectories and selects the best one after evaluation (see \Cref{fig:cand}).
This approach closely resembles black-box model predictive control frameworks, which utilize a simulator to propose and refine candidate motion plans before selection.
The effectiveness of this approach is evident in \Cref{fig:selection_vs_argmax}, where rollouts with candidate trajectory selection consistently outperform argmax rollouts, indicating stronger generalization.
Notably, larger $N_{\text{traj}}$ results in better performance in general.
Using Wilcoxon's signed-rank test, we obtain $p < 0.005$ comparing the percentiles between argmax and $N_{\text{traj}}=32$.
These results are consistent with the findings of \citen{starke2024categorical}, who also examined inference-time candidate selection. While their work focused on the continuity of generated full-body motion, our approach specifically targets optimizing gameplay.

\begin{figure}
    \centering
    \includegraphics[width=0.99\linewidth,clip,trim={0 10 0 10}]{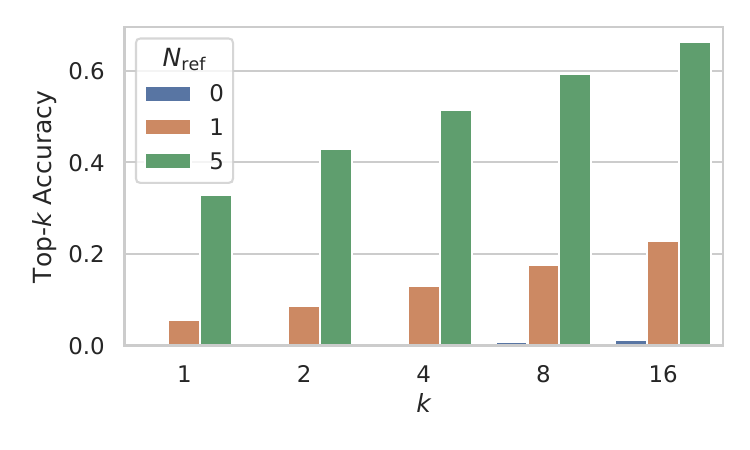}
    \caption{
    Quantifying Robo-Saber's ability to produce trajectories consistent with the style reference.
    The oracle player classifier's top-$k$ accuracy measures how well the generated $3p$ trajectories are recognized.
    Adding style reference segments clearly improves the recognizability.
    }
    \label{fig:sticking}
\end{figure}

\makeatletter
\renewcommand{\thesubsection}{Q3}
\makeatother

\subsection{\ul{Can Robo-Saber reproduce reference skill and style?}} \label{sec:Q3}

\begin{figure}
    \centering
    \includegraphics[width=0.99\linewidth]{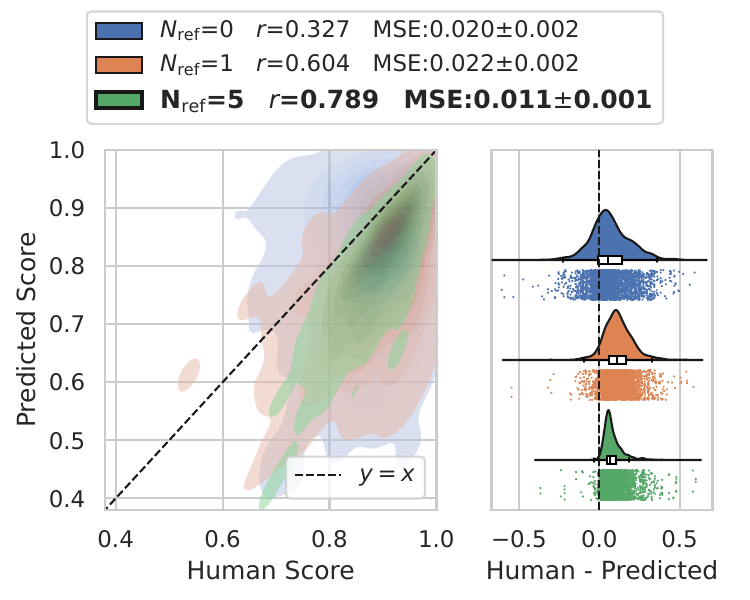}
    \caption{Quantifying Robo-Saber variants' calibration to the skill levels of human players, measured by the Pearson correlation ($r$) between Robo-Saber and human players' performance on held-out maps. \textbf{Left:} Densities of points comparing robot (predicted) scores against ground truth human scores. The reference-aware Robo-Saber with $\NRefs=5$ achieves a strong correlation of $r=0.789$. \textbf{Right:} Distributions of score differences between humans and Robo-Saber variants.}
    \label{fig:skill-calibration}
\end{figure}

Our model is tasked with recognizing player-specific characteristics from the input references and incorporating them into gameplay.
A well-calibrated model would allow one to simulate a player's gameplay on entirely new game content from using a few reference gameplay segments.
In other words, we aim to validate whether the model emulates a player's skill level and movement patterns, replicating the \emph{personality} of that player.

\noindent \textit{Calibration to skill level.}
Our results suggest that our model calibrates to the skill level represented by the input $\RefSegments$.
For a given player-map, a well-calibrated model should perform similarly to the given player; that is, conditioning on novice examples should yield low scores and expert examples high scores.
This is illustrated in \Cref{fig:skill-calibration}: compared to the reference-agnostic baseline ($\NRefs=0$), our reference-aware model variants more accurately approximate the ground truth human performance, resulting in moderate correlations and residuals distributed closer to 0.
Unsurprisingly, the signals in $\RefSegments$ get clearer with a larger $\NRefs$, resulting in a stronger correlation in TS scores and generally smaller score differences.

\noindent \textit{Calibration to movement patterns.}
Individual players' movement patterns contain rich information, such as spatial positioning habits and various biases in movements--\citen{nair2023unique} and \citen{nair2024deep}, for example, showed that these traits can identify which player has produced the $3p$ trajectory.
Inspired by this, we systematically assess whether the generated output resembles the input exemplars.
We train an \emph{oracle player classifier}, which predicts held-out player labels from held-out gameplay segments.
The oracle's recognition indicates whether an output gameplay sequence resembles that of a specified player and thereby whether the generator utilizes the signals inherent to $\RefSegments$ to produce output consistent with them.

We quantify the output trajectories' recognizability with the oracle's top-$k$ classification accuracy; \textit{i.e.}, we sample $\RefSegments$ for a held-out player, generate a $3p$ trajectory using $\RefSegments$, compute the oracle classifier's logits from the trajectory, and assess whether the top $k$ values in the logits include the category corresponding to the held-out player's ID.
Intuitively, a higher top-$k$ accuracy value indicates higher similarity between the model's and the referenced player's movements.
The oracle uses a Transformer architecture similar to $\StyleEncoder$ in \Cref{eq:style_encoder}; see \Cref{appendix:architecture} in the Appendix for more details.
As shown in \Cref{fig:sticking}, the oracle's top-$k$ accuracy, across different values of $k$, is significantly higher when we condition Robo-Saber with style information, and the effect increases with $\NRefs$.

\begin{figure*}
\centering
\begin{tcolorbox}[colback=white, colframe=black, sharp corners, boxrule=1pt, after skip=0pt]
\par\vspace{1em}
\textline{\text{5 reference segments (4th includes obstacles)}}
\par\vspace{1em}
\includegraphics[width=0.14\linewidth,clip,trim={0 0 0 0}]{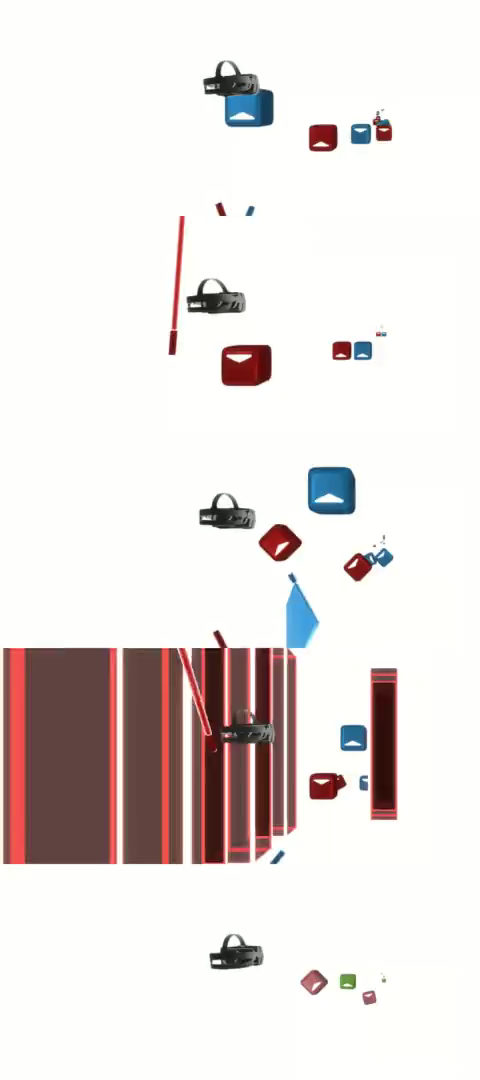}\hfill
\includegraphics[width=0.14\linewidth,clip,trim={0 0 0 0}]{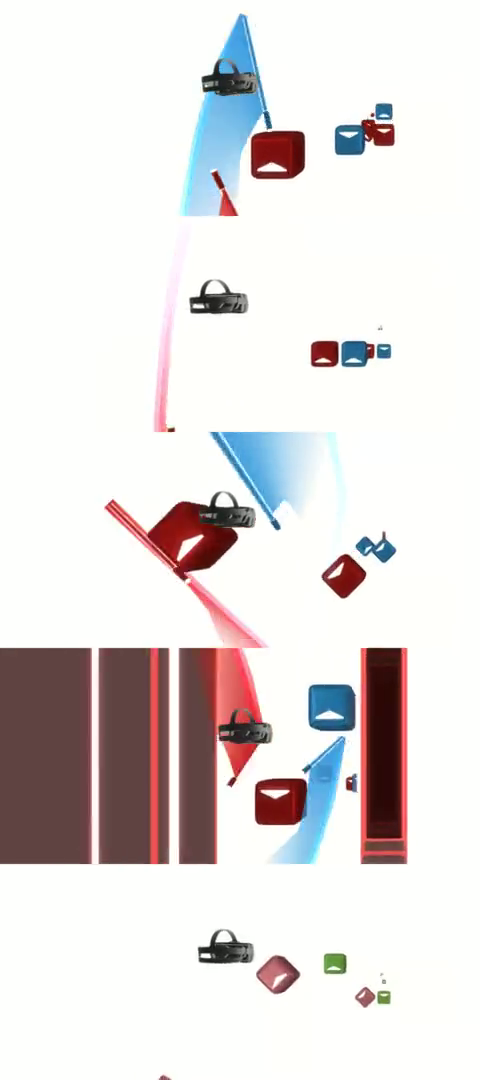}\hfill
\includegraphics[width=0.14\linewidth,clip,trim={0 0 0 0}]{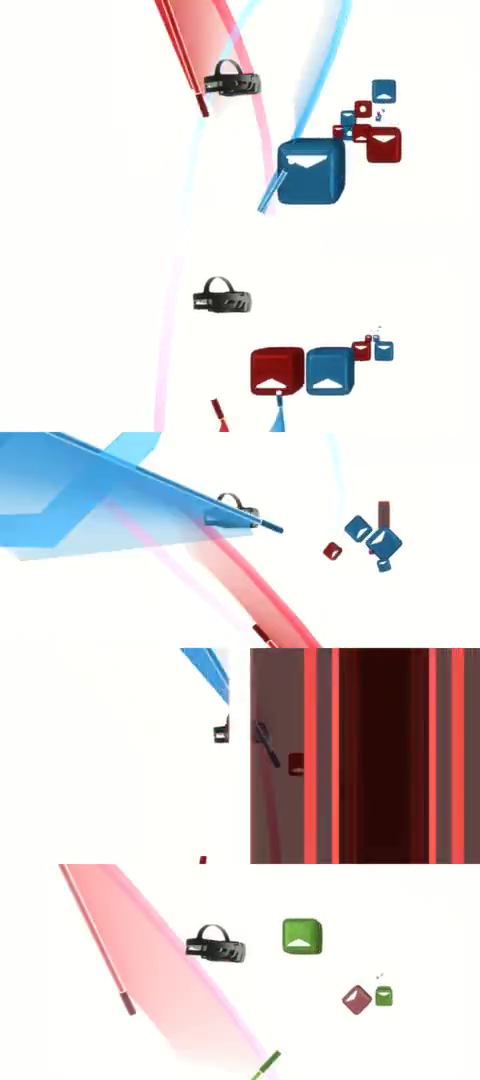}\hfill
\includegraphics[width=0.14\linewidth,clip,trim={0 0 0 0}]{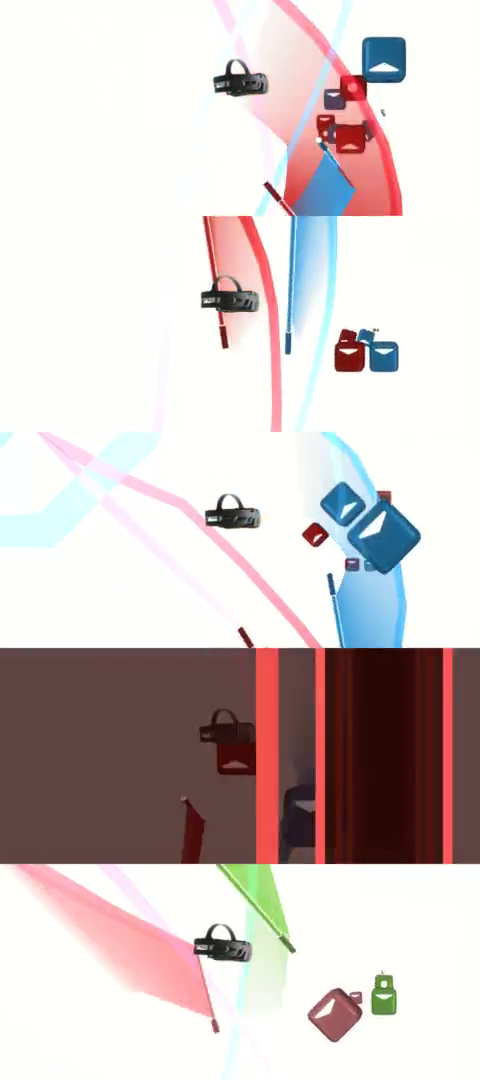}\hfill
\includegraphics[width=0.14\linewidth,clip,trim={0 0 0 0}]{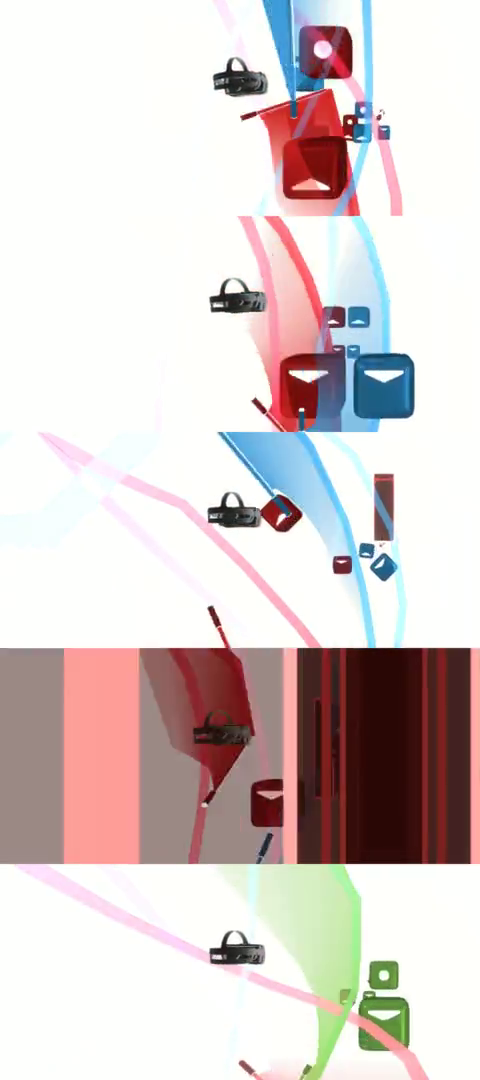}\hfill
\includegraphics[width=0.14\linewidth,clip,trim={0 0 0 0}]{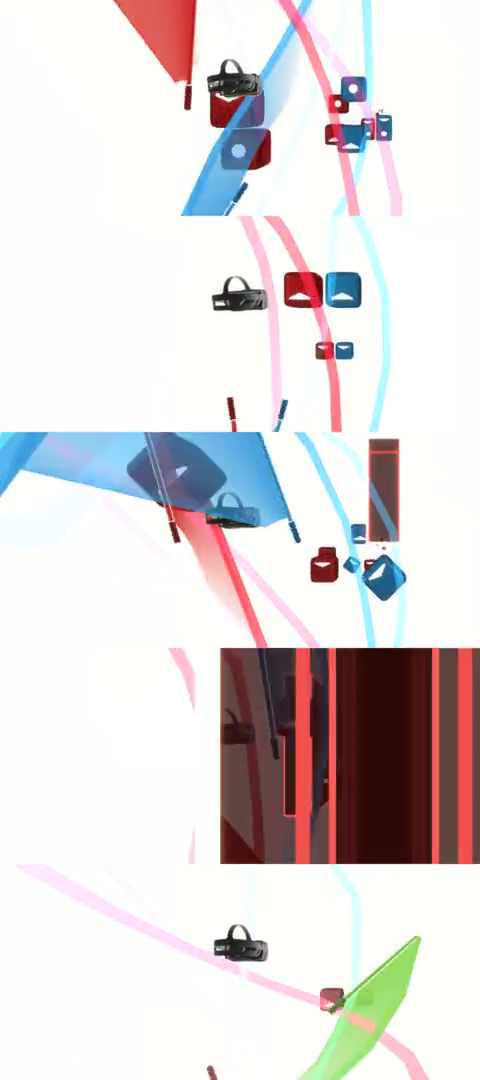}\hfill
\includegraphics[width=0.14\linewidth,clip,trim={0 0 0 0}]{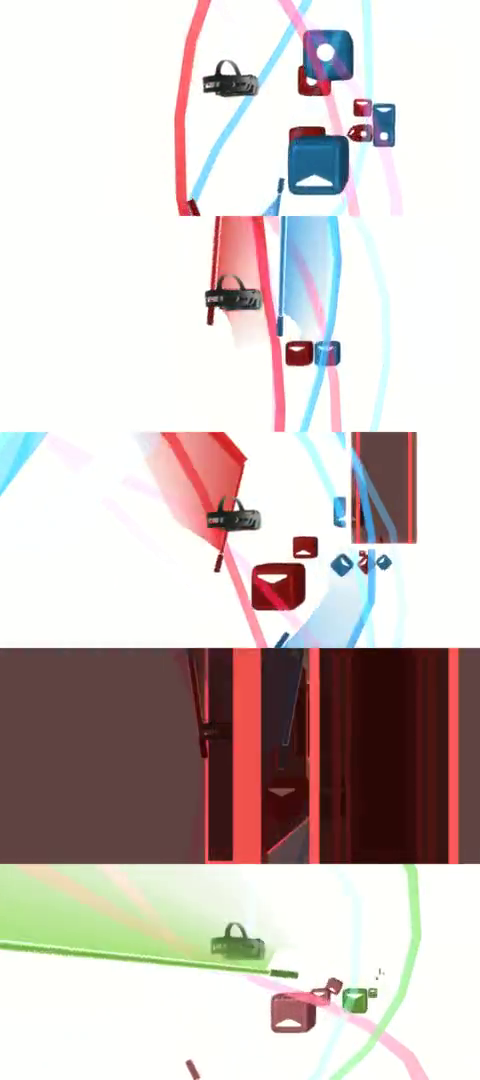}
\par\vspace{1em}
\textline{\text{Generated trajectory}}
\par\vspace{1em}
\includegraphics[width=0.14\linewidth,clip,trim={0 0 0 0}]{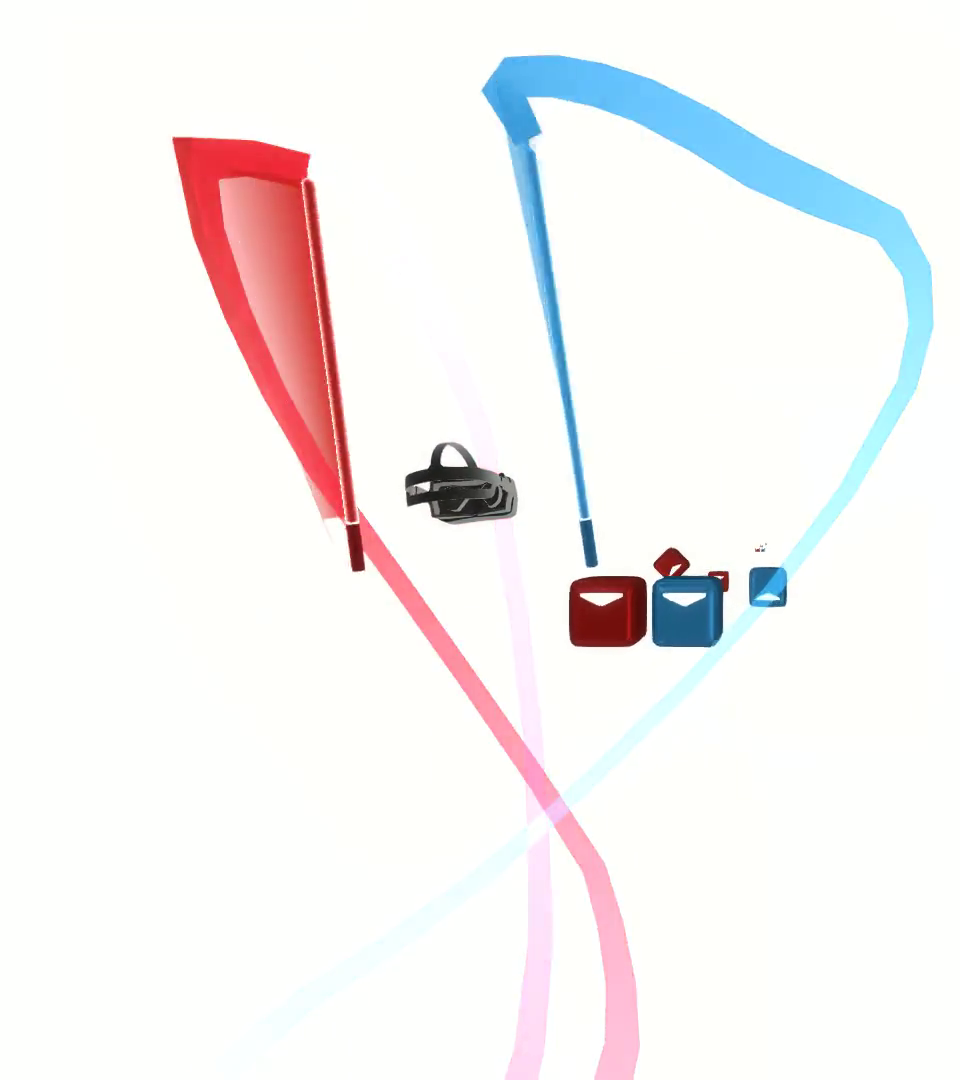}\hfill
\includegraphics[width=0.14\linewidth,clip,trim={0 0 0 0}]{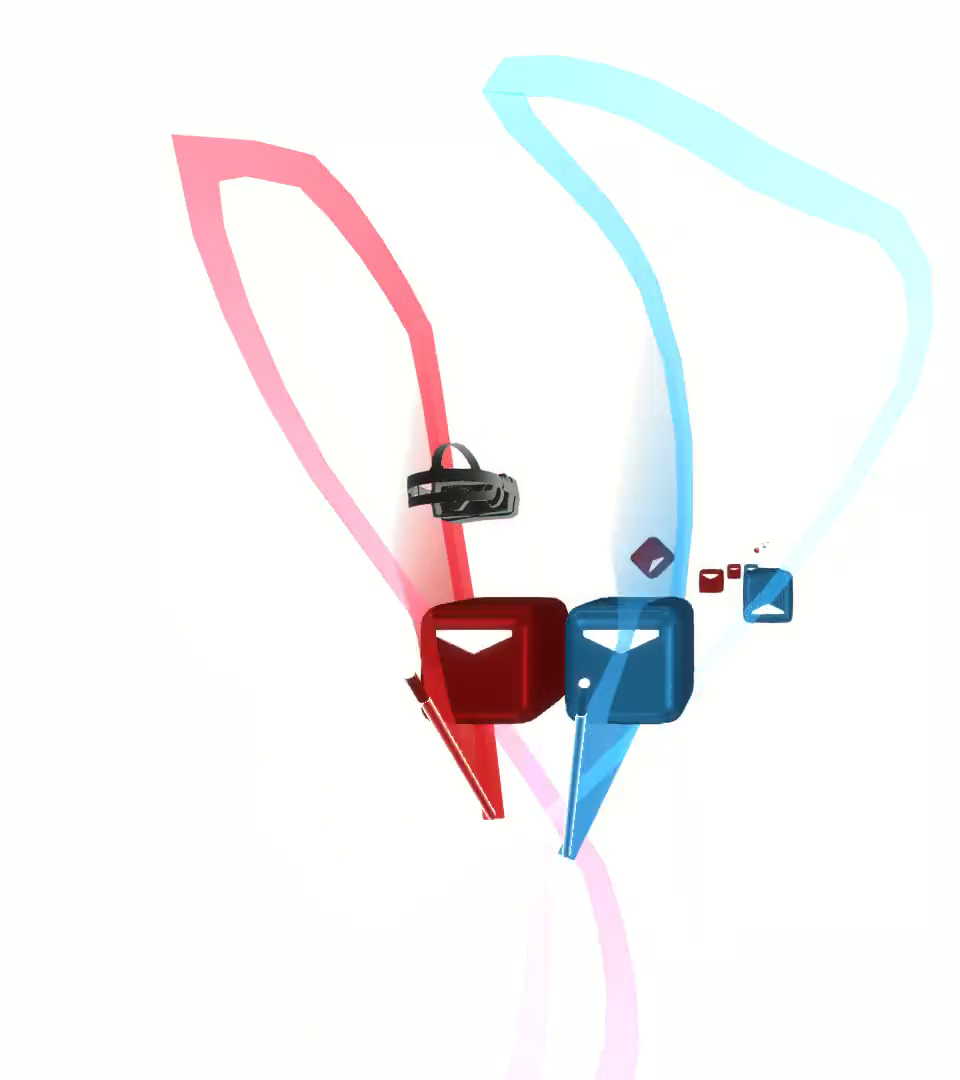}\hfill
\includegraphics[width=0.14\linewidth,clip,trim={0 0 0 0}]{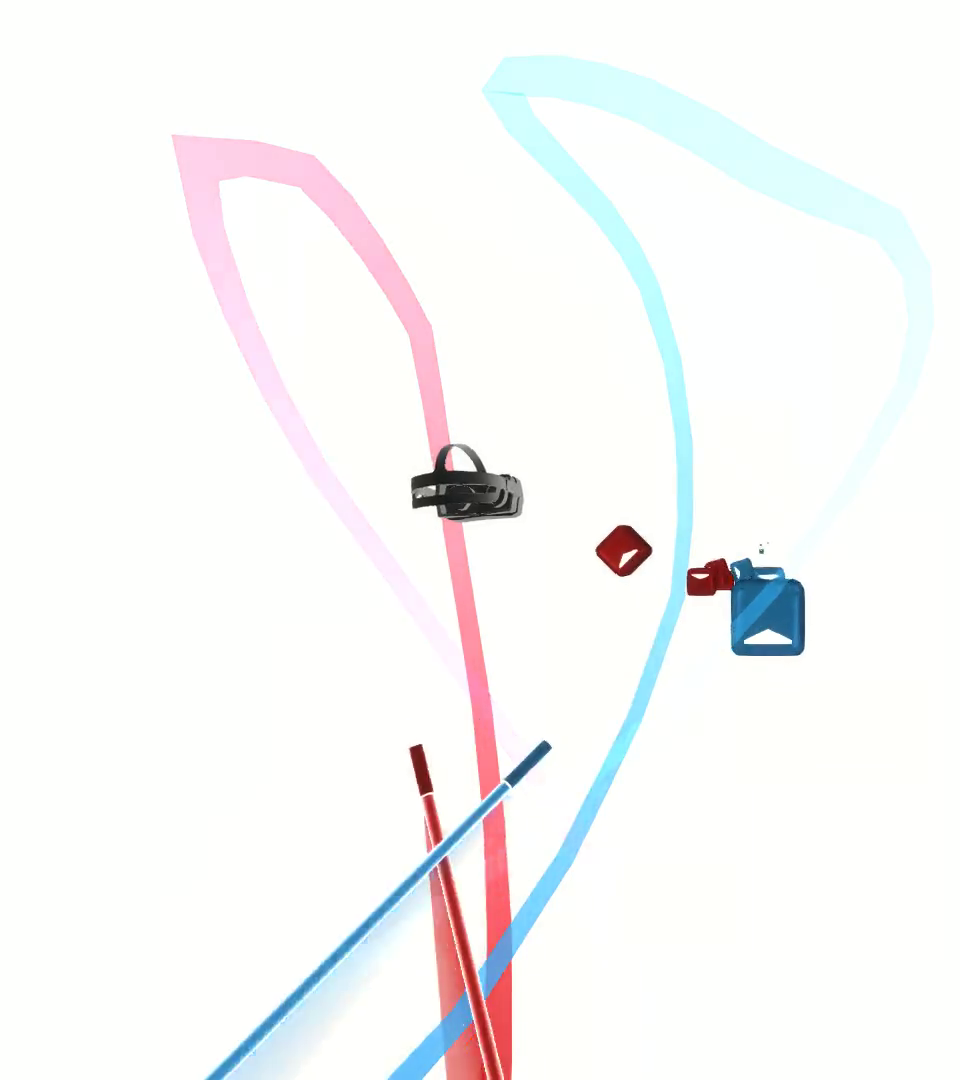}\hfill
\includegraphics[width=0.14\linewidth,clip,trim={0 0 0 0}]{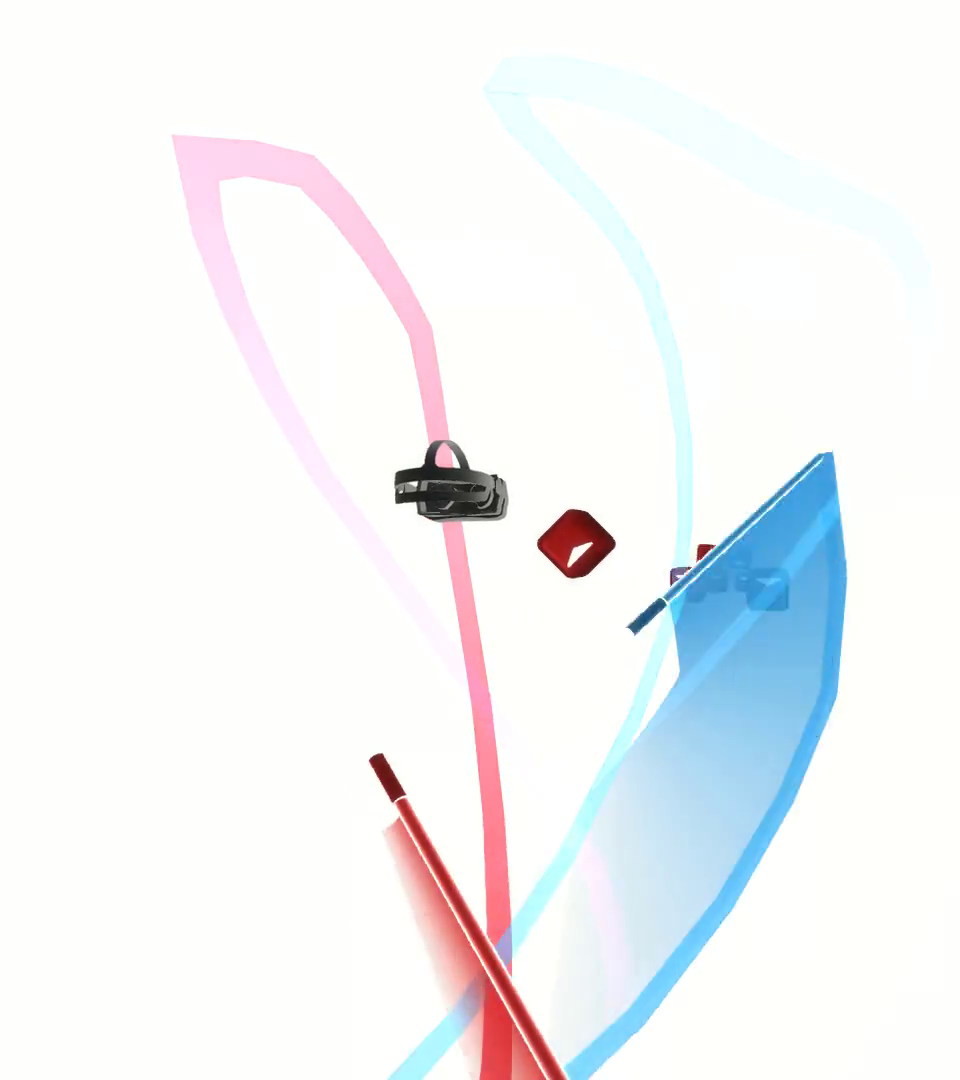}\hfill
\includegraphics[width=0.14\linewidth,clip,trim={0 0 0 0}]{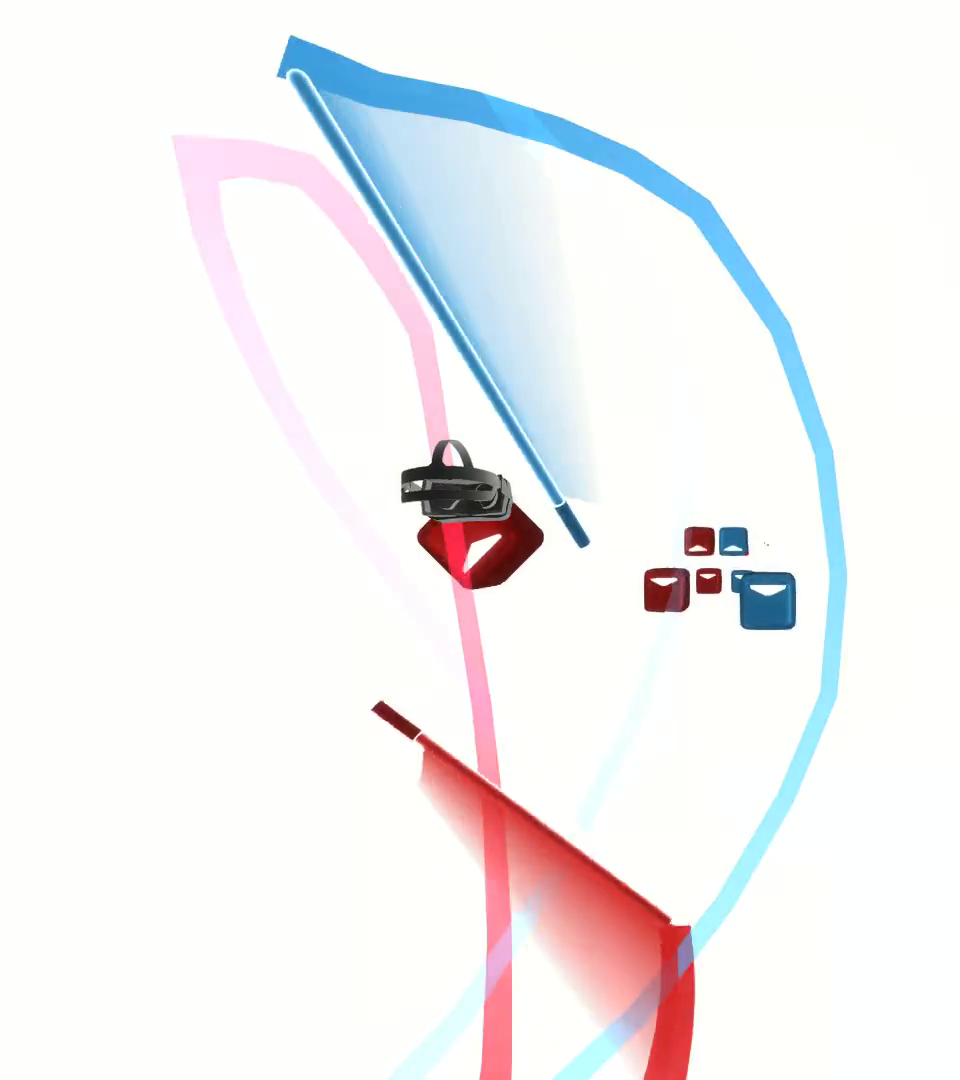}\hfill
\includegraphics[width=0.14\linewidth,clip,trim={0 0 0 0}]{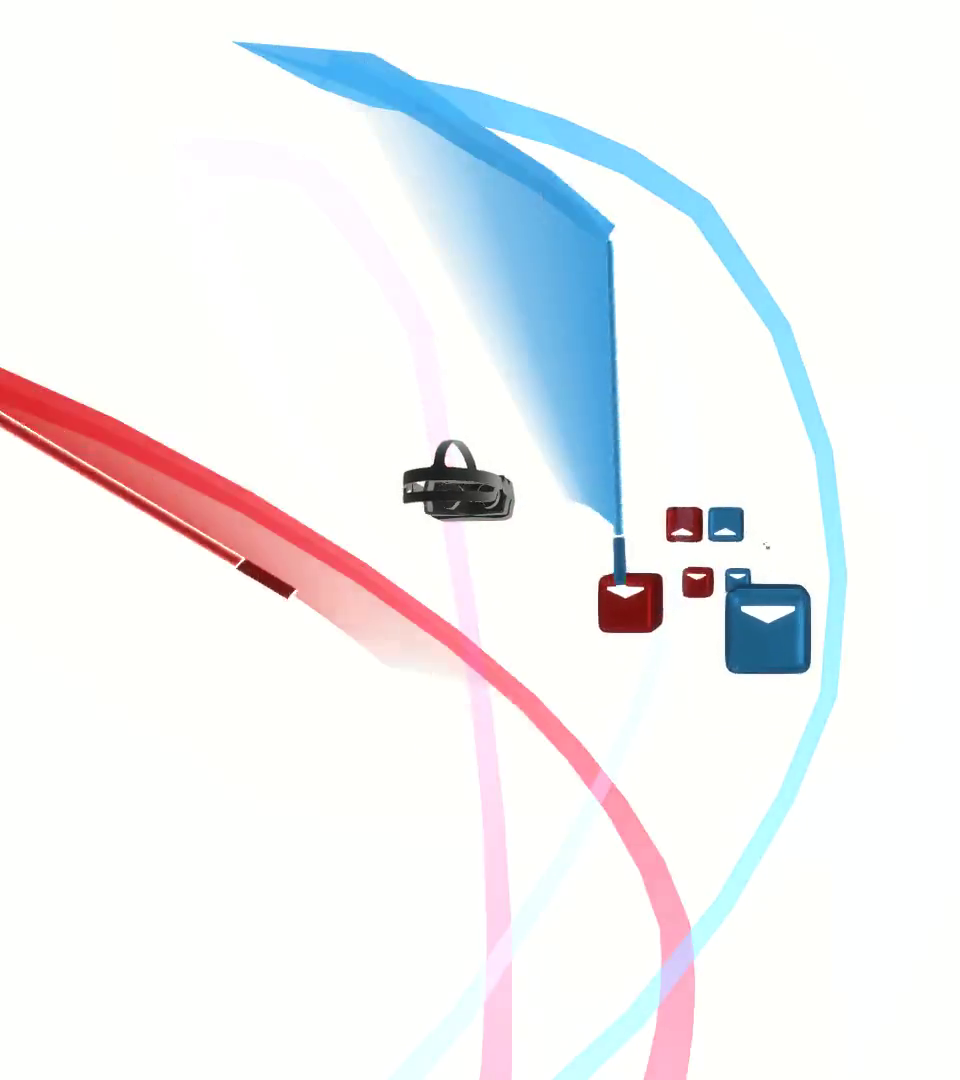}\hfill
\includegraphics[width=0.14\linewidth,clip,trim={0 0 0 0}]{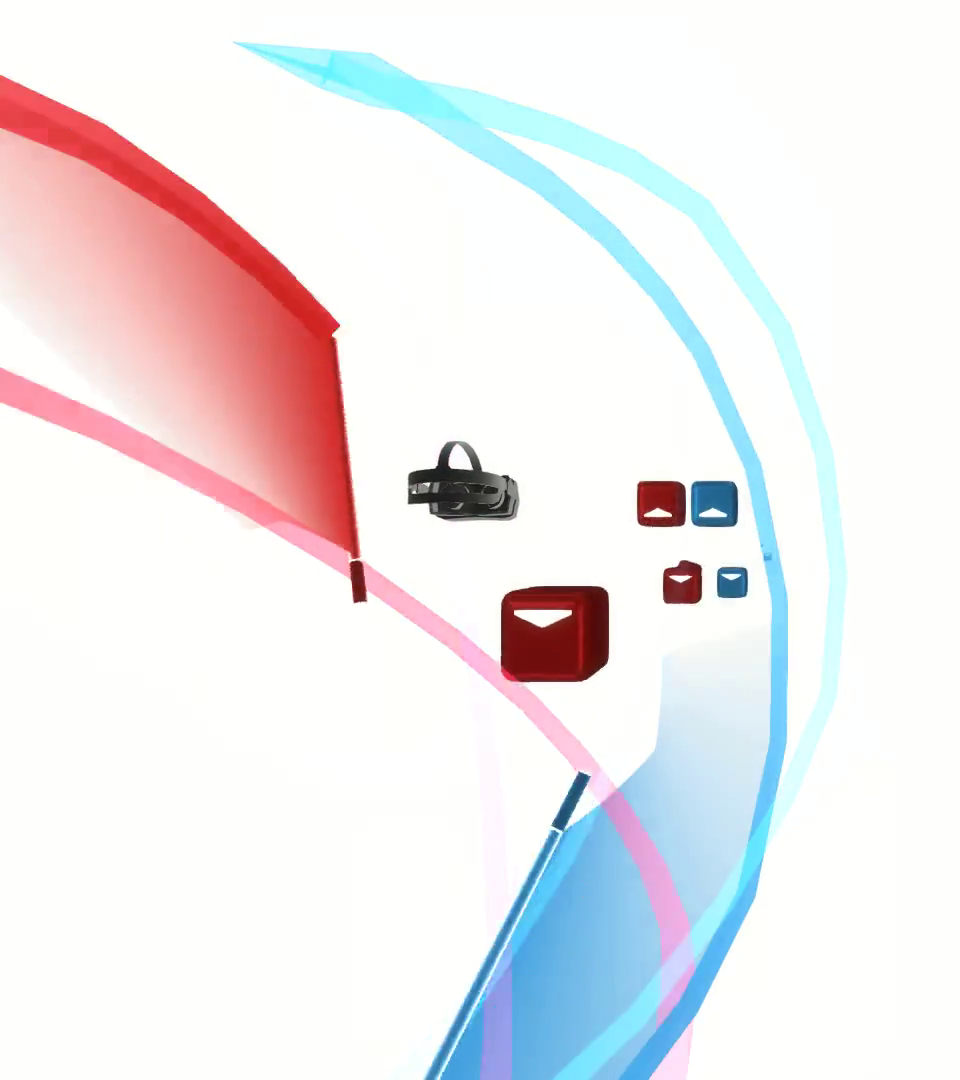}
\end{tcolorbox}
\begin{tcolorbox}[colback=white, colframe=black, sharp corners, boxrule=1pt, after skip=0pt]
\par\vspace{1em}
\textline{\text{5 reference segments (bottom includes obstacles)}}
\par\vspace{1em}
\includegraphics[width=0.14\linewidth,clip,trim={0 0 0 0}]{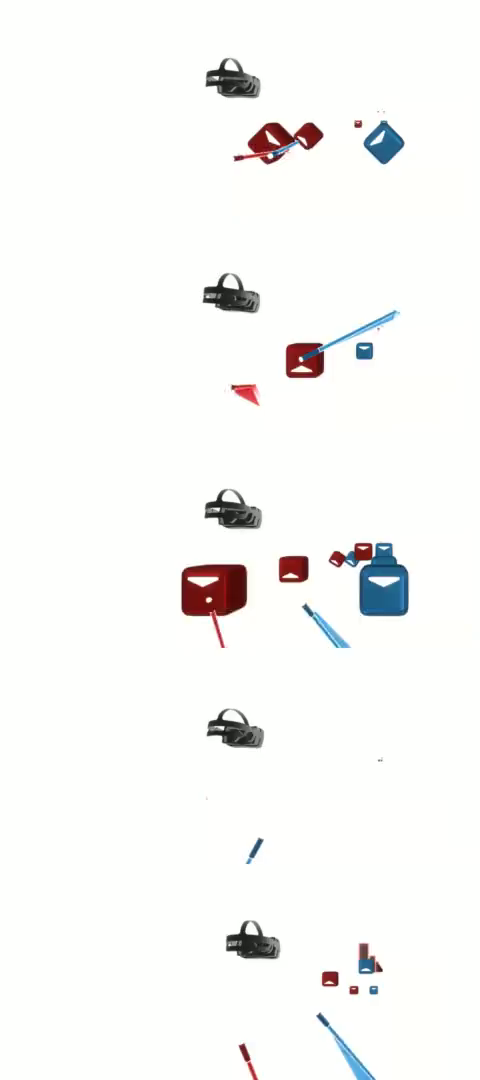}\hfill
\includegraphics[width=0.14\linewidth,clip,trim={0 0 0 0}]{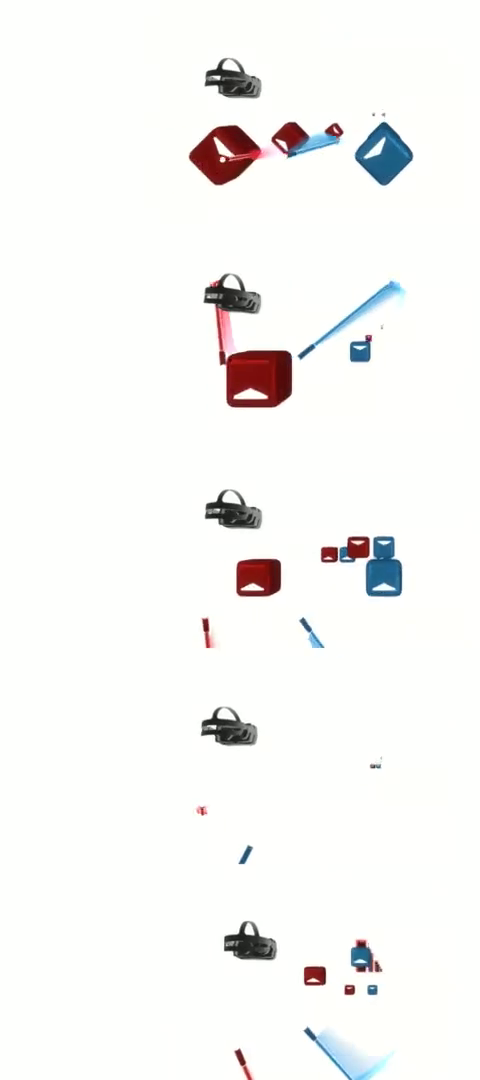}\hfill
\includegraphics[width=0.14\linewidth,clip,trim={0 0 0 0}]{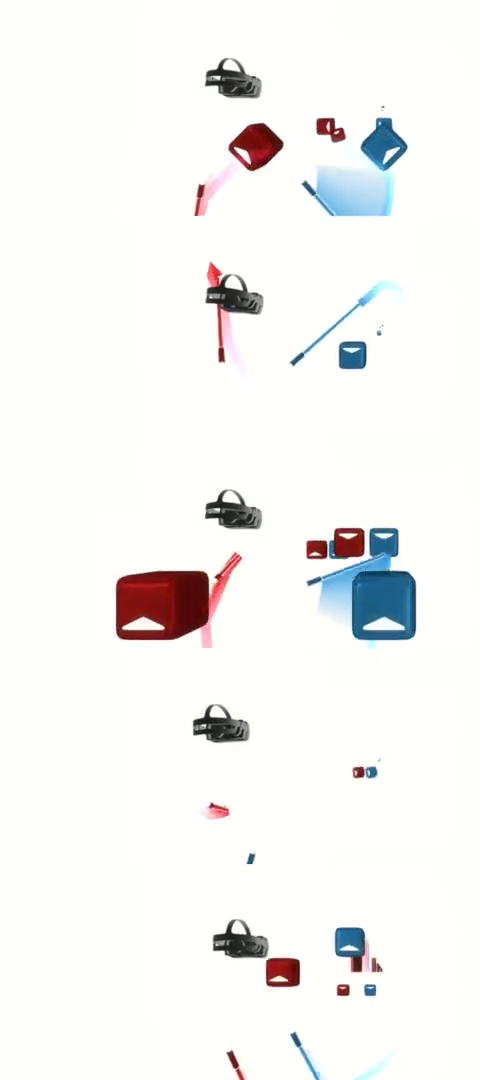}\hfill
\includegraphics[width=0.14\linewidth,clip,trim={0 0 0 0}]{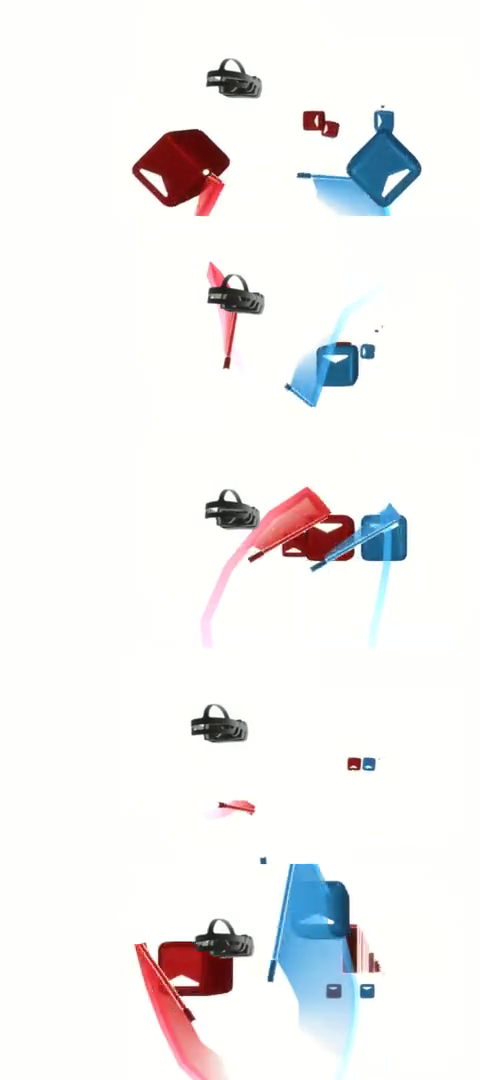}\hfill
\includegraphics[width=0.14\linewidth,clip,trim={0 0 0 0}]{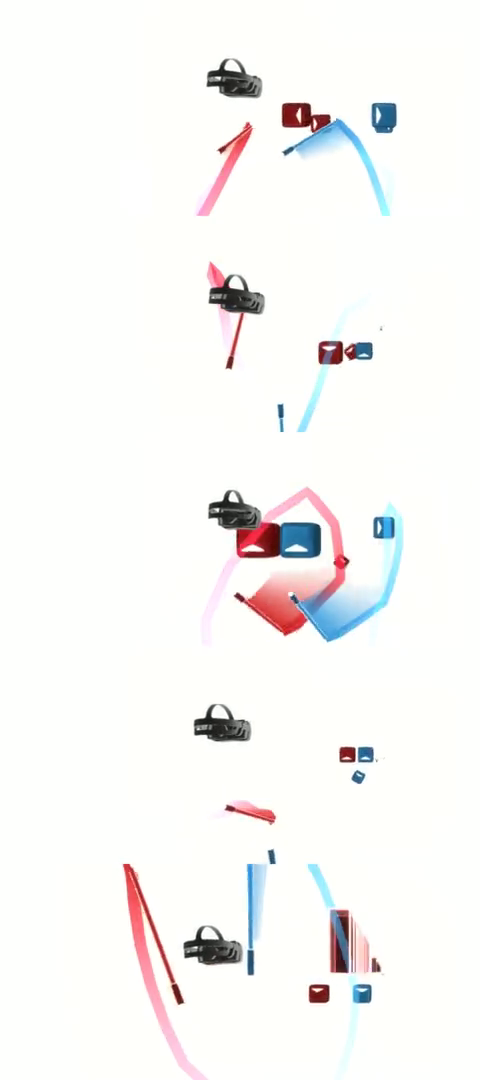}\hfill
\includegraphics[width=0.14\linewidth,clip,trim={0 0 0 0}]{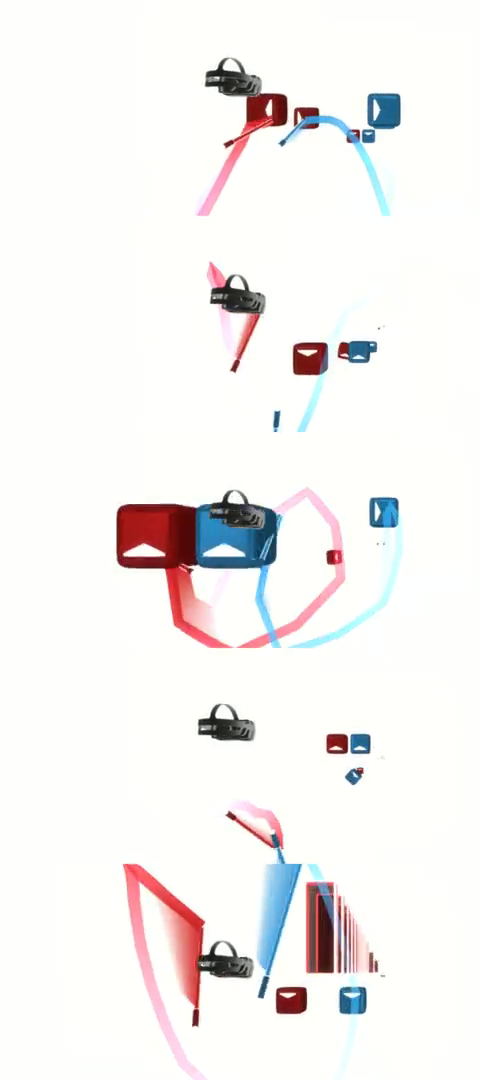}\hfill
\includegraphics[width=0.14\linewidth,clip,trim={0 0 0 0}]{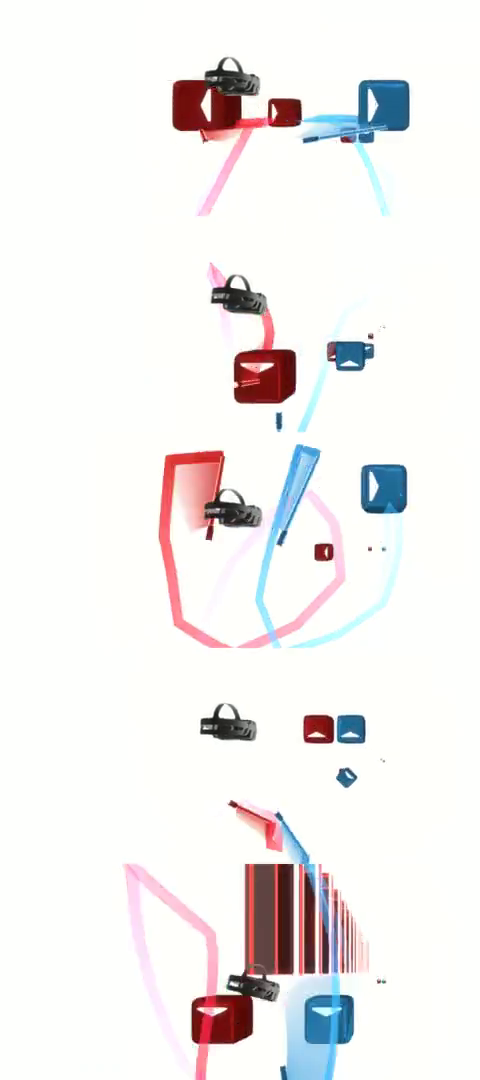}
\par\vspace{1em}
\textline{\text{Generated trajectory}}
\par\vspace{1em}
\includegraphics[width=0.14\linewidth,clip,trim={0 0 0 0}]{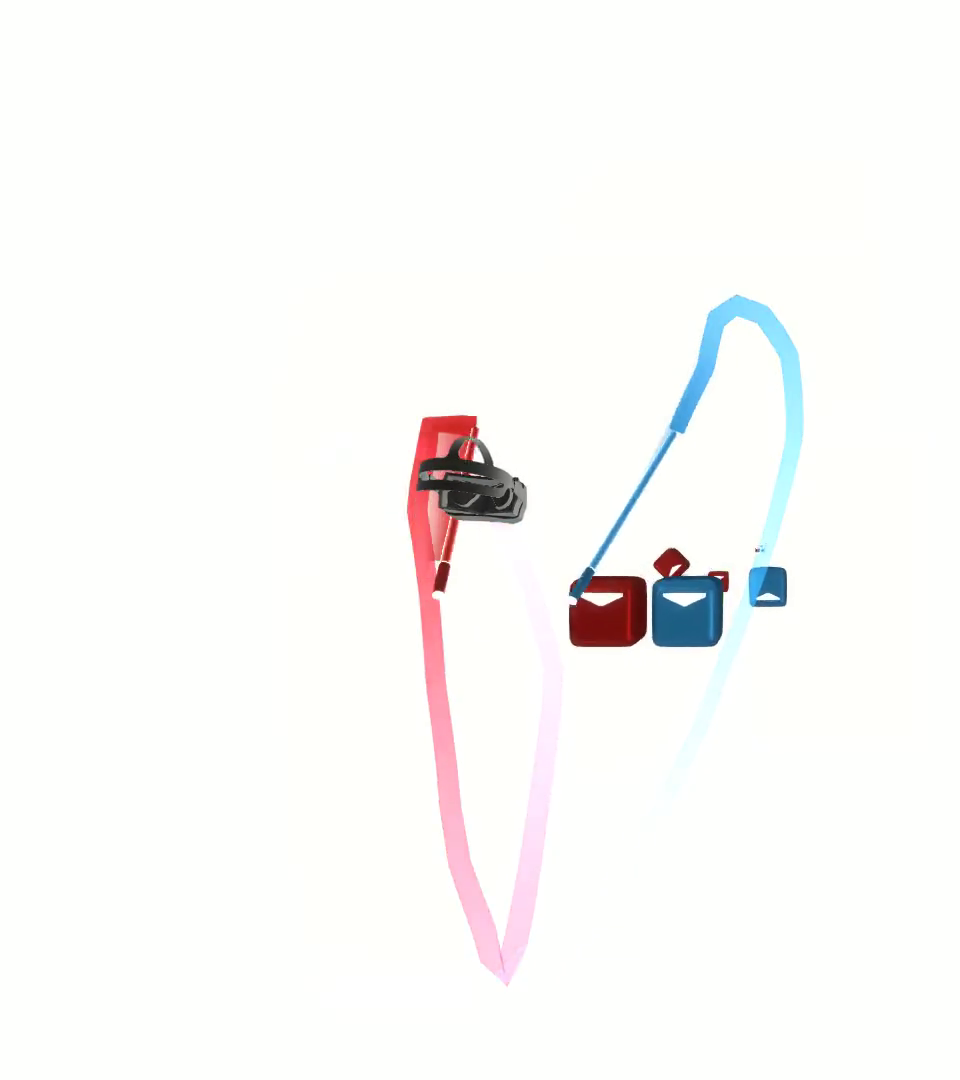}\hfill
\includegraphics[width=0.14\linewidth,clip,trim={0 0 0 0}]{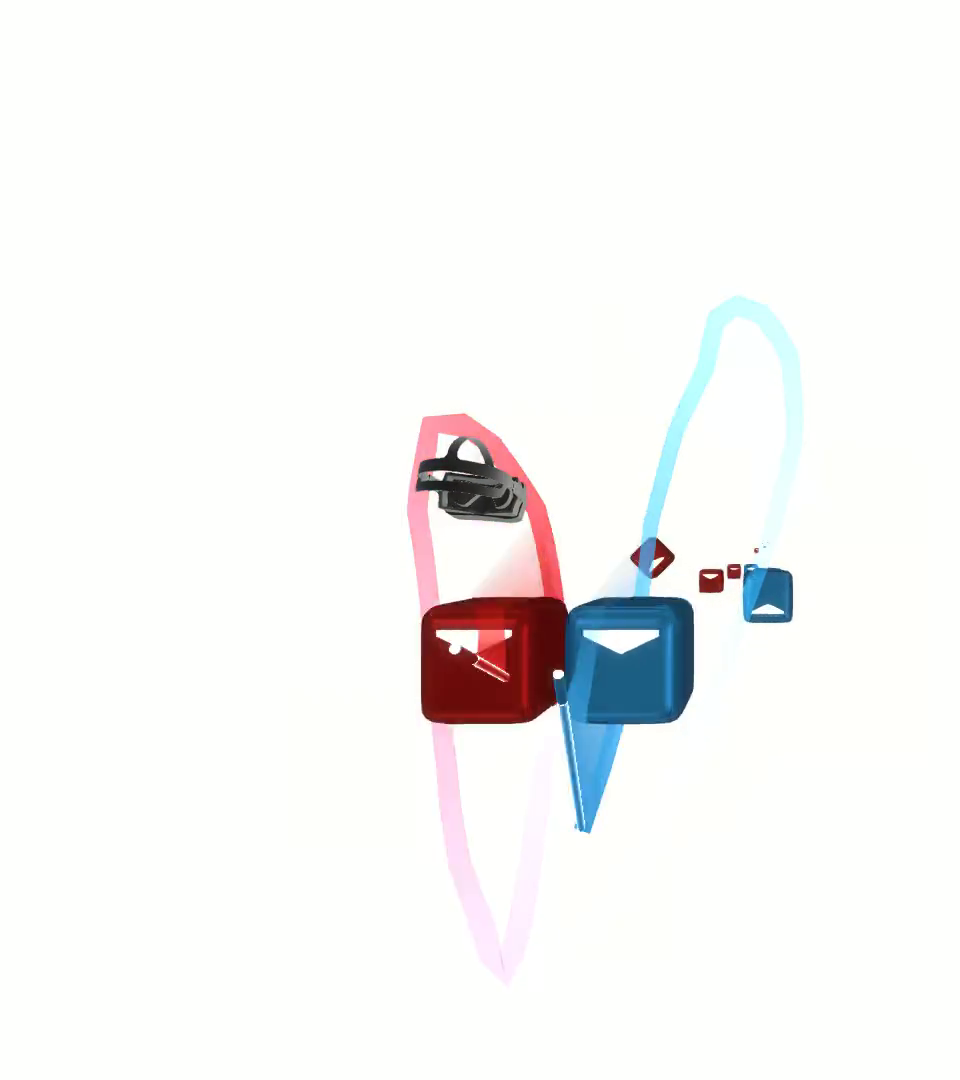}\hfill
\includegraphics[width=0.14\linewidth,clip,trim={0 0 0 0}]{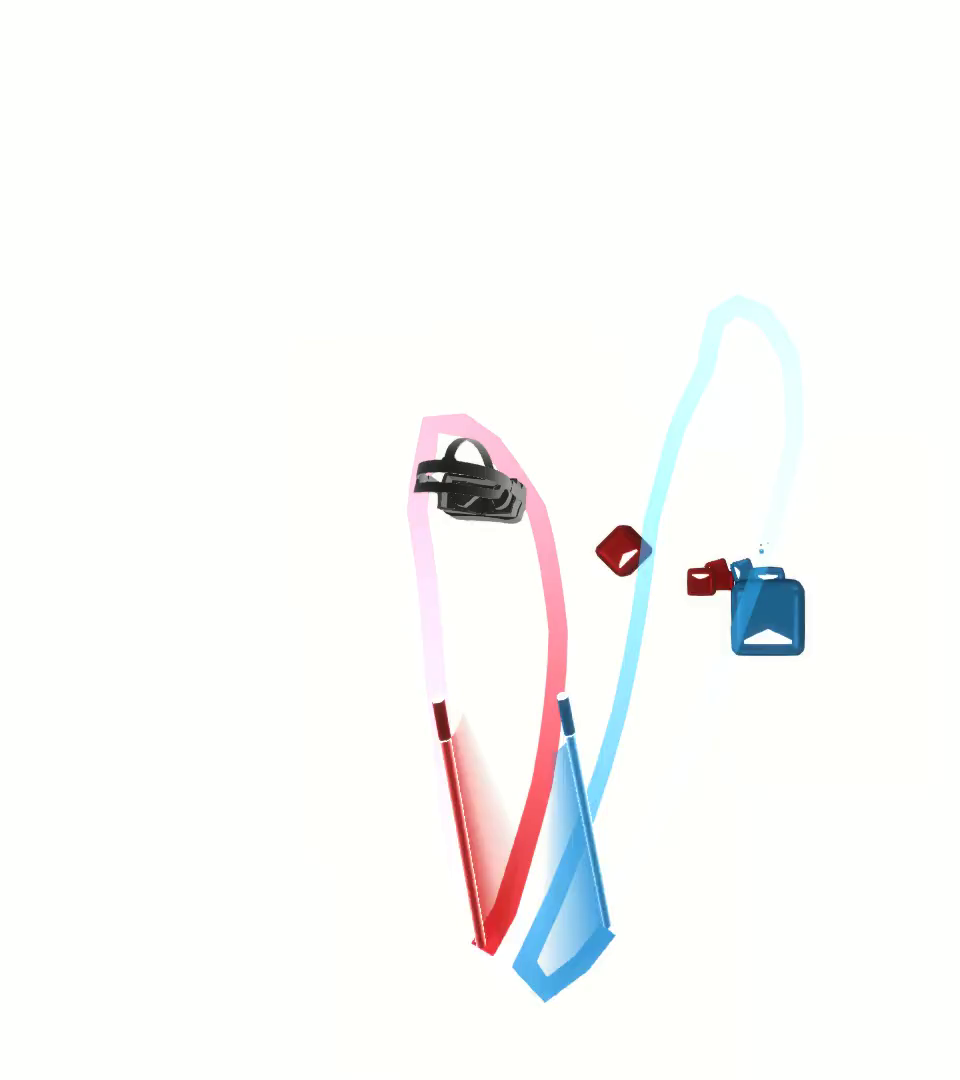}\hfill
\includegraphics[width=0.14\linewidth,clip,trim={0 0 0 0}]{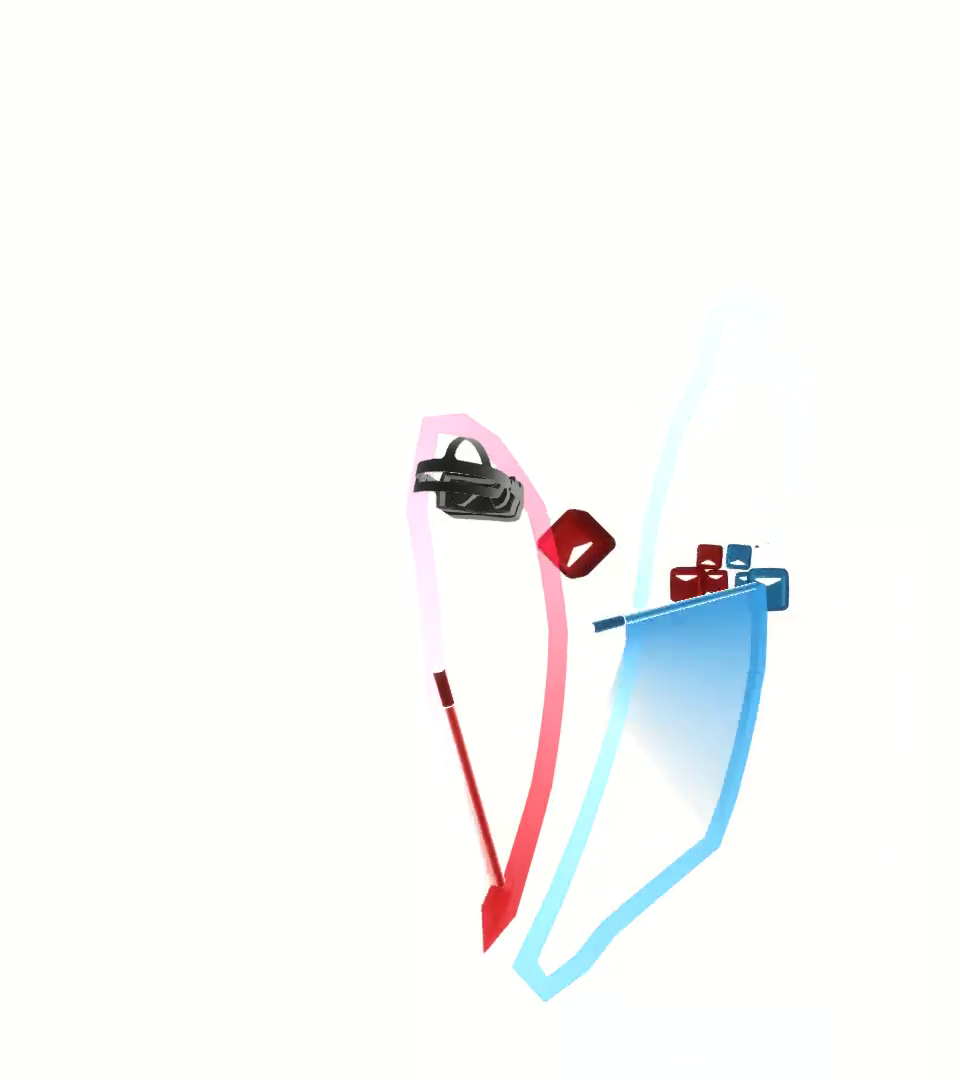}\hfill
\includegraphics[width=0.14\linewidth,clip,trim={0 0 0 0}]{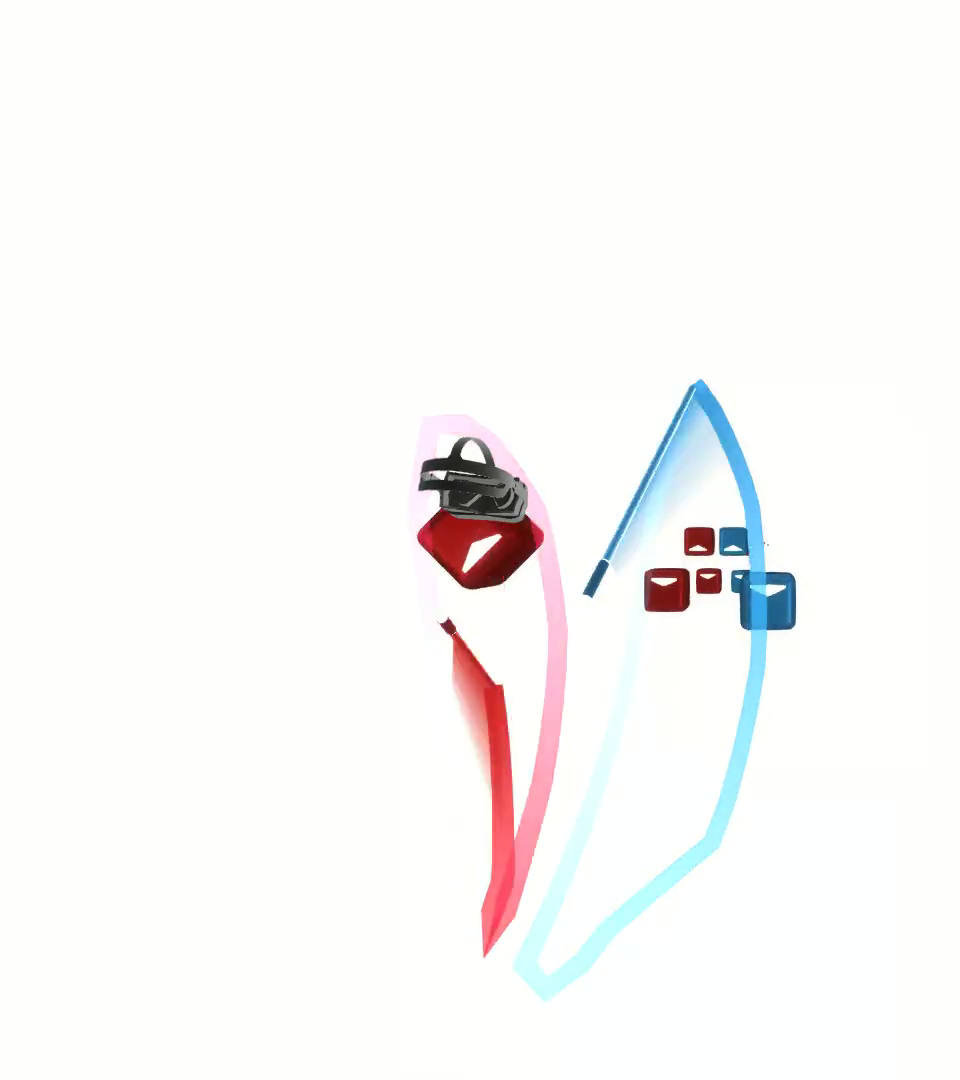}\hfill
\includegraphics[width=0.14\linewidth,clip,trim={0 0 0 0}]{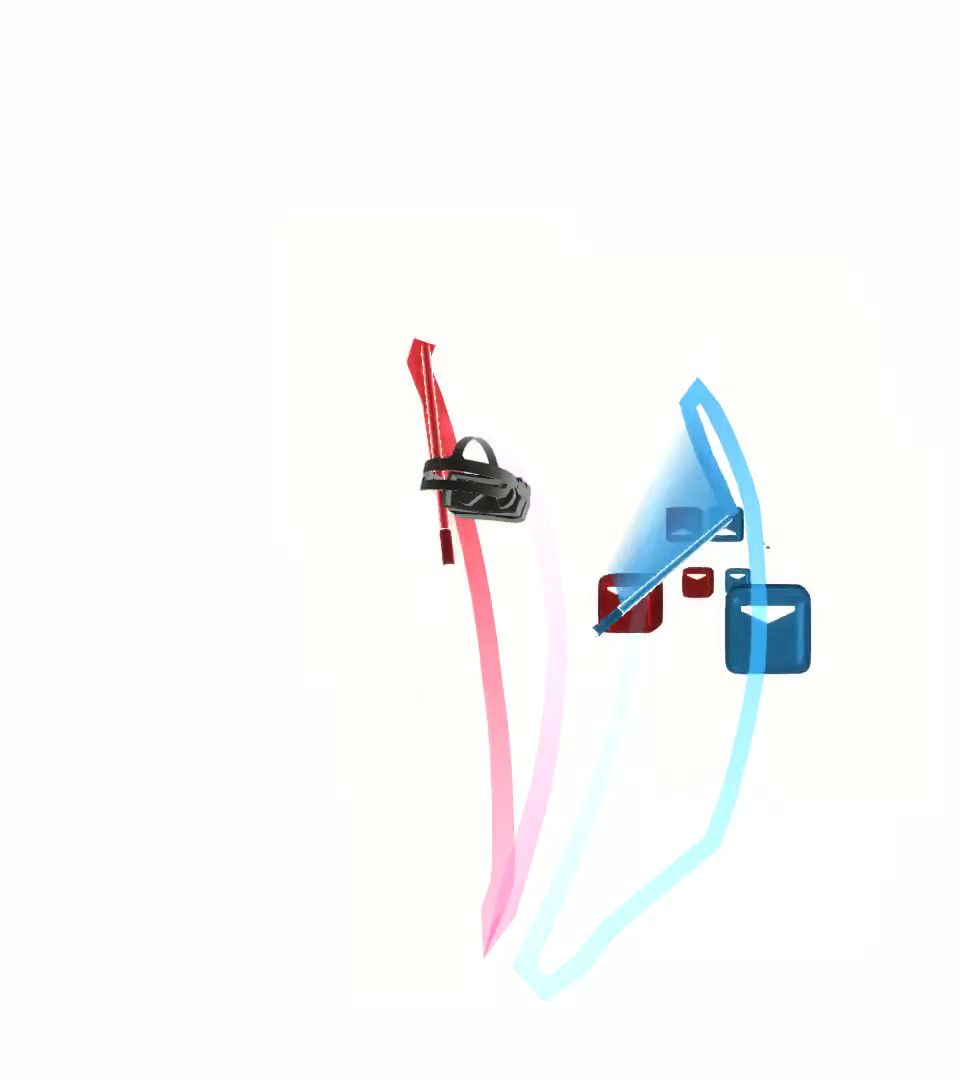}\hfill
\includegraphics[width=0.14\linewidth,clip,trim={0 0 0 0}]{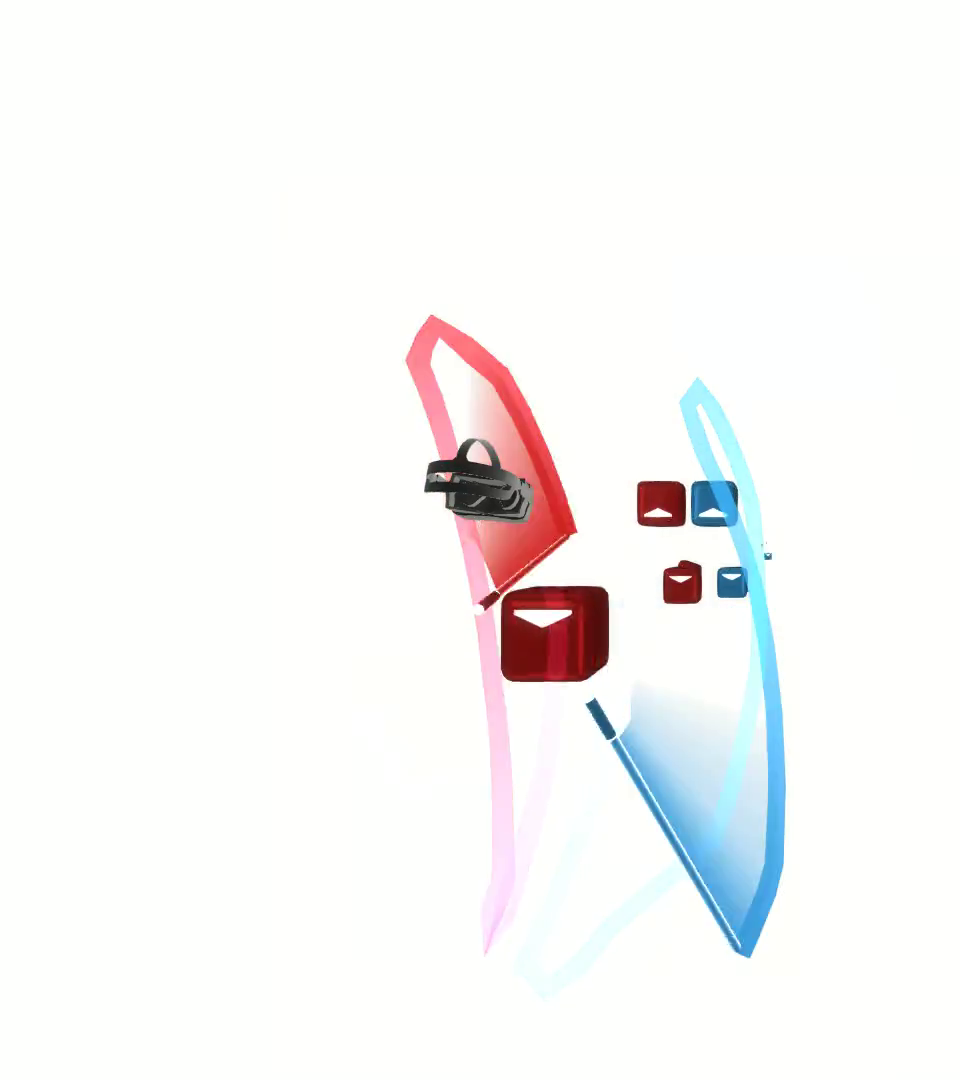}
\end{tcolorbox}
\caption{
Two image sequences, each comparing input reference segments (first 5 rows) and the resulting $3p$ generation (last row). \textbf{Top:} Conditioning on an expert's skilled gameplay results in finessed and confident movements, featuring anticipation and fast swing speed. \textbf{Bottom:} Conditioning on a novice's reference segments yields hesitant and cautious positioning. The two are evaluated on the same map.
}
\label{fig:sticking_qualitative}
\end{figure*}

\noindent \textit{Qualitative result.}
As shown qualitatively in \Cref{fig:sticking_qualitative} and the supplementary video, the input $\RefSegments$ can instruct what the generated movements should look like.
\Cref{fig:sticking_qualitative} shows two generated trajectories on the same held-out map, one conditioned on the reference segments $\RefSegments$ of an elite-level player and the other on a novice player.
The output trajectories reflect visible characteristics of the reference movements, such as swinging speed, anticipatory movements, and the positioning of the handhelds relative to the headset.

\makeatletter
\renewcommand{\thesubsection}{Q4}
\makeatother

\subsection{\ul{Can Robo-Saber predict personalized scores?}} \label{sec:Q4}

\Cref{sec:Q3} establishes a strong correlation between ground-truth human performance and Robo-Saber's player simulation.
While directly simulating user behavior is a viable way to predict scores, we seek to leverage the strong correlation by refining the \emph{in silico} player simulation results into signals for predictive models.
Thus, we investigate a collaborative filtering approach for personalized score prediction (PSP), inspired by \citen{kristensen2022personalized}, and examine the viability of Robo-Saber as a synthetic data augmentation mechanism for predictive applications.

\subsubsection{Setup: PSP via collaborative filtering}
Collaborative filtering (CF) enables predicting variables for user-item pairs, such as a target player's difficulty playing a target game level, based on how difficult other players found it \cite{kristensen2022personalized}.
In our case, we seek to predict TS scores for player-map pairs.
As a novel approach to leverage synthetic gameplay data, we augment CF training data with Robo-Saber scores. We use a \emph{diverse population of Robo-Saber agents}, with diversity elicited by the reference-aware $3p$ trajectories representing different players using $\NRefs=5$.

\noindent \textit{Data for collaborative filtering.} 
For testing, we generate a dataset $\mathbf{N}$, representing new maps that have never been played, each paired with a single unique player in the population.
Our aim is to predict the scores for each player-map pair in $\mathbf{N}$ by learning from some existing performance data of the player population.
To construct $\mathbf{N}$, we pair 1,000 held-out maps with 1,000 held-out players in a one-to-one match. 
Then, we construct $\mathbf{R}$, representing the existing population-wise performance data, including a wide pool of player population, including those in $\mathbf{N}$ (the maps do not overlap).
$\mathbf{R}$ is produced by removing the maps in $\mathbf{N}$ from the remaining holdout and comprises 11,357 player-map pairs.

\noindent \textit{Synthetic data augmentation.}
While \citen{kristensen2022personalized} only used human player data, we augment the CF training data with Robo-Saber scores, \textit{i.e.}, deploy Robo-Saber on $\mathbf{N}$ and $\mathbf{R}$ to produce additional datapoints for CF--we refer to them as $\mathbf{\hat N}$ and $\mathbf{\hat R}$.
For each player-map pair, we emulate the player as in \Cref{sec:Q2}, but additionally indicate that the data is synthetic.
Consequently, the brand-new maps in $\mathbf{N}$ are now associated with the simulated players' performance, providing a basis for training a score-prediction model.
Then, we attempt to predict real players' scores on those brand-new maps.

\subsubsection{Results}
\begin{figure}
    \centering
    \includegraphics[width=0.99\linewidth]{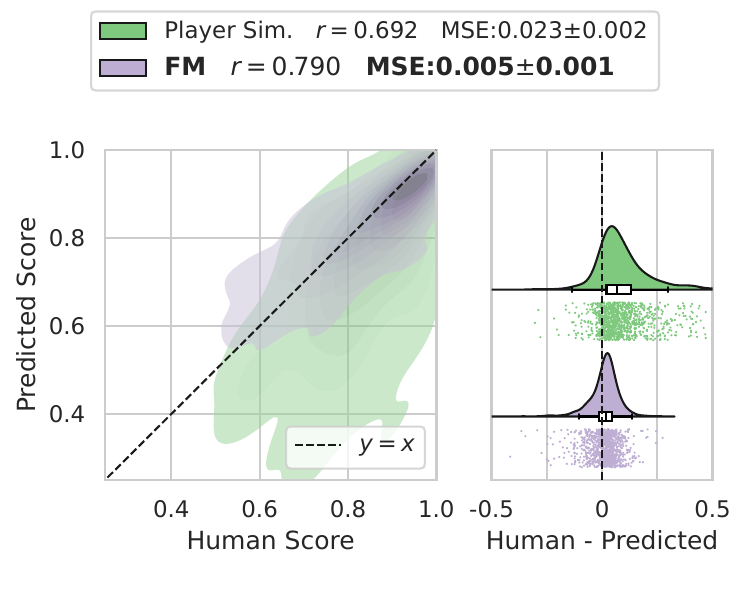}
    \caption{
    \textbf{Left:} 2D density plot comparing predicted and ground truth TS scores evaluated on $\mathbf{N}$. We compare our FM configuration (violet) with the direct player simulation results (green). A significant reduction in MSE and improvement in Pearson's $r$ is detected. \textbf{Right:} A raincloud plot comparing the residuals. We report the MSE $\pm$ standard error in the legend.
    }
    \label{fig:gt_vs_pr}
\end{figure}

From the player-map pairs annotated with human TS scores in $\mathbf{R}$ and synthetic TS scores in $\mathbf{\hat N}$ and $\mathbf{\hat R}$, we learn factorization machines (FMs, \citen{rendle2010factorization}) to perform collaborative filtering.
We validate our FM's predictions on the player-map pairs in $\mathbf{N}$.
Briefly, FMs learn player and map embeddings, such that their dot product can predict the score of the player-map pair;
\Cref{appendix:fm} and \Cref{appendix:hyperparameters} in the Appendix provides more detailed description of FMs, our customization, and the hyperparameters used in training our FM model.
We evaluate the PSP success in terms of mean squared error (MSE) and Pearson's correlation ($r$) across the real player-map pairs in $\mathbf{N}$.
The baseline for our comparison is directly using Robo-Saber's ability to emulate players in $\mathbf{N}$, without using an FM (referred to as \emph{Player Sim.} in \Cref{fig:gt_vs_pr}).

We report the result in \Cref{fig:gt_vs_pr}, which indicates that FM's prediction greatly increases the PSP performance in both MSE ($0.006 \pm 0.001$) and $r$ (0.794).
Notably, as $\mathbf{N}$ is not identically distributed as the entire held-out population, we observe a discrepancy in the player simulation's performance vs. \Cref{sec:Q3}.
Nonetheless, the degraded prediction performance is remedied fully by learning meaningful FM embeddings.
Using Wilcoxon's signed-rank test, we compare the squared errors of player simulation and FM and detect statistical significance ($p\ll0.00001$).

\section{From $3p$ Generation to Full-body Movement} \label{sec:robotic_agent}
In this section, we examine whether our generative model can help enable a physics-based full-body VR user model and report the performance of our resulting system.
Computational user models thus far feature limited or no embodiment, \textit{e.g.}, predicting the difficulty of mobile games using a player model that directly generates touchscreen interaction events without simulating the human hand \cite{roohi2020predicting, roohi2021predicting,kristensen2020estimating}, or simulating touchscreen typing using a simplified model of finger movements \cite{shi2025simulating}.
In the rare systems featuring intelligent control of an actual physical or biomechanical simulation model, the focus has been on simulating only a single arm \cite{cheema2020predicting,ikkala2022breathing,fischer2024sim2vr}. 
Consequently, developing a physically simulated full-body user model for embodied interaction remains an open challenge.

We propose and evaluate a solution that leverages Robo-Saber's $3p$ motion generation combined with physics-based full-body motion tracking. 
Namely, we interface our $3p$ generative model with a popular physics-based humanoid tracking controller from Perpetual Humanoid Control foundation model (PHC, \citen{luo2023perpetual}).
\Cref{appendix:tracking} in the Appendix describes the details of our physics-based tracking implementation.
We then investigate the following questions: 

\begin{itemize}[leftmargin=*,label={}]
    \item \ref{sec:Q5}.~\nameref{sec:Q5}
    \item \ref{sec:Q6}.~\nameref{sec:Q6}
\end{itemize}

\begin{figure}
    \centering
    \begin{tabular}{c|c}
        Kinematic & Physics-based \\
        \includegraphics[width=0.45\linewidth,clip,trim={0 20 0 0}]{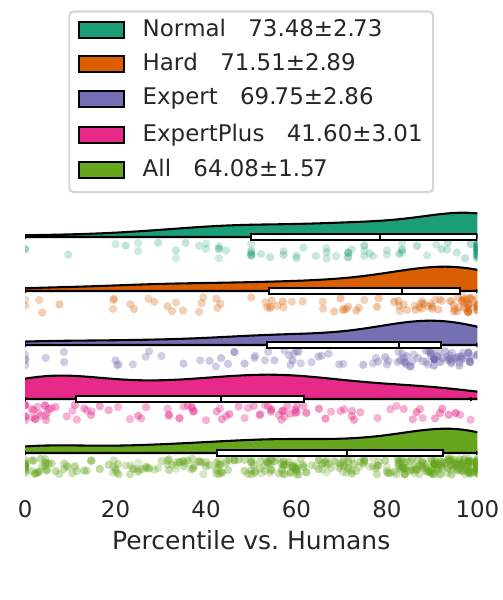} & \includegraphics[width=0.45\linewidth,clip,trim={0 20 0 0}]{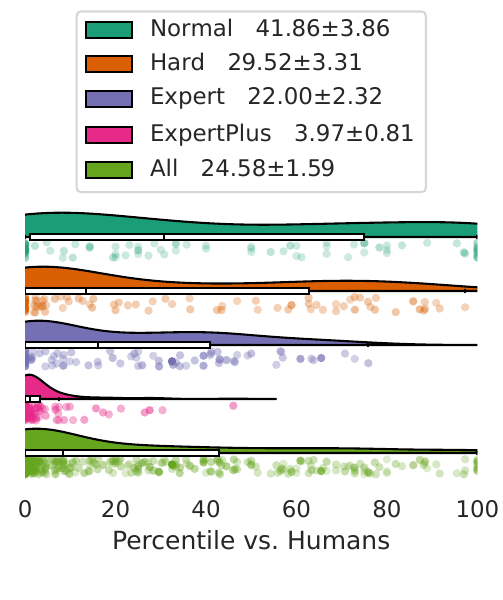}
    \end{tabular}
    \caption{
    Evaluating the performances of kinematic \textbf{(left)} and physics-based \textbf{(right)} versions of Robo-Saber ($\NRefs=5$ for both).
    The distribution of Robo-Saber's TS score percentiles per map is shown across difficulty levels.
    Physics-based trajectories are produced by tracking the generated $3p$ trajectories.
    As expected, absolute performance degrades due to physical/embodied constraints and tracking errors.
    The kinematic agent achieves respectable percentiles across all difficulty levels.
    While the physics-based agent achieves percentiles above 40\% on average on Normal maps, the performance is poor on Expert and above.
    }
    \label{fig:physics_addendum}
\end{figure}

\makeatletter
\renewcommand{\thesubsection}{Q5}
\makeatother

\subsection{\ul{How performant is physics-based tracking?}} \label{sec:Q5}

\noindent \textit{Quantitative results.}
Tracking with the whole body would naturally degrade and alter $3p$ movements, due to tracking error and additional constraints introduced by physics simulation; this is an expected consequence of added physical realism rather than an anomaly.
The right half of \Cref{fig:physics_addendum} shows the effect of tracking the $3p$ trajectories for the left half.
Introducing physical constraints indeed results in lower scores across difficulty levels.
The physics-based Robo-Saber does not keep up with humans on Expert and Expert+ maps, although it performs reasonably well on Normal and Hard.
This is presumably due to the tracking controller being unable to achieve sufficient movement speed and precision for difficult gameplay; we envision further research to enable higher physics-based performance as a promising direction, as noted in \Cref{sec:conclusion}.

\begin{figure*}
    \centering
    \includegraphics[width=0.16\linewidth,clip,trim={560 0 560 0}]{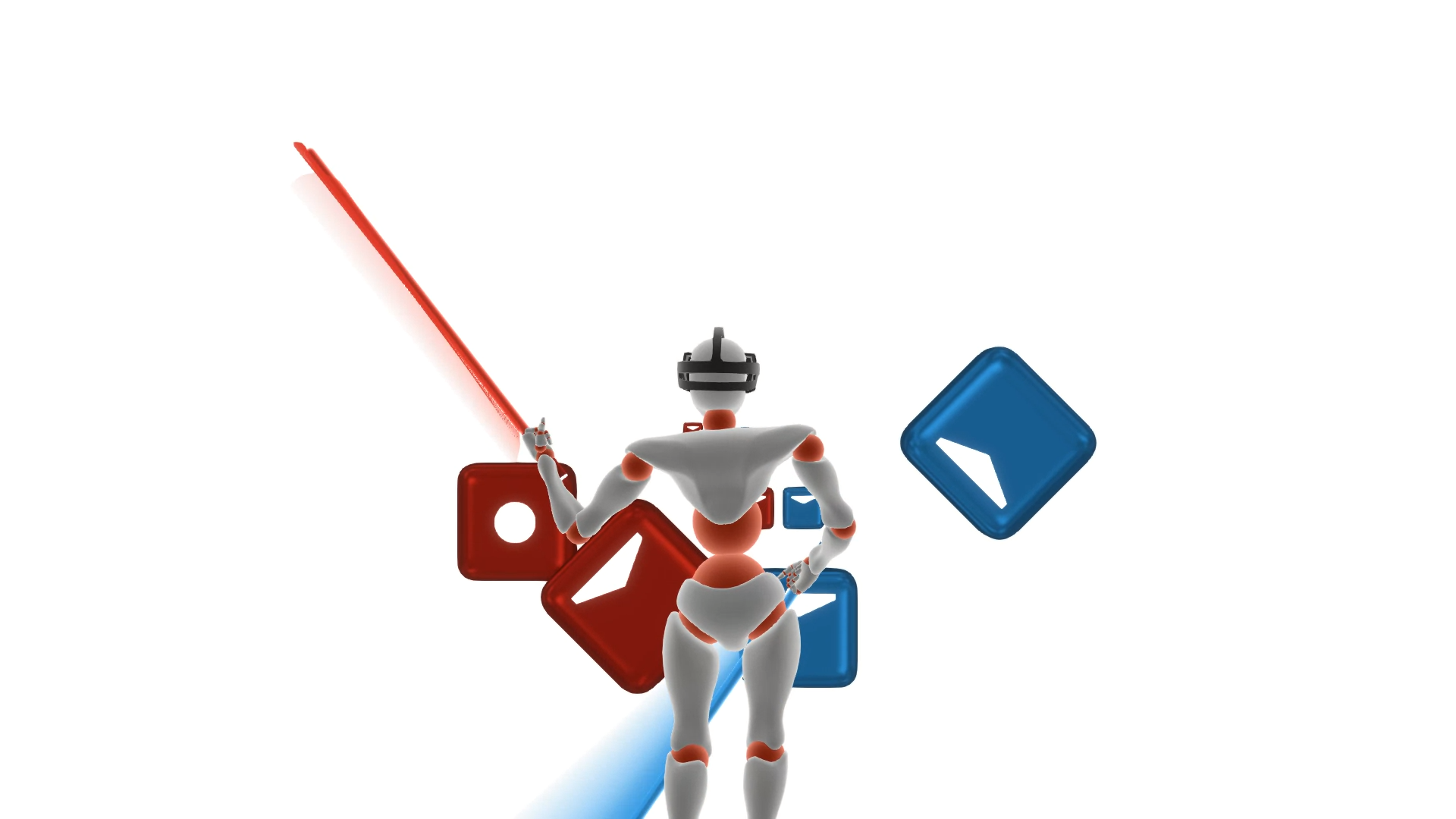}
    \hfill
    \includegraphics[width=0.16\linewidth,clip,trim={560 0 560 0}]{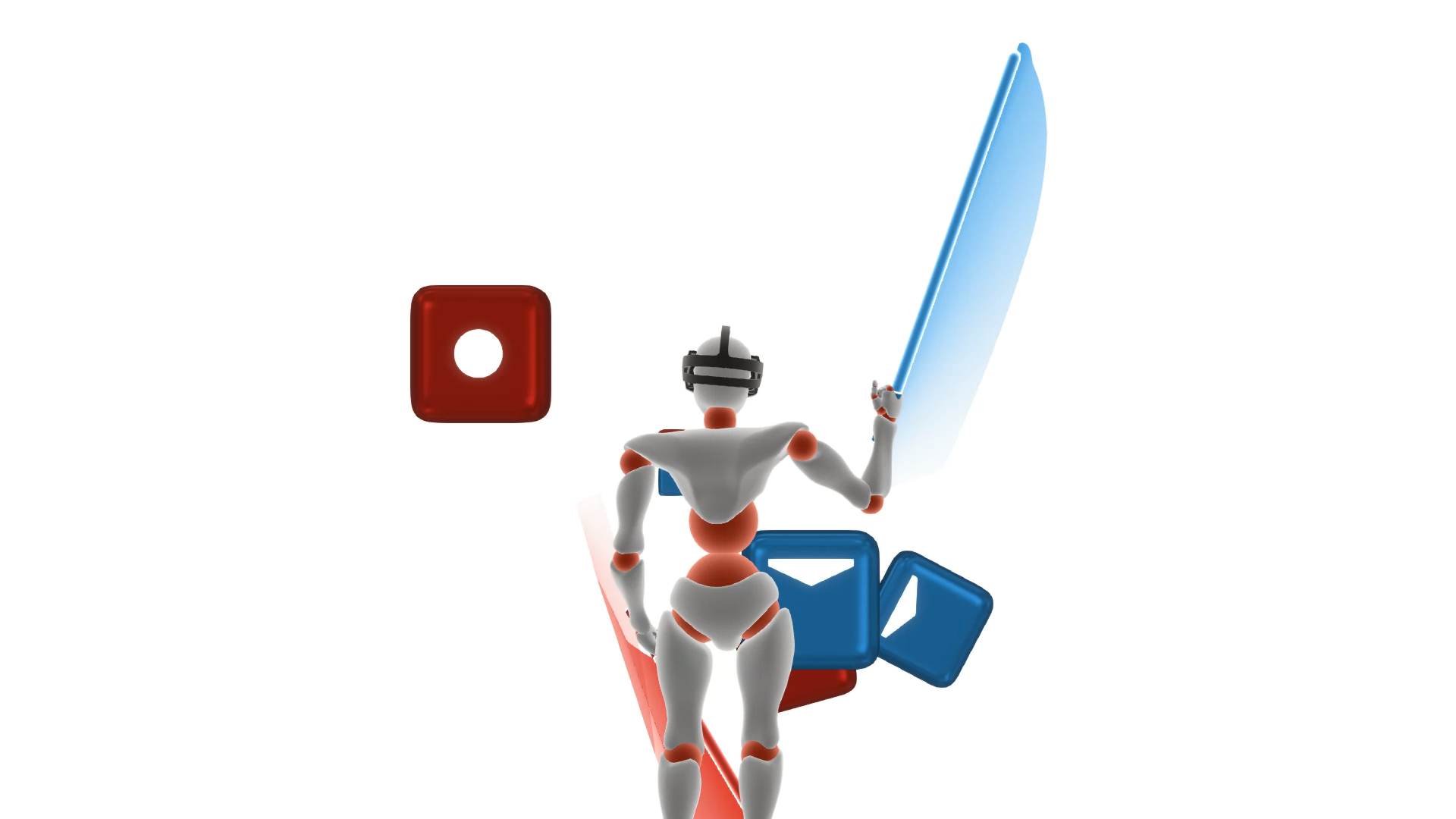}
    \hfill
    \includegraphics[width=0.16\linewidth,clip,trim={560 0 560 0}]{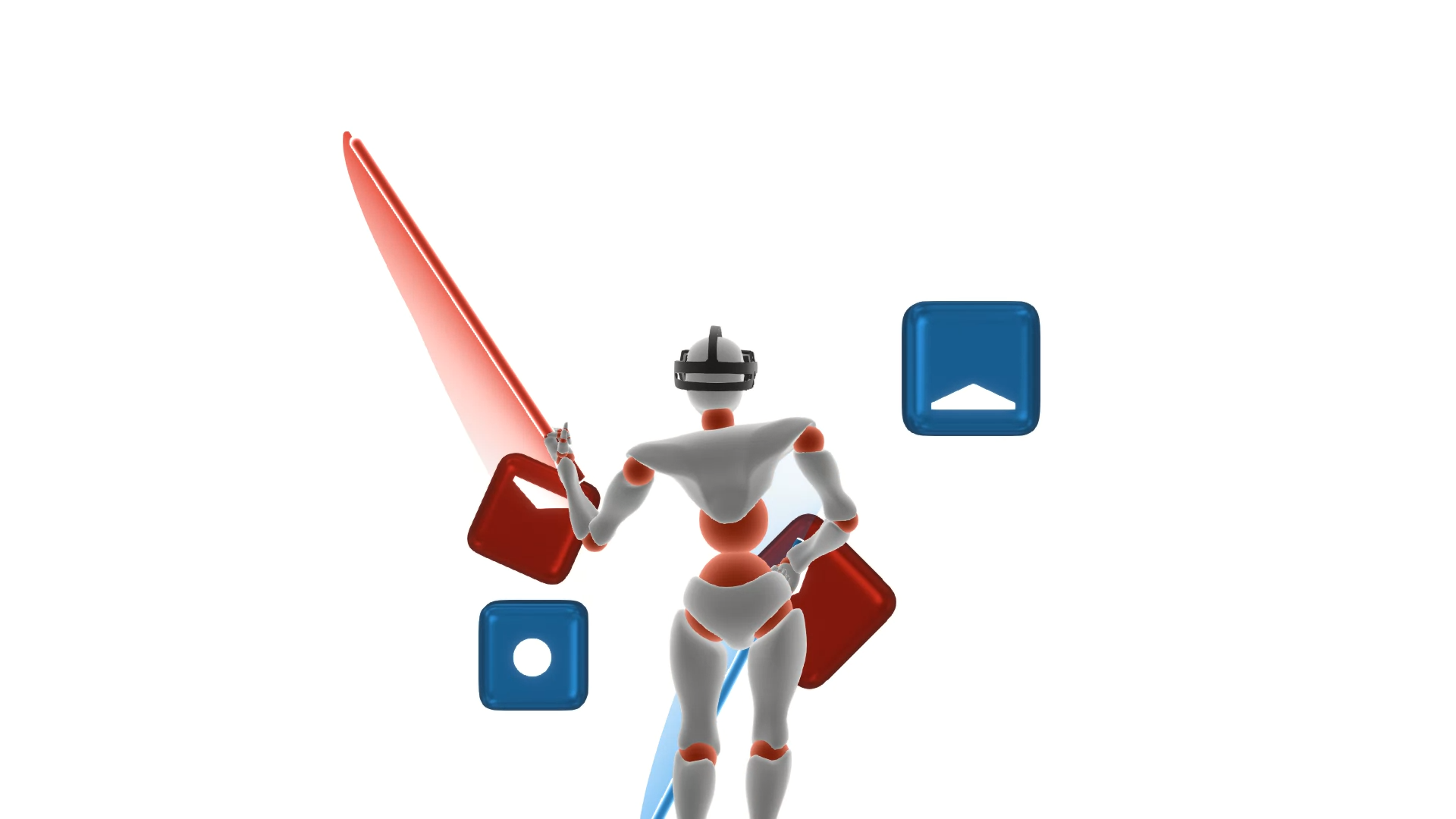}
    \hfill
    \includegraphics[width=0.16\linewidth,clip,trim={560 0 560 0}]{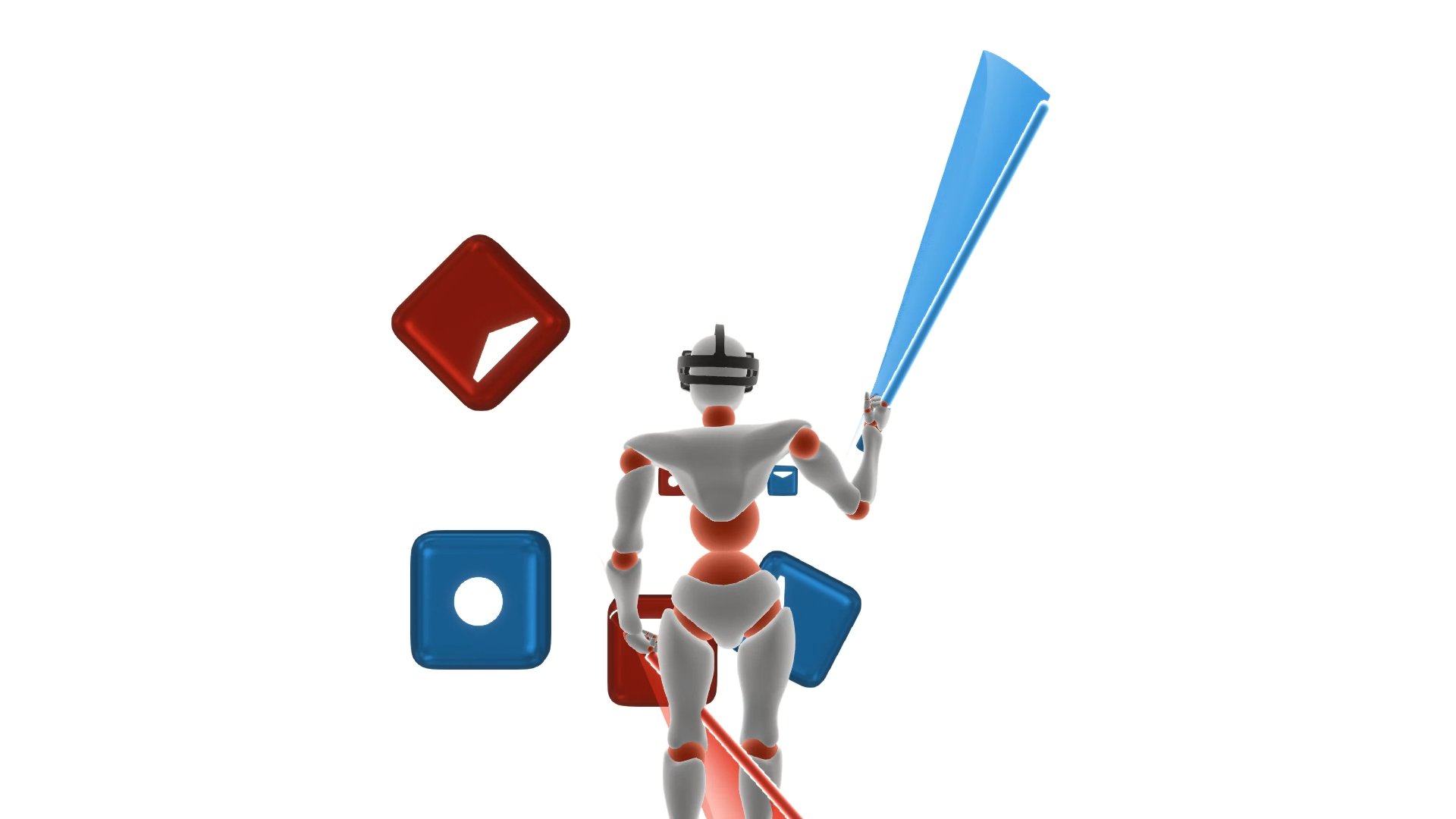}
    \hfill
    \includegraphics[width=0.16\linewidth,clip,trim={560 0 560 0}]{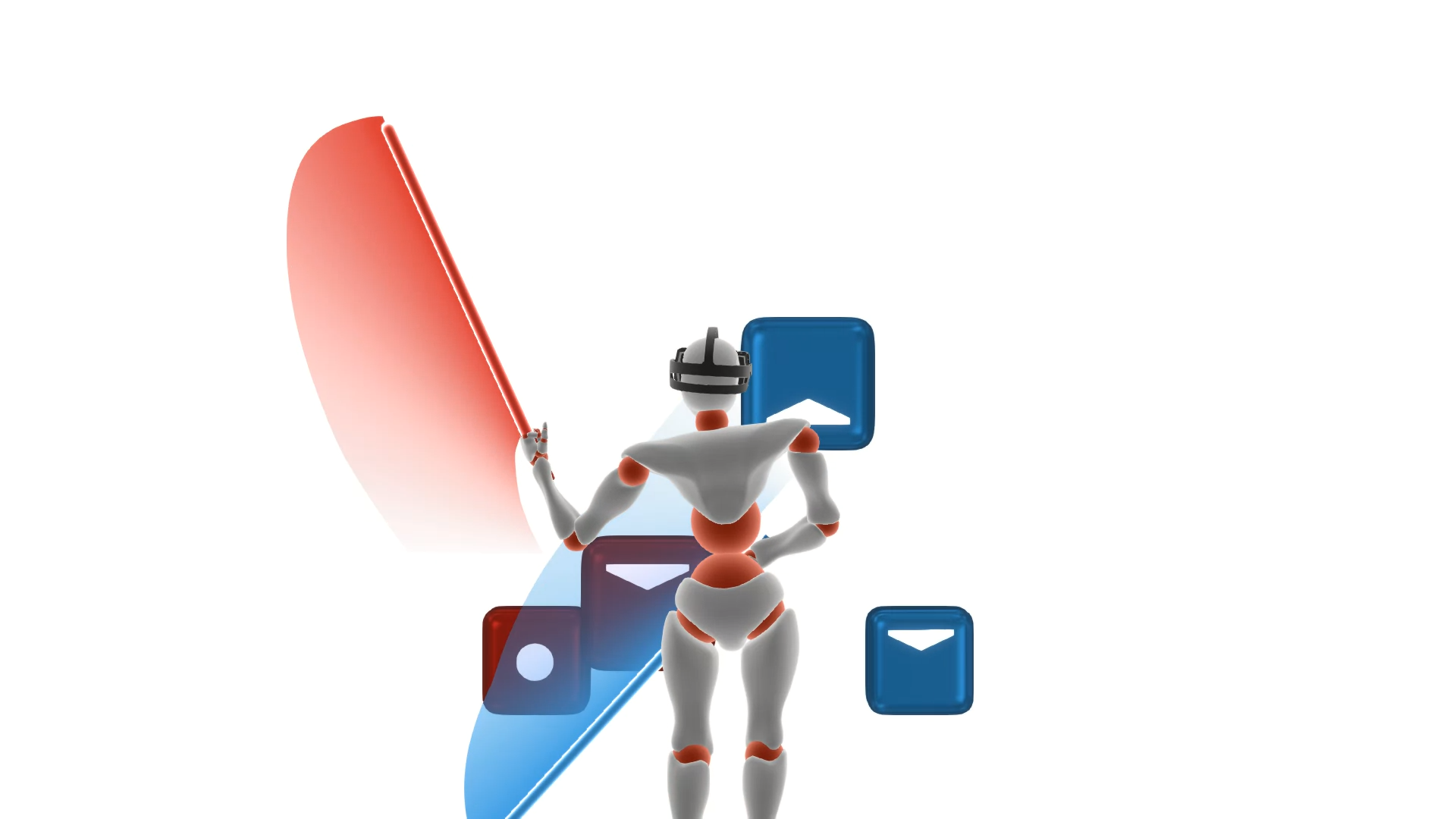}    
    \caption{The physics-based version of Robo-Saber plays colored notes. The red and blue notes are correctly cut in sequence, with matching saber colors and directions. The dotted notes can be hit from any direction.}
    \label{fig:robo-saber-plays}
\end{figure*}

\begin{figure*}
    \centering
    \includegraphics[width=0.2\linewidth,clip,trim={300 0 960 0}]{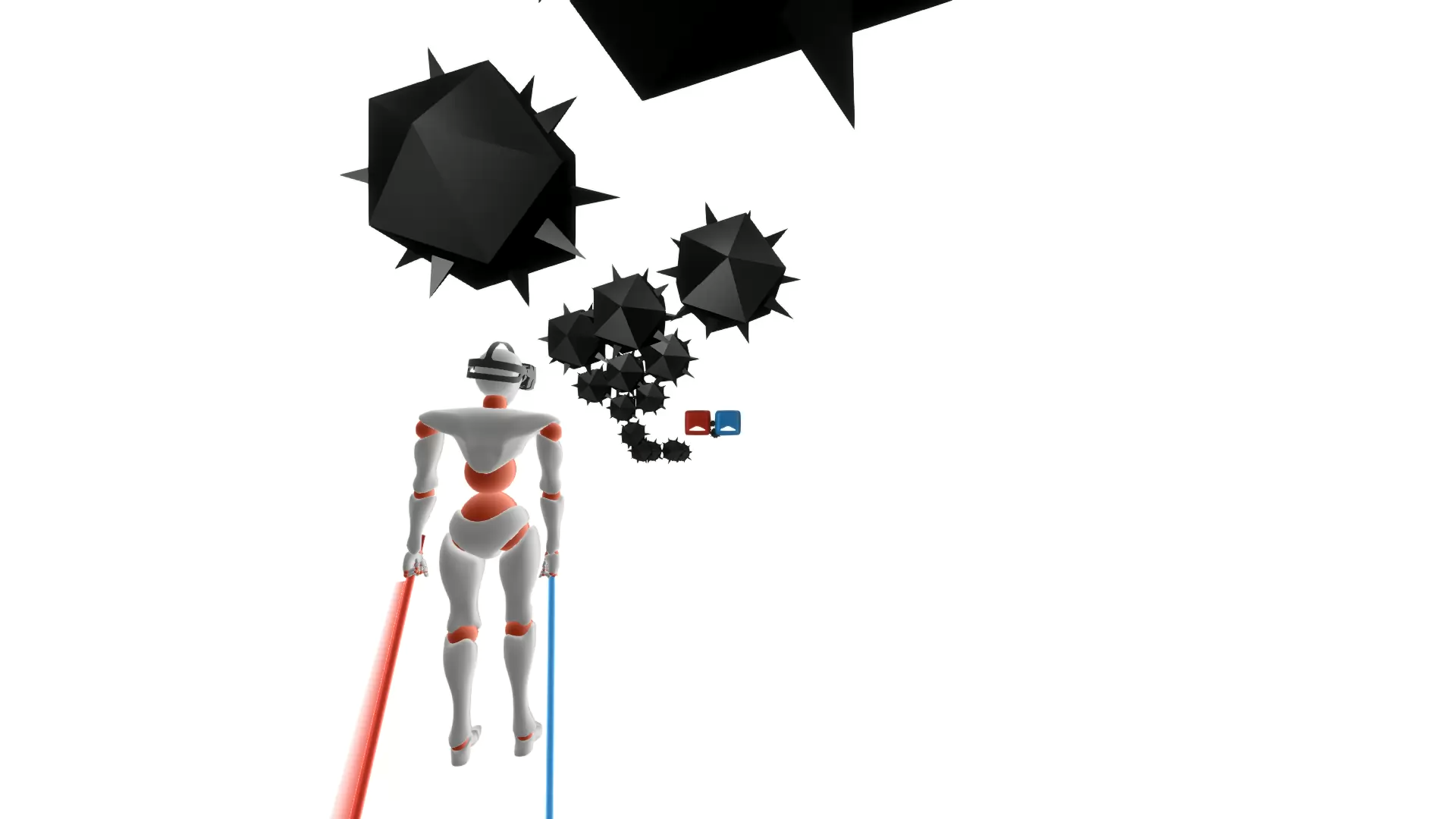}
    \hfill
    \includegraphics[width=0.2\linewidth,clip,trim={300 0 960 0}]{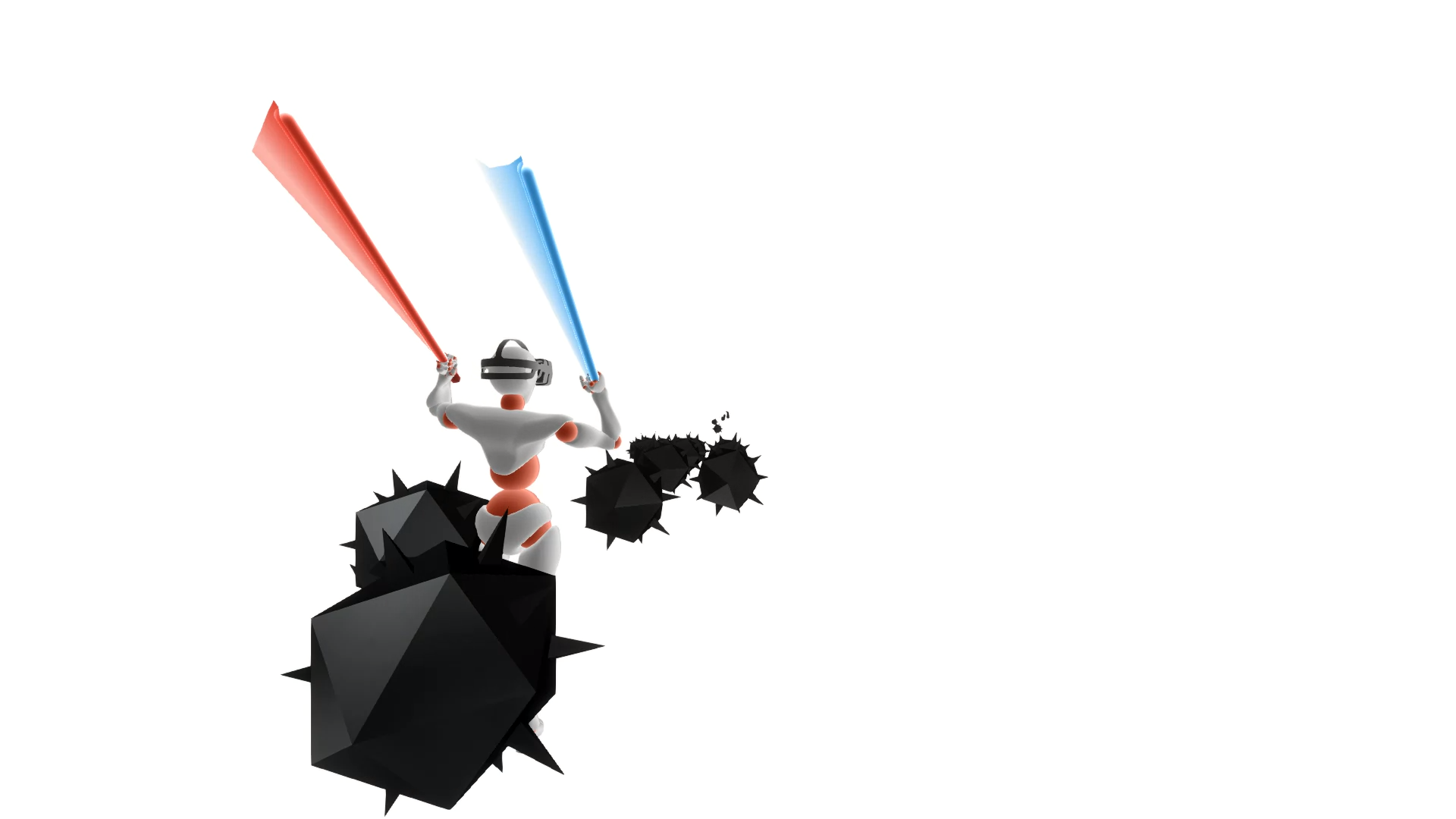}
    \hfill
    \includegraphics[width=0.2\linewidth,clip,trim={300 0 960 0}]{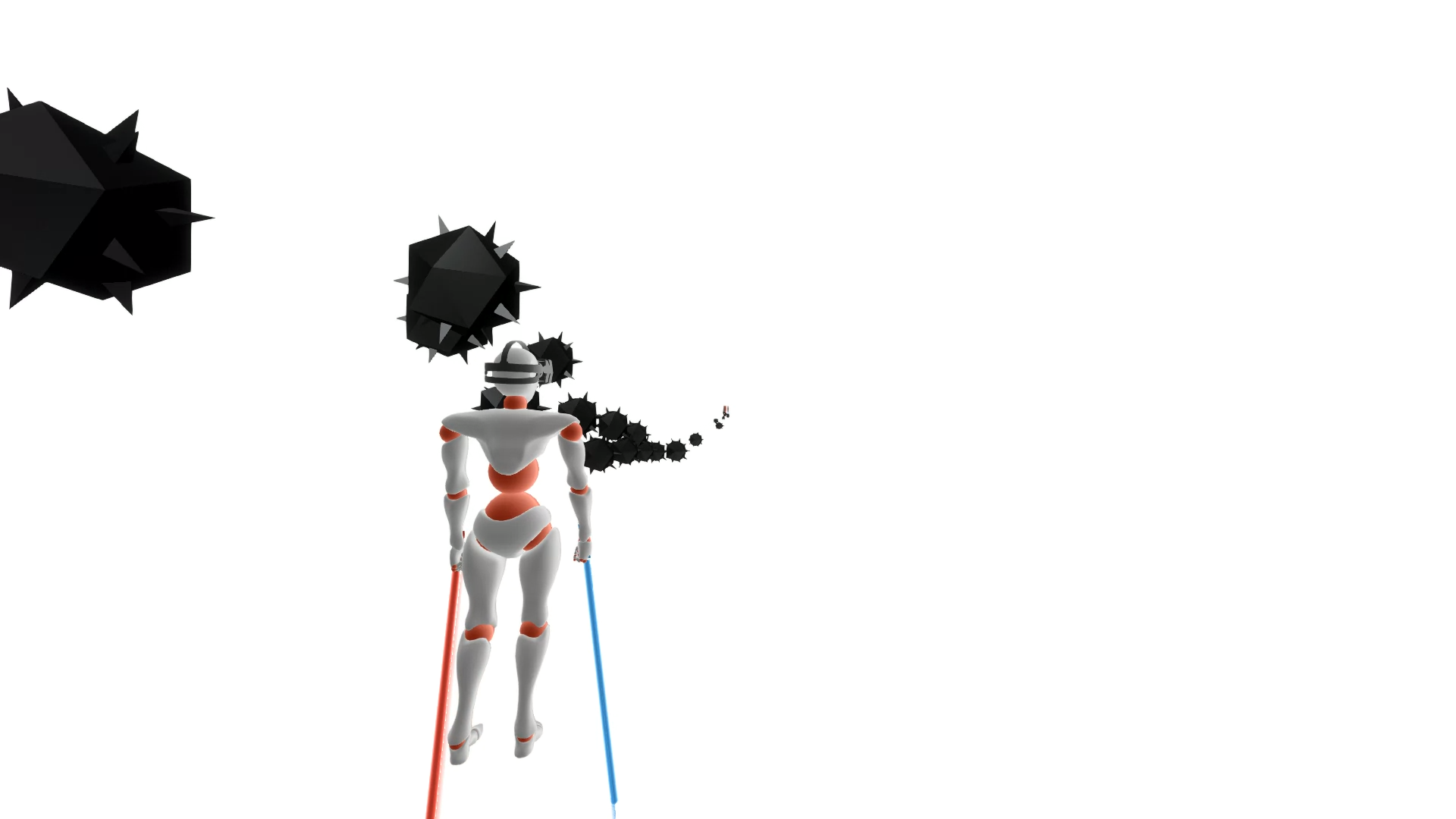}
    \hfill
    \includegraphics[width=0.2\linewidth,clip,trim={300 0 960 0}]{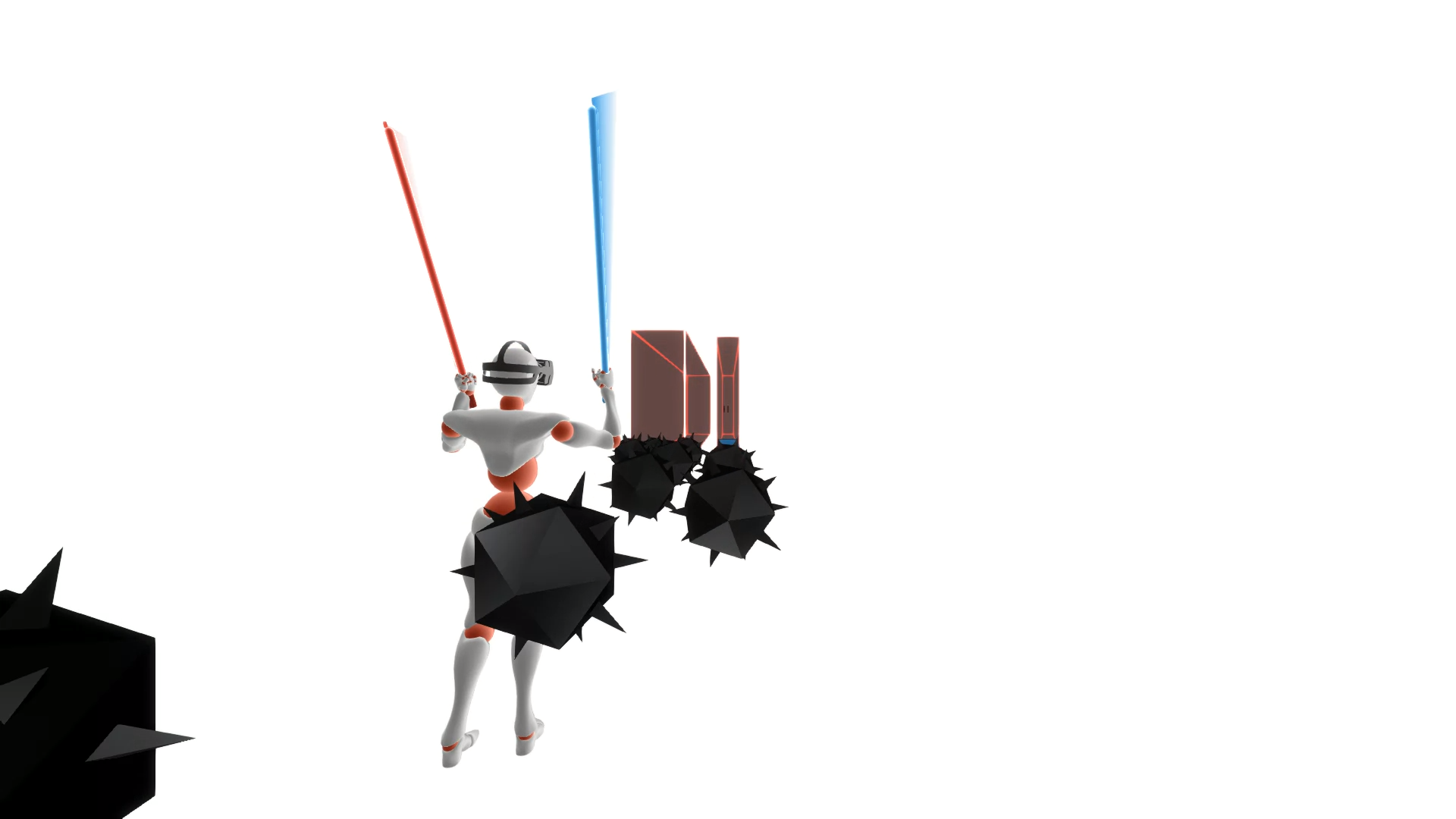}
    \caption{Robo-Saber avoids a long sequence of bomb notes by moving its hands up and down and orienting the sabers away from them.}
    \label{fig:robo-saber-bombs}
\end{figure*}

\begin{figure*}
    \centering
    \includegraphics[width=0.25\linewidth,clip,trim={400 0 400 0}]{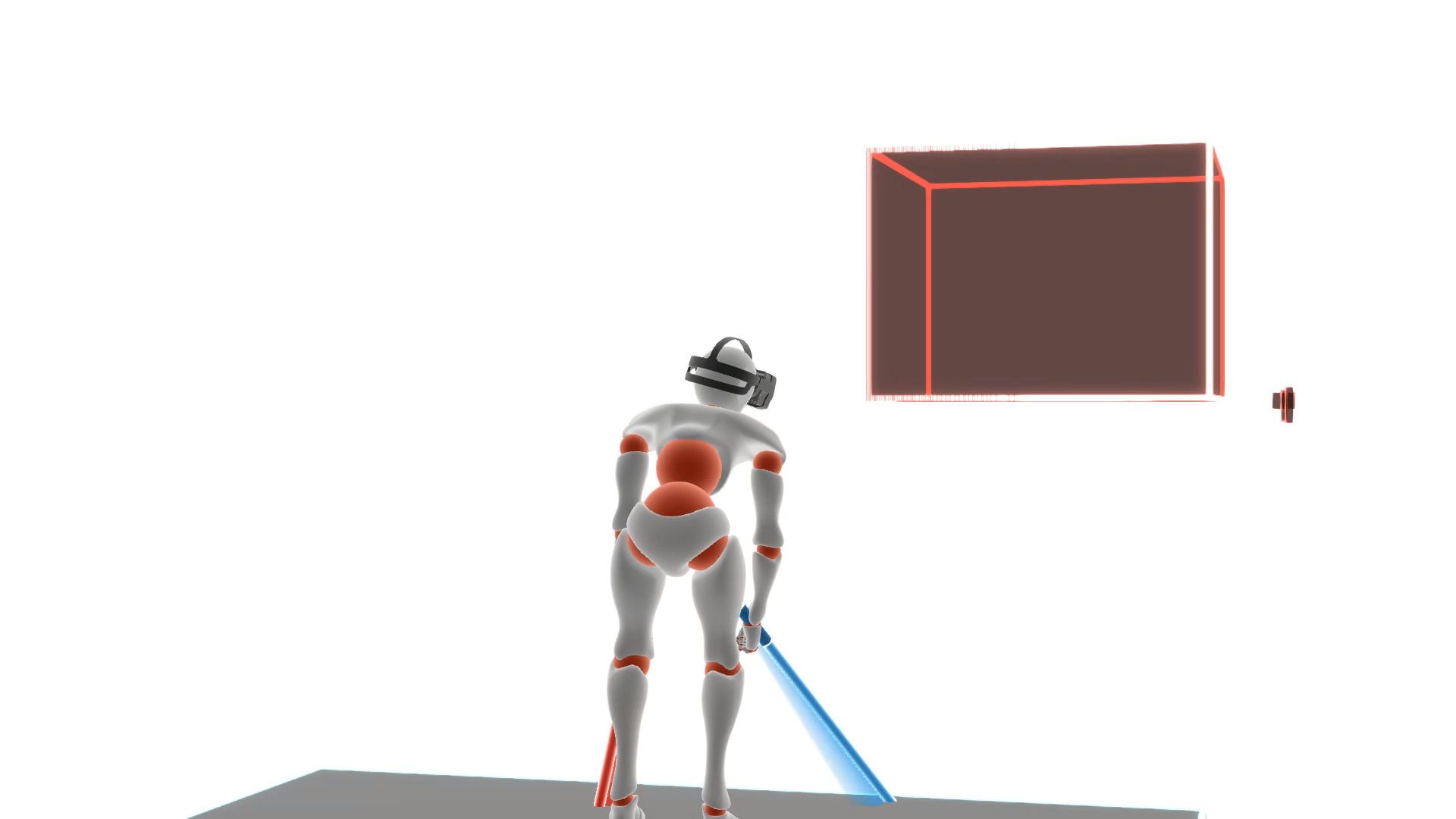}
    \hfill
    \includegraphics[width=0.25\linewidth,clip,trim={400 0 400 0}]{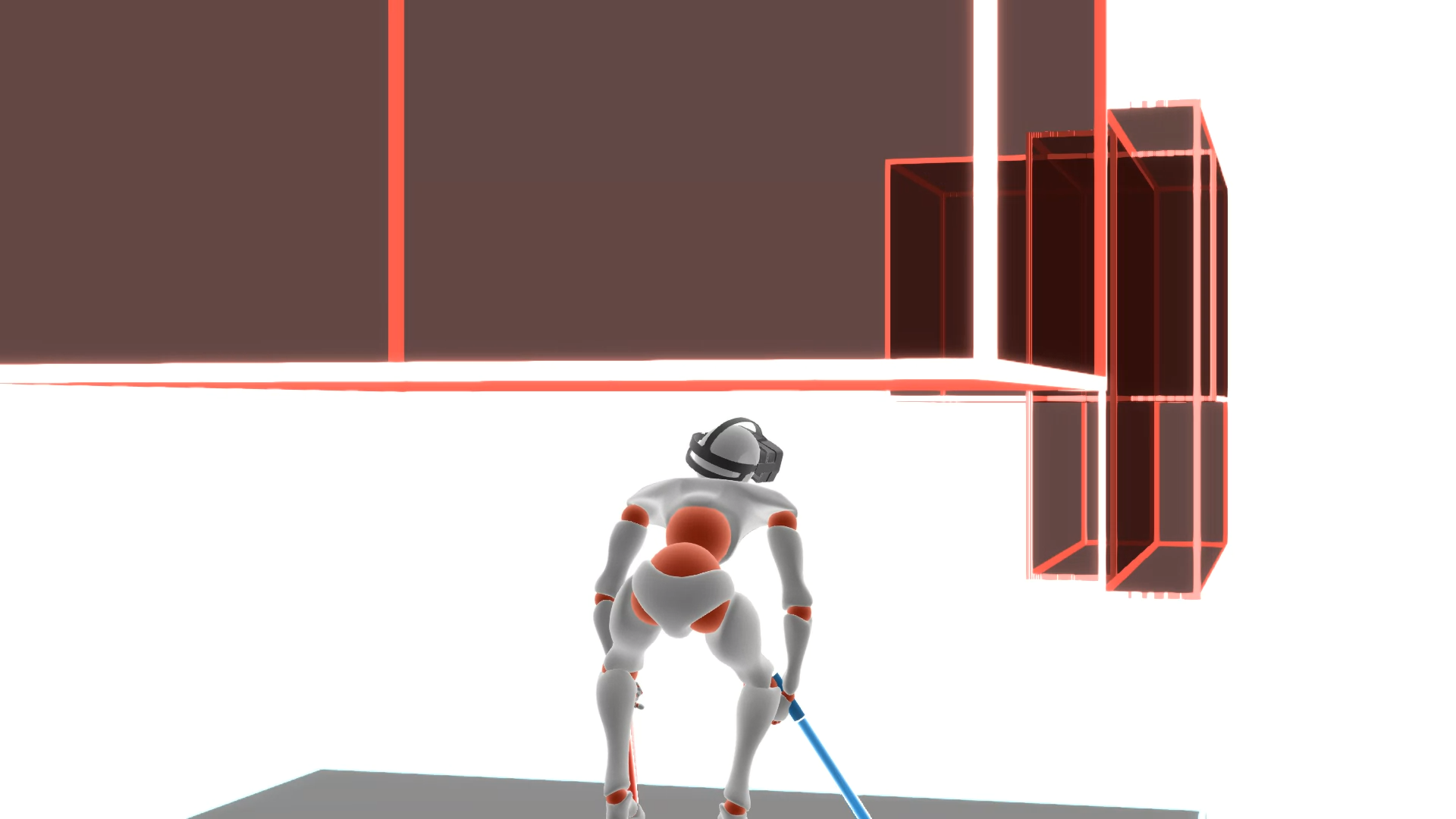}
    \hfill    \includegraphics[width=0.25\linewidth,clip,trim={400 0 400 0}]{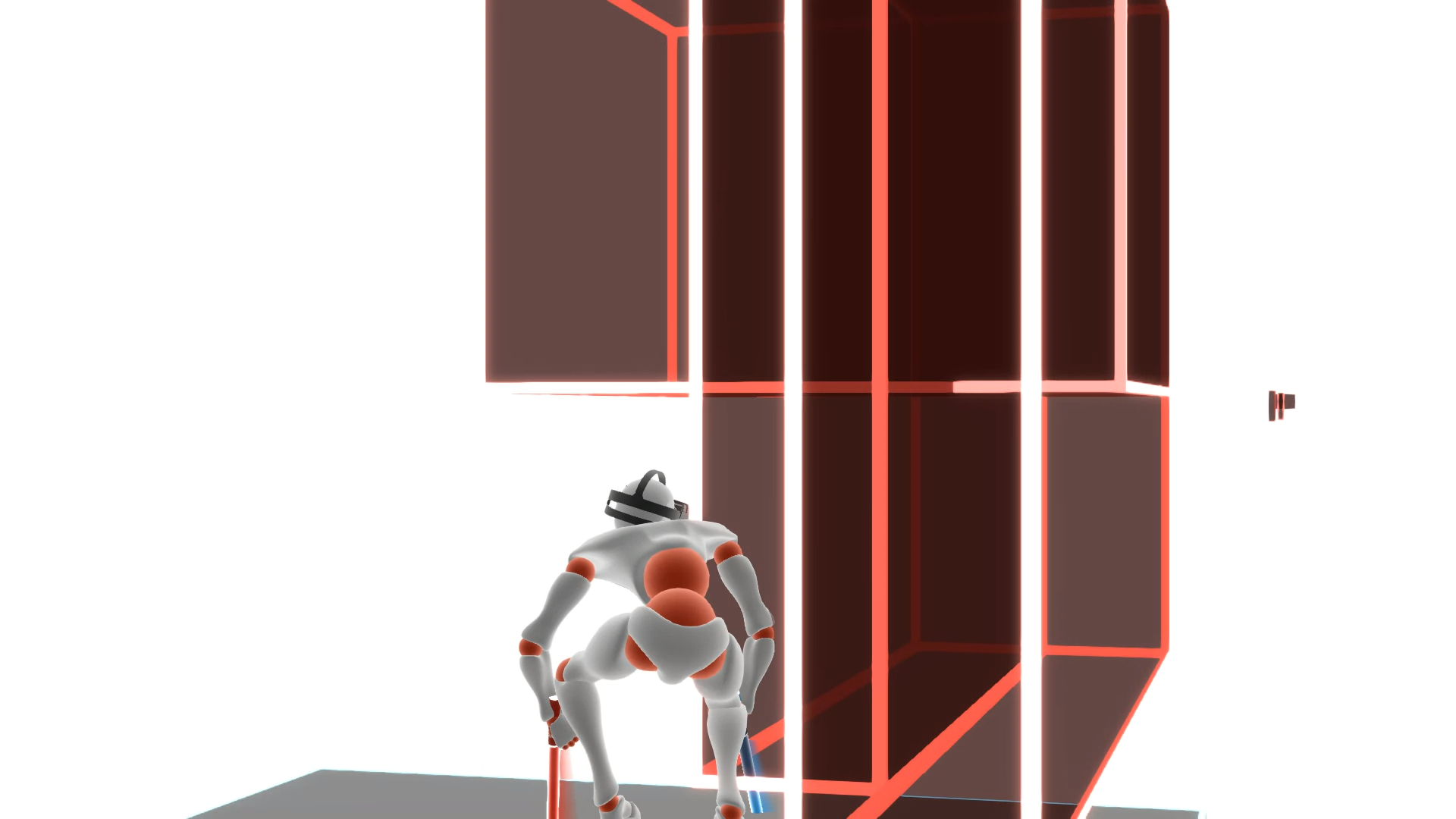}
    \caption{Robo-Saber avoids obstacles (red boxes) by ducking to lower its head and swaying away from them. Robo-Saber is first in performing articulated whole-body gameplay movements for VR.}
    \label{fig:robo-saber-obstacles}
\end{figure*}

\begin{figure*}
    \centering
\includegraphics[width=0.24\linewidth,clip,trim={200 0 860 0}]{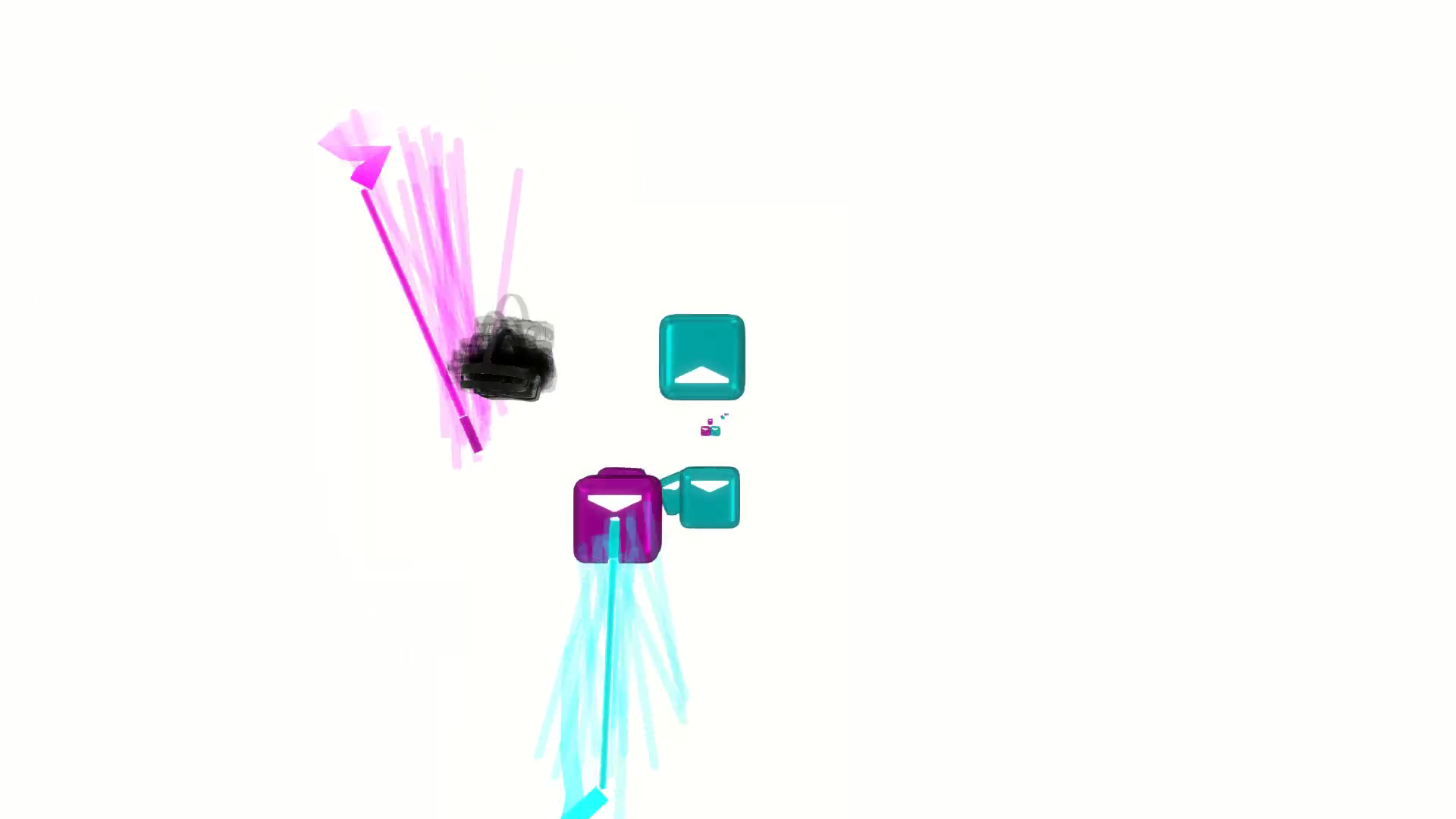}\hfill
\includegraphics[width=0.24\linewidth,clip,trim={200 0 860 0}]{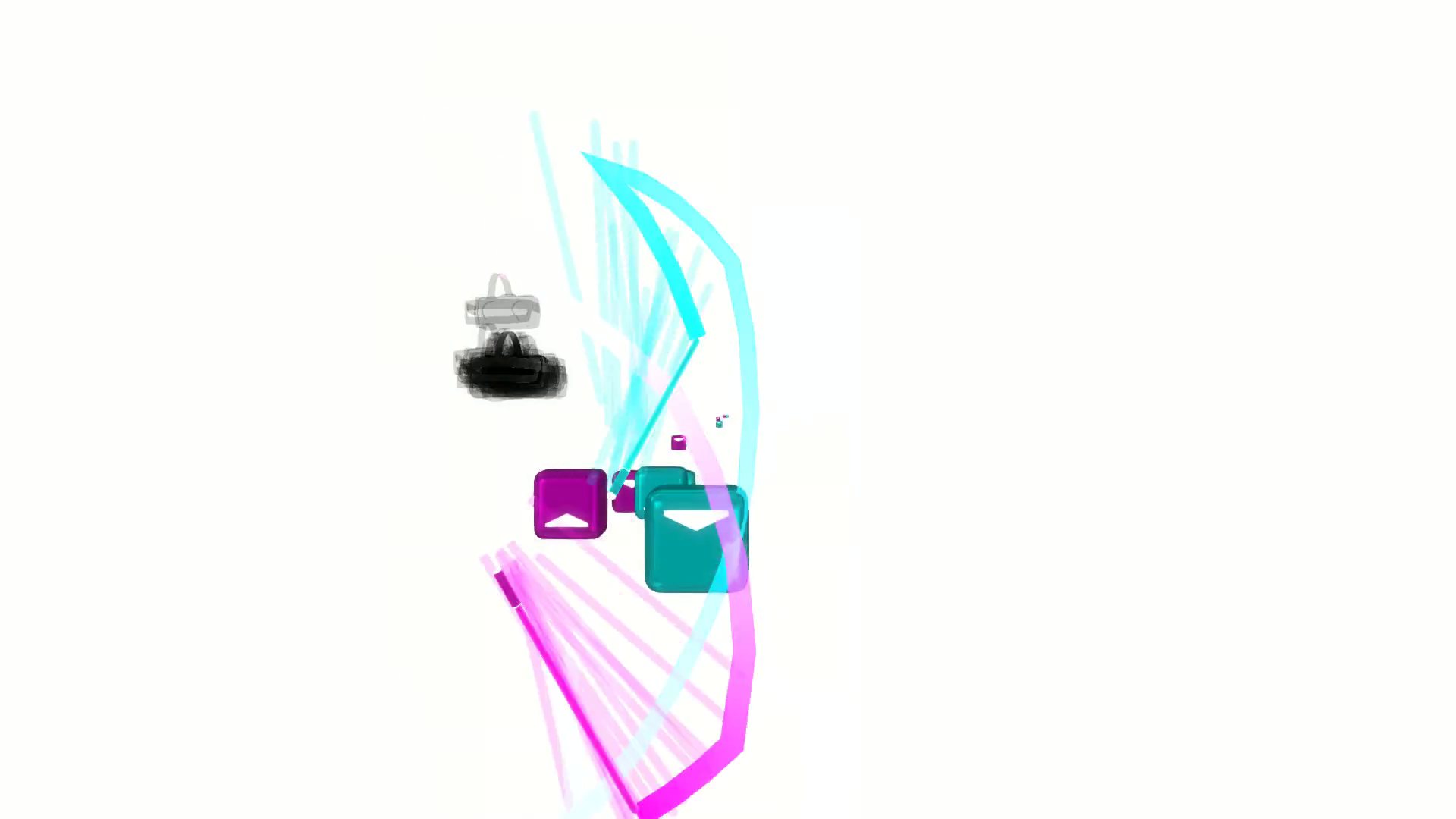}\hfill
\includegraphics[width=0.24\linewidth,clip,trim={200 0 860 0}]{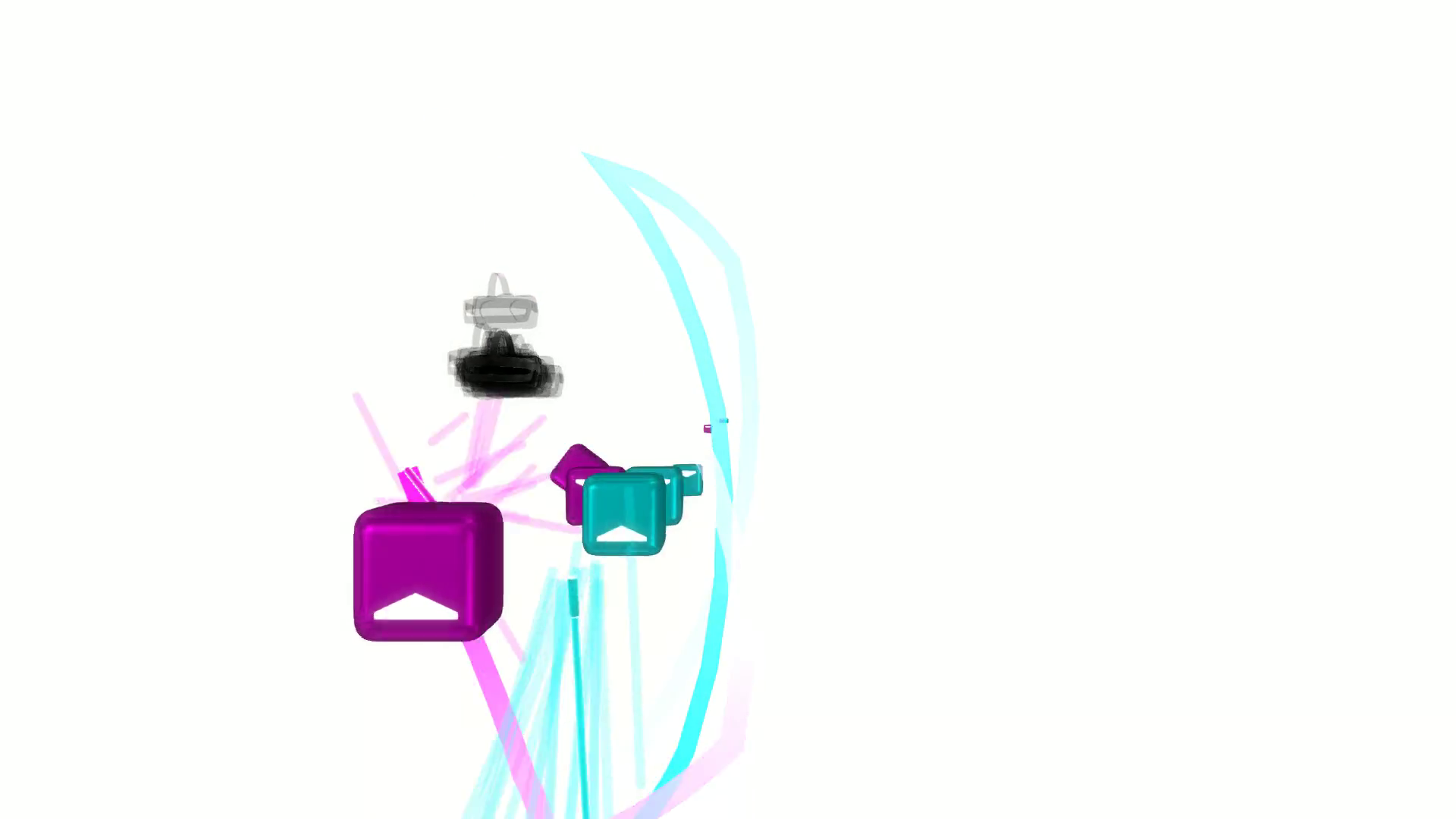}\hfill \includegraphics[width=0.24\linewidth,clip,trim={200 0 860 0}]{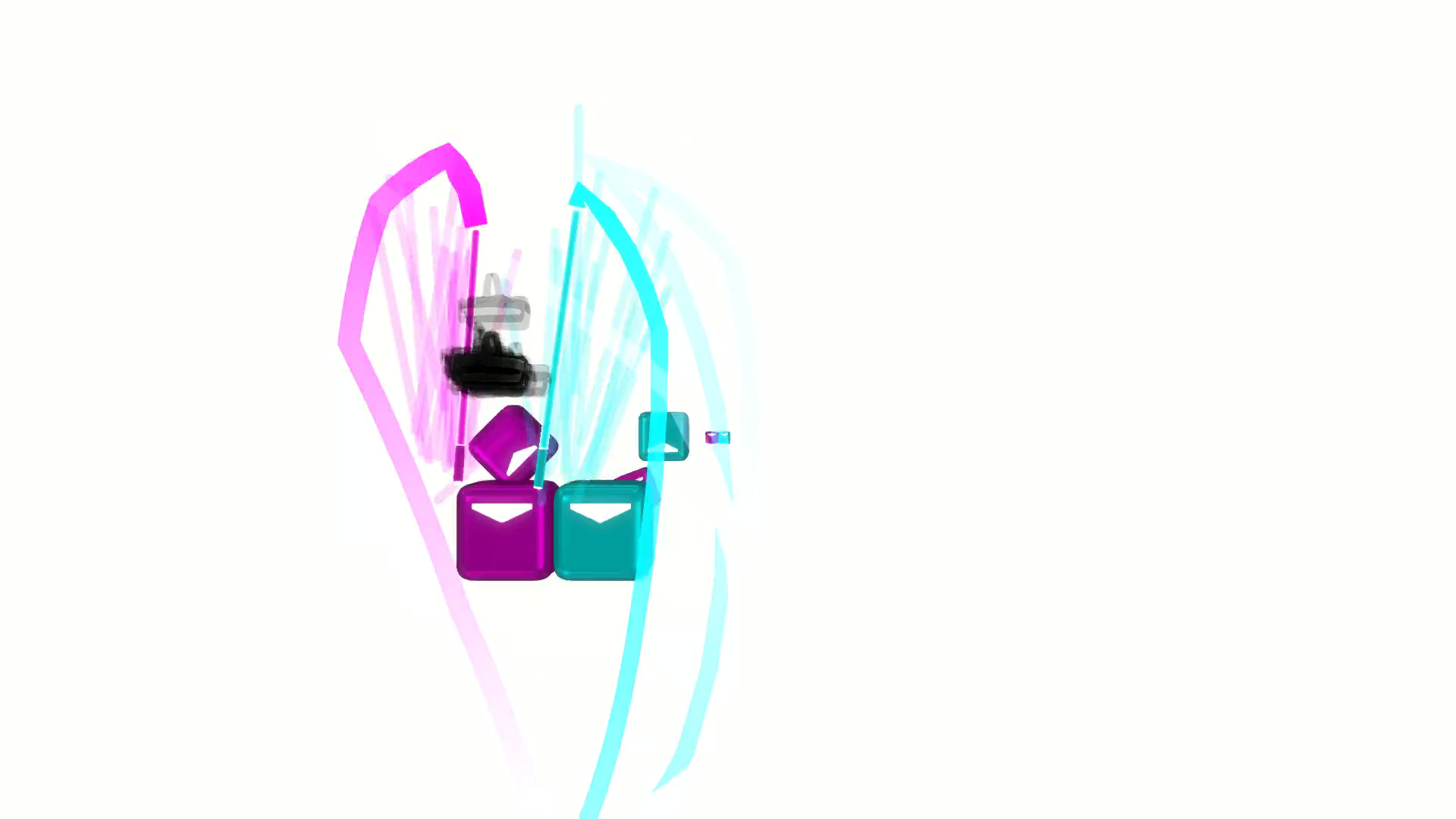}
\caption{Robo-Saber samples viable $3p$ trajectory samples (shown as semi-transparent headset and sabers) for the same game state input, from which the most optimal one is selected. Many possible trajectories for cutting the colored notes appear, and after evaluation, the one with the largest pre- and post-swing angle is selected. }
    \label{fig:cand}
\end{figure*}

\begin{figure*}
    \centering
\includegraphics[width=0.2\linewidth,clip,trim={200 0 960 0}]{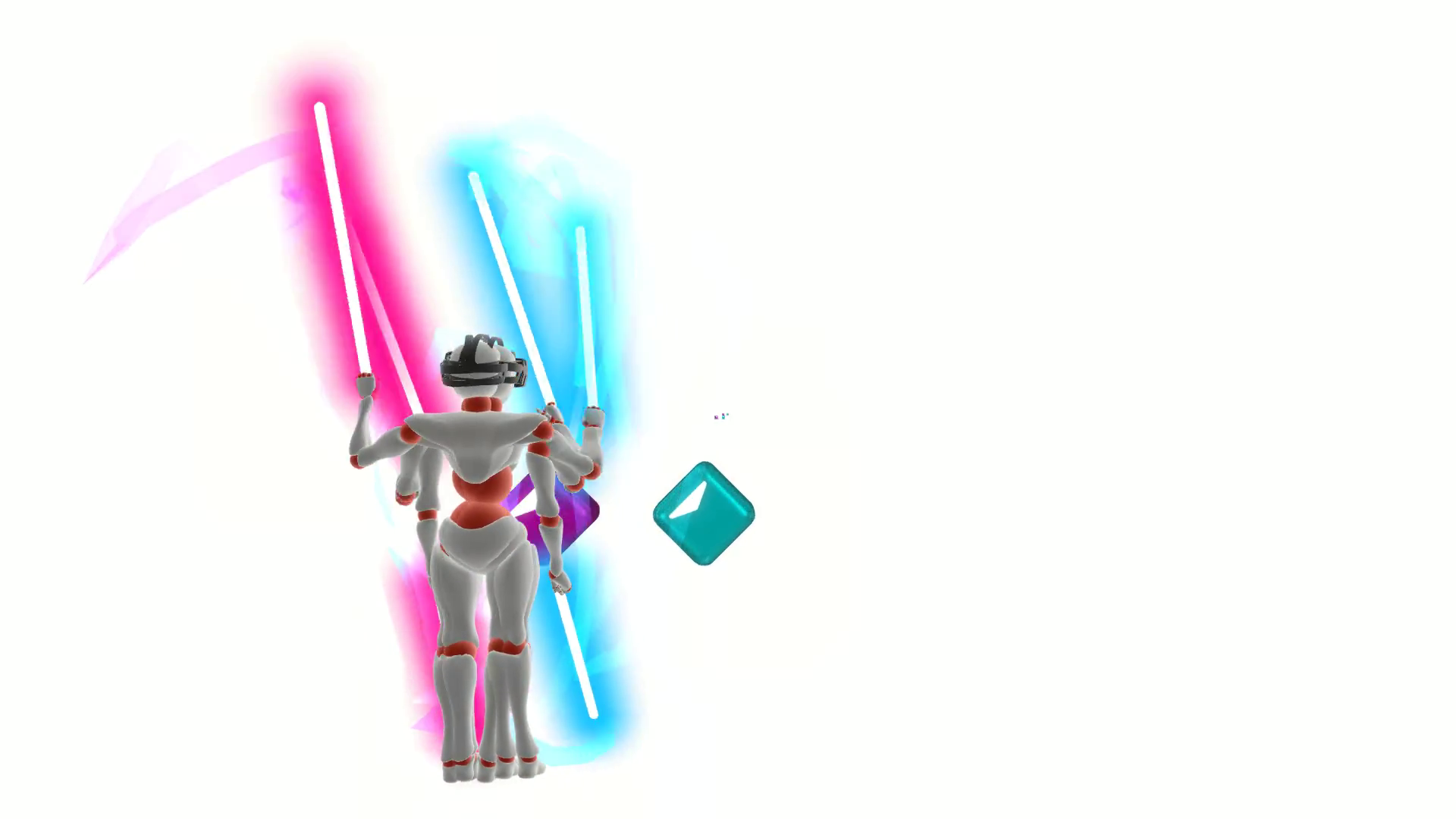}\hfill
 \includegraphics[width=0.2\linewidth,clip,trim={200 0 960 0}]{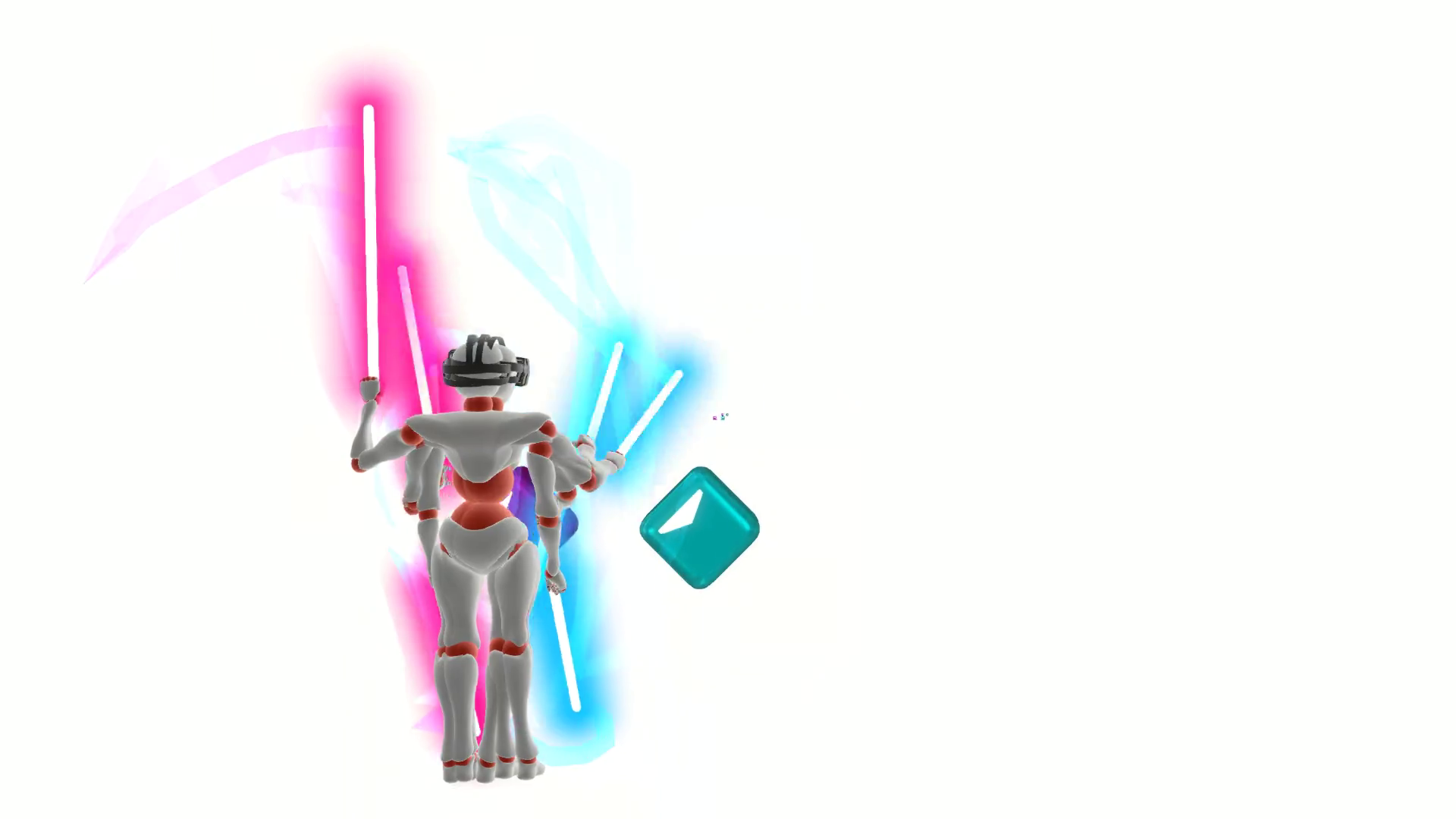}\hfill
 \includegraphics[width=0.2\linewidth,clip,trim={200 0 960 0}]{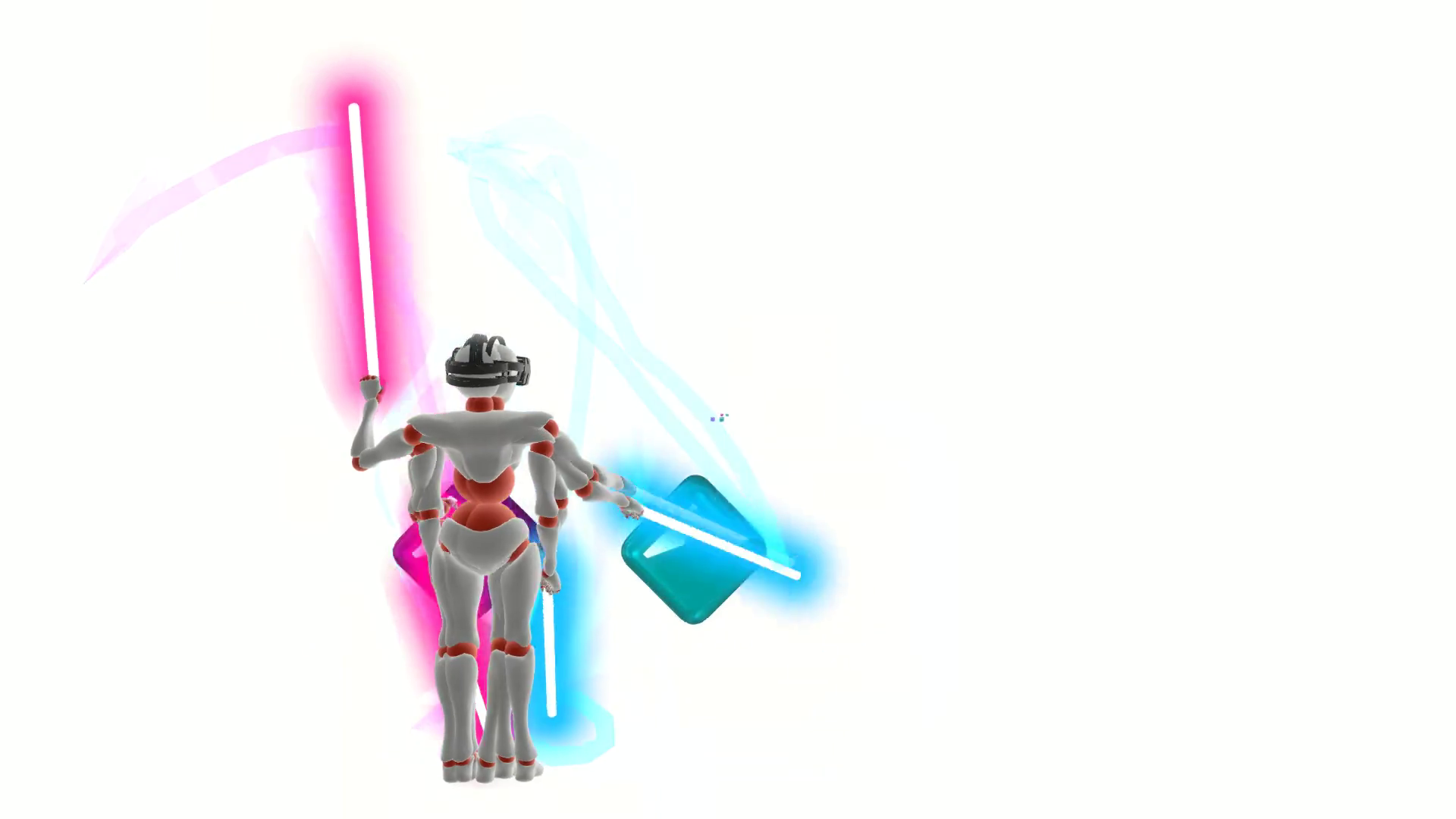}\hfill
 \includegraphics[width=0.2\linewidth,clip,trim={200 0 960 0}]{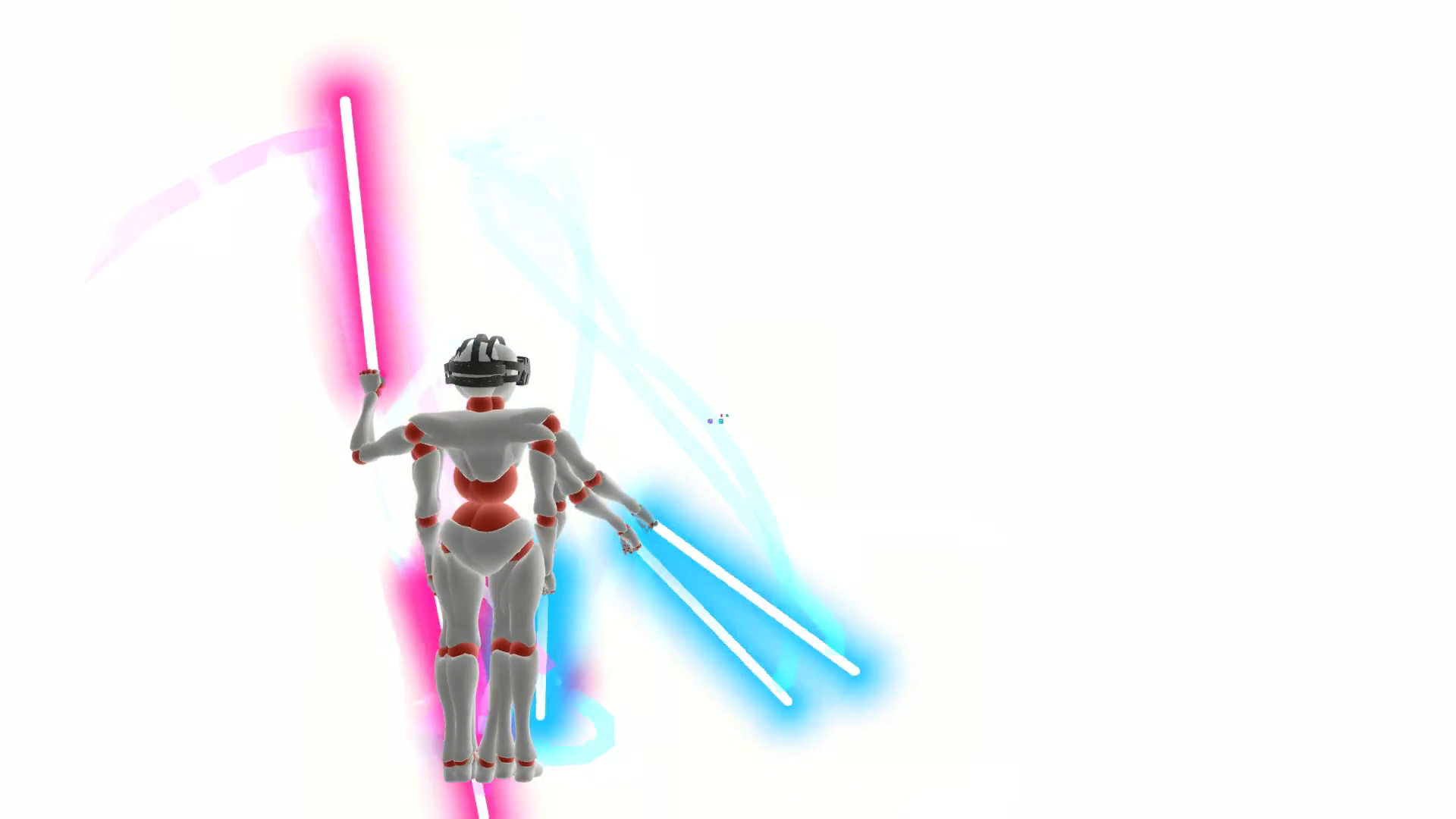}
 \includegraphics[width=0.2\linewidth,clip,trim={200 0 960 0}]{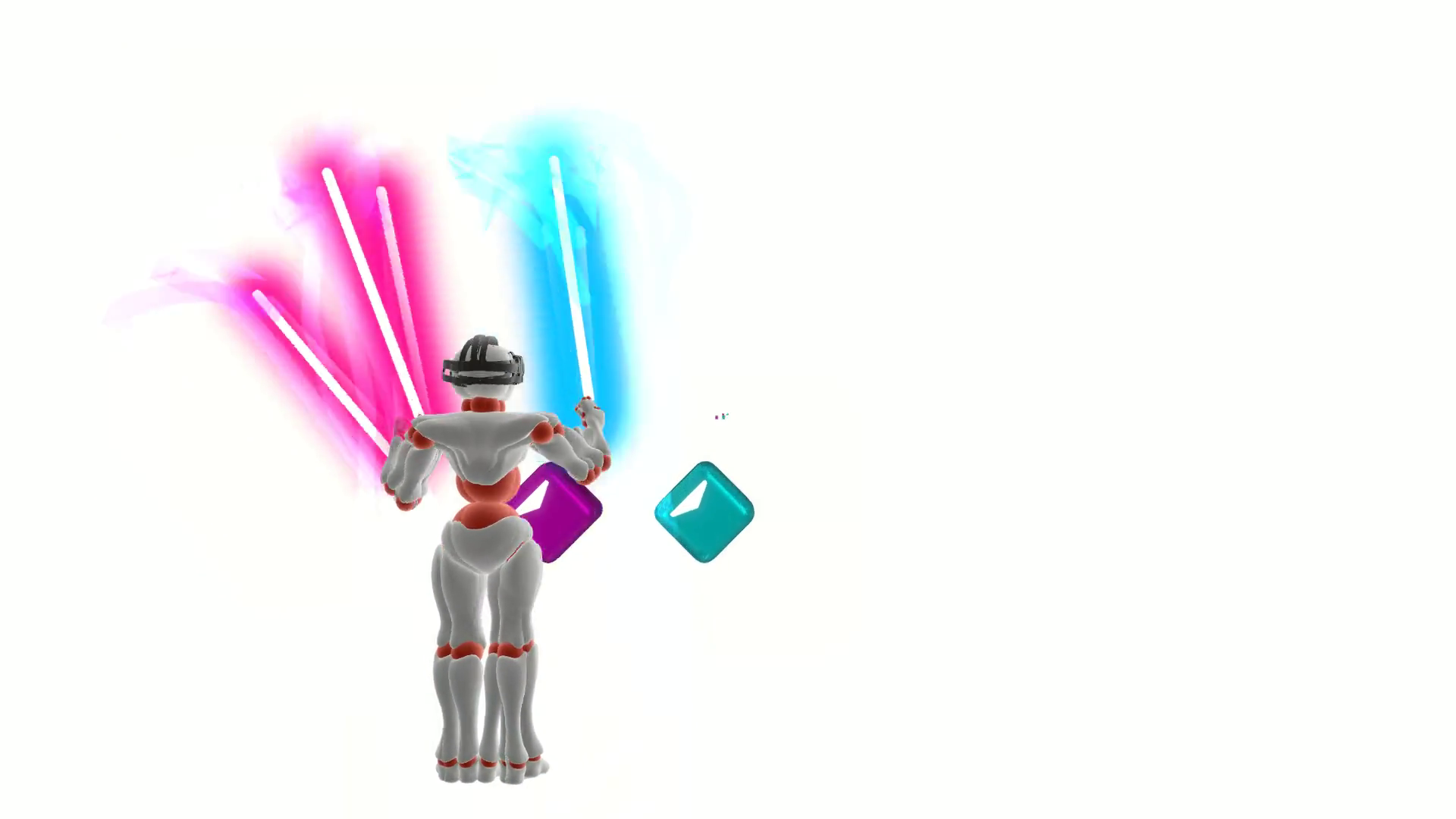}\hfill
 \includegraphics[width=0.2\linewidth,clip,trim={200 0 960 0}]{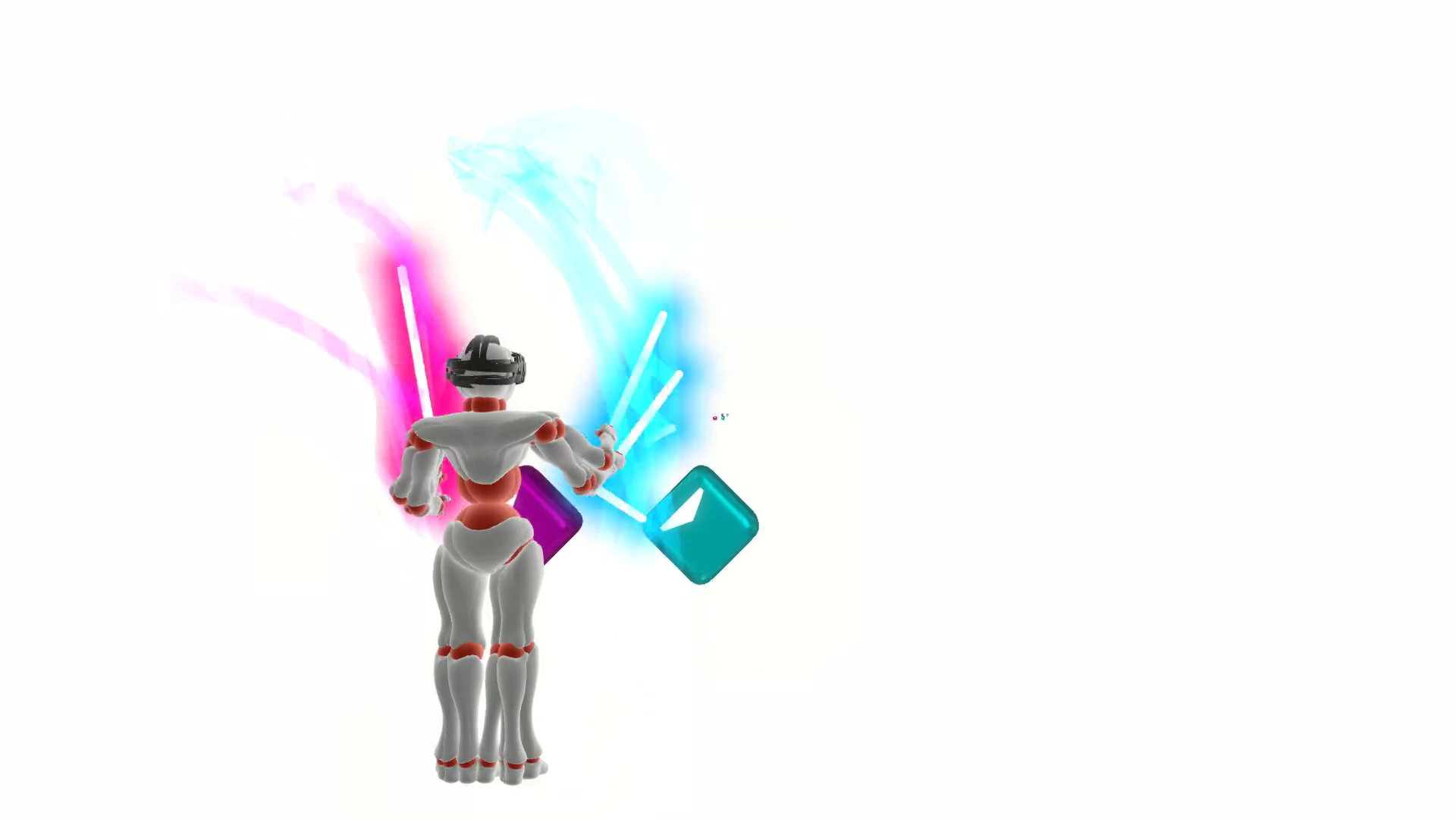}\hfill
 \includegraphics[width=0.2\linewidth,clip,trim={200 0 960 0}]{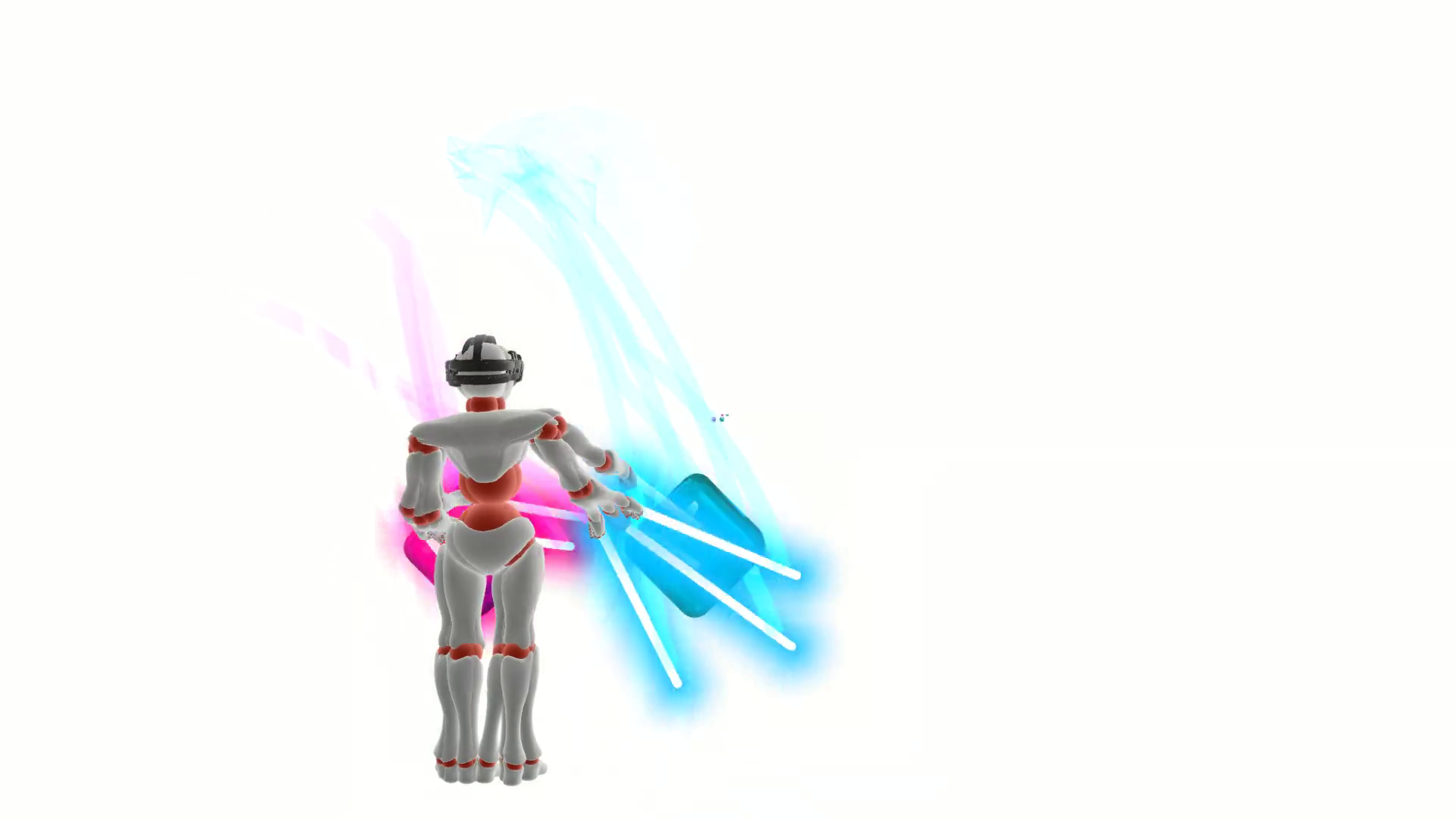}\hfill
 \includegraphics[width=0.2\linewidth,clip,trim={200 0 960 0}]{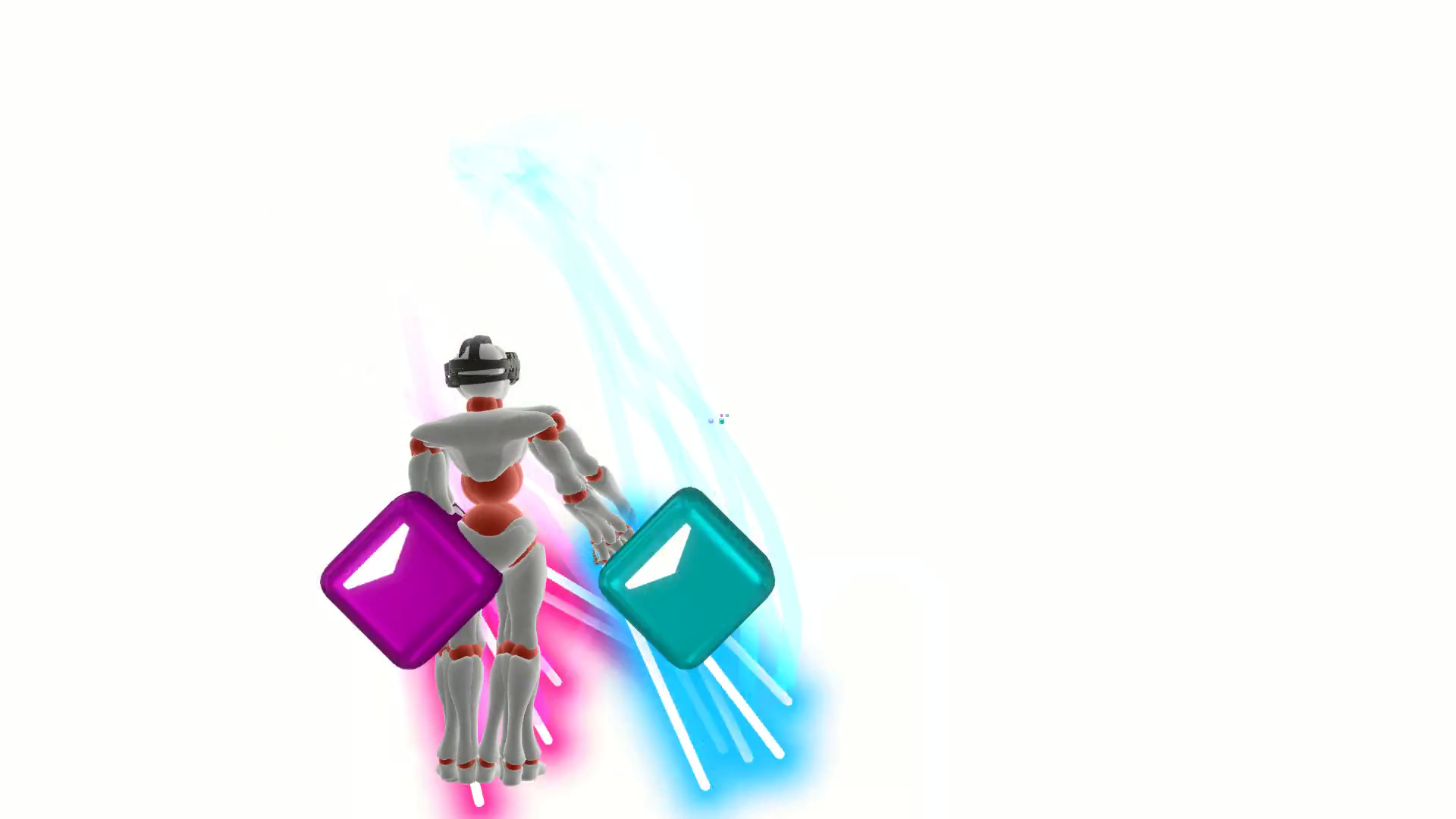}
    \caption{\textbf{Top:} Conditioning Robo-Saber on different $\RefSegments$ induces variations in behavior. Some correct and incorrect saber movements emerge, reflecting skill level variations present in conditioning signals. \textbf{Bottom:} Varying the random seed instead while conditioned on the same reference, the movement variation is less noticeable.}
    \label{fig:diversity}
\end{figure*}

\noindent \textit{Qualitative results.} 
As shown in the supplementary video, the full-body gameplay performance remains respectable when deployed on the most popular 1\% of maps. 
\Cref{fig:robo-saber-plays,fig:robo-saber-bombs,fig:robo-saber-obstacles} show image sequences demonstrating how the robot player moves in response to the observed colored notes, bombs, and obstacles. Unlike in previous physics-based user models, whole-body movements such as swaying and ducking can be observed.
As a native feature of our GS-VAE architecture, a diverse set of whole-body trajectories can be generated as a response to the same input (\Cref{fig:cand,fig:diversity}).
While random seed variations on the style-agnostic baseline exhibit some diversity in output, we observe significantly more visible diversity when varying the style references. 

\makeatletter
\renewcommand{\thesubsection}{Q6}
\makeatother

\subsection{\ul{How does physics-based tracking impact PSP?}}\label{sec:Q6}

\begin{figure*}
    \centering
    \includegraphics[width=0.99\linewidth, clip,trim={0 0 0 0}]{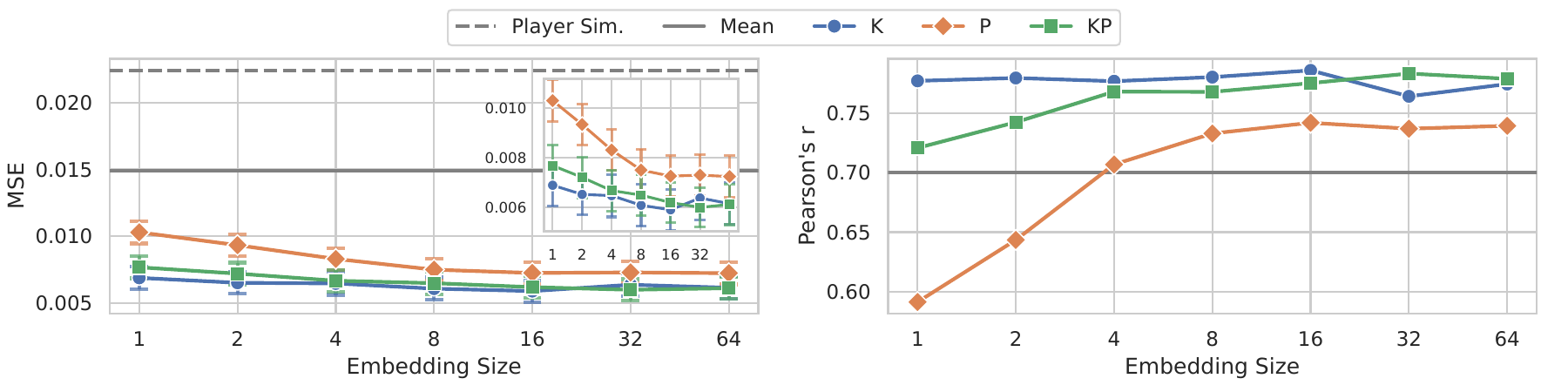}
    \caption{
    Examining the effect of including physics-based tracking results in factorization machine (FM) training.
    Each point corresponds to a converged FM model's report of the validation metric in question.
    K=kinematic, P=physics-based, KP=both kinematic and physics-based, annotating the synthetic dataset included in FM training.
    ``Mean'' is the baseline of using the mean score in $\mathbf{R}$ as a constant predictor.
    Standard error of the mean is visualized with bars at each point.
    \textbf{Left:} As the FM embedding size grows, MSE on $\mathbf{N}$ for KP converges to lower values than those of others, while possibly underfitting at smaller embedding sizes. P nearly competes with K, suggesting the presence of some meaningful signal for PSP.
    The inset figure zooms into the curves at the bottom.
    \textbf{Right:} Pearson's $r$ for $\mathbf{N}$ converges to higher values for KP at larger embedding sizes but shows diminishing returns.}
    \label{fig:fm_k_vs_kp}
\end{figure*}

As previously discussed, physics-based tracking introduces errors and constraints that degrade gameplay performance.
Rather than random, however, this degradation is owed to gameplay difficulty; prompted by this, we examine whether the physics-based player model's scores, though much less performant than the kinematic player model, can provide meaningful additional signals for user modeling.
To this end, we replicate the CF approach for PSP as above, but we further augment the synthetic Robo-Saber data with scores obtained with physics-based tracking.
Given that we need sufficiently performant and consistent scores, we elect to obtain synthetic physics-based data by simulating 3 randomly-chosen expert reference players across all maps in $\mathbf{N}$ and $\mathbf{R}$, instead of simulating the ground truth players as in \Cref{sec:Q4}.
We add the resulting physics-based data and the corresponding kinematic data in FM training, thereby isolating the signals introduced by tracking.

\Cref{fig:fm_k_vs_kp} illustrates evidence that the additional data might contain signals that can improve PSP performance.
The curves indicate that the FM model including physics-based data (KP) may underfit at smaller embedding sizes, while accuracy improves at larger ones.
Intuitively, learning meaningful embeddings for the maps in $\mathbf{N}$ is the key objective of our FM, and the additional signals added by physics-based synthetic data may help refine them.
However, as shown in the figure, the absolute PSP performance improvement by using additional scores alone is statistically weak; more experimentation may be necessary to ascertain the significance of physics-based gameplay data more concretely.
As we note in \Cref{sec:conclusion}, incorporating additional physical features such as energy expenditure and fatigue into the FM formulation may be beneficial to this end.

\section{Conclusion} \label{sec:conclusion}

This work introduces a generative framework for producing a VR player model that exhibits diverse, skilled gameplay behavior.
By calibrating to individual skill levels and movement patterns, Robo-Saber demonstrates that \emph{a player's gameplay can be simulated on entirely new game content} using a few example gameplay segments.
While the precise concept of difficulty is elusive, estimating how well a particular player would perform on a wide range of maps paves the way for practical applications.
Map designers could leverage Robo-Saber to simulate diverse user gameplay and assess their designs automatically.
Players could discover new content tailored to their skill levels and personal play styles, enhancing their overall experience.
These capabilities position the incorporation of style-aware player models in user modeling as a fertile research direction for enhancing VR application design and user experience, with our work contributing both a benchmark and baseline solution.

\noindent \textit{Applicability for other VR scenarios.}
We envision VR platforms' inherent telemetry and data-collection capabilities to enable developers of both large and small games to collect the motion data needed to train with our method.
In our work, Robo-Saber's ability to extract style signals relies on the behavioral diversity accumulated over 70k players.
How well this performs at smaller scales (\textit{e.g.}, dozens/hundreds of players) is an important direction for future work and may require improved feature engineering and model architectures.
Developing and evaluating a player model requires either programmatically accessing official gameplay or building a sufficiently faithful proxy for it (such as TorchSaber in this work).
Given such an environment and suitable $3p$ example data, our framework should generalize to VR scenarios beyond \emph{Beat Saber}.
Behavioral variation across players is ubiquitous in VR games due to morphological and habitual factors, and our work validates contextual exemplar embeddings as a general approach to capture this diversity.
As \emph{tabula-rasa} DRL remains infeasible for modeling diverse player behaviors, learning from data-mined play sequences will likely remain a promising direction.
Currently, \emph{Beat Saber} is the only VR game with substantial open-source gameplay data.
We hope to spur further data collection and follow-up work on generative player models for other VR games.

\noindent \textit{Utility of automated playtesting.}
We emphasize that our goal is to augment, not replace, costly human playtesting.
Presently, only human playtesters can provide reliable information about subjective experience like fun.
Robo-Saber's primary utility is to test the feasibility of a new map and predict its difficulty relative to other maps, helping craft a smooth difficulty curve.
Evaluating the general utility of automated playtesting for design and personalization would be an orthogonal yet exciting direction for future work.

\noindent \textit{Performance of physics-based tracking.}
Further work is needed before a physics-based player model can fully realize its benefits.
A primary limitation of our system is the degraded gameplay performance due to physics-based tracking.
As some generated $3p$ goals are too challenging or dynamically infeasible for the tracker, a tighter communication between the kinematic generator and the tracking controller could alleviate the issue.
Additionally, as seen in the video, the produced whole-body motions remain far from human-like or realistic.
We highlight bridging these gaps as a compelling direction for future work.
To maximize physics-based gameplay performance, one might integrate the simulation of not only the game mechanics but also physics and tracking performance into candidate trajectory selection, or fine-tune the $3p$ generation and/or physics-based tracking end-to-end using DRL.

In potential follow-up work, we also propose utilizing quantities from physics simulations to enhance the modeling of player diversity and to estimate variables beyond game score. For instance, the cumulative fatigue modeling presented in \citen{cheema2020predicting,cheema2023discovering} may be useful. A simulation model with sufficiently realistic anatomy could be used for evaluating the safety of evoked player movements.
We are also interested in modeling variations in body proportions, weight, and strength, which our current physics-based controller cannot handle.

\section*{Acknowledgements}

We would like to thank Adas Slezas and Markus Laattala for their generous help with custom motion capture collection, Vivek Nair and Džiugas Ramonas (aka CyberRamen) for their guidance during the early phases of this project, and Pauli Kemppinen, Heikki Timonen, Erik H\"ark\"onen, and Michiel van de Panne for the insightful discussion.
Our game visualizations are built on AllPoland's wonderful ArcViewer project.
Finally, we thank \emph{Beat Saber}'s custom mapping community and the original authors of the maps and music featured in our project.
We acknowledge the following funding sources: NSERC Postgraduate Scholarship--Doctorate (PGS-D, 567794-2022) and FCAI Doctoral Open Application Grant.

\bibliographystyle{eg-alpha-doi}  
\bibliography{egbibsample}        

@inproceedings{nair2023unique,
  title={Unique identification of 50,000+ virtual reality users from head \& hand motion data},
  author={Nair, Vivek and Guo, Wenbo and Mattern, Justus and Wang, Rui and O'Brien, James F and Rosenberg, Louis and Song, Dawn},
  booktitle={32nd USENIX Security Symposium (USENIX Security 23)},
  pages={895--910},
  year={2023}
}

@article{xie2023hierarchical,
  title={Hierarchical planning and control for box loco-manipulation},
  author={Xie, Zhaoming and Tseng, Jonathan and Starke, Sebastian and van de Panne, Michiel and Liu, C Karen},
  journal={Proceedings of the ACM on Computer Graphics and Interactive Techniques},
  volume={6},
  number={3},
  pages={1--18},
  year={2023},
  publisher={ACM New York, NY, USA}
}

@inproceedings{xu2025parc,
  title={Parc: Physics-based augmentation with reinforcement learning for character controllers},
  author={Xu, Michael and Shi, Yi and Yin, KangKang and Peng, Xue Bin},
  booktitle={Proceedings of the Special Interest Group on Computer Graphics and Interactive Techniques Conference Conference Papers},
  pages={1--11},
  year={2025}
}

@article{tevet2024closd,
  title={Closd: Closing the loop between simulation and diffusion for multi-task character control},
  author={Tevet, Guy and Raab, Sigal and Cohan, Setareh and Reda, Daniele and Luo, Zhengyi and Peng, Xue Bin and Bermano, Amit H and van de Panne, Michiel},
  journal={arXiv preprint arXiv:2410.03441},
  year={2024}
}

@incollection{wilcoxon1992individual,
  title={Individual comparisons by ranking methods},
  author={Wilcoxon, Frank},
  booktitle={Breakthroughs in statistics: Methodology and distribution},
  pages={196--202},
  year={1992},
  publisher={Springer}
}

@inproceedings{kim2025personabooth,
  title={PersonaBooth: Personalized Text-to-Motion Generation},
  author={Kim, Boeun and Jeong, Hea In and Sung, JungHoon and Cheng, Yihua and Lee, Jeongmin and Chang, Ju Yong and Choi, Sang-Il and Choi, Younggeun and Shin, Saim and Kim, Jungho and others},
  booktitle={Proceedings of the Computer Vision and Pattern Recognition Conference},
  pages={22756--22765},
  year={2025}
}

@article{sawdayee2025dance,
  title={Dance Like a Chicken: Low-Rank Stylization for Human Motion Diffusion},
  author={Sawdayee, Haim and Guo, Chuan and Tevet, Guy and Zhou, Bing and Wang, Jian and Bermano, Amit H},
  journal={arXiv preprint arXiv:2503.19557},
  year={2025}
}

@inproceedings{raab2024monkey,
  title={Monkey see, monkey do: Harnessing self-attention in motion diffusion for zero-shot motion transfer},
  author={Raab, Sigal and Gat, Inbar and Sala, Nathan and Tevet, Guy and Shalev-Arkushin, Rotem and Fried, Ohad and Bermano, Amit Haim and Cohen-Or, Daniel},
  booktitle={SIGGRAPH Asia 2024 Conference Papers},
  pages={1--13},
  year={2024}
}

@article{jiao2025bach,
  title={BACH: Bi-Stage Data-Driven Piano Performance Animation for Controllable Hand Motion},
  author={Jiao, Jihui and Zeng, Rui and Dai, Ju and Pan, Junjun},
  journal={Computer Animation and Virtual Worlds},
  volume={36},
  number={3},
  pages={e70044},
  year={2025},
  publisher={Wiley Online Library}
}

@inproceedings{chen2025midi,
  title={From MIDI to Motion: Learning to Play the Piano at Scale with Bi-Manual Dexterous Robot Hands},
  author={Chen, Le and Zhao, Yi and Schneider, Jan and Gao, Quankai and Kannala, Juho and Sch{\"o}lkopf, Bernhard and Pajarinen, Joni and B{\"u}chler, Dieter},
  booktitle={ICRA 2025 Workshop''Handy Moves: Dexterity in Multi-Fingered Hands''Paper Submission}
}

@inproceedings{wang2024furelise,
  title={F{\"u}rElise: Capturing and physically synthesizing hand motion of piano performance},
  author={Wang, Ruocheng and Xu, Pei and Shi, Haochen and Schumann, Elizabeth and Liu, C Karen},
  booktitle={SIGGRAPH Asia 2024 Conference Papers},
  pages={1--11},
  year={2024}
}

@article{ao2023gesturediffuclip,
  title={Gesturediffuclip: Gesture diffusion model with clip latents},
  author={Ao, Tenglong and Zhang, Zeyi and Liu, Libin},
  journal={ACM Transactions on Graphics (TOG)},
  volume={42},
  number={4},
  pages={1--18},
  year={2023},
  publisher={ACM New York, NY, USA}
}

@article{alexanderson2023listen,
  title={Listen, denoise, action! audio-driven motion synthesis with diffusion models},
  author={Alexanderson, Simon and Nagy, Rajmund and Beskow, Jonas and Henter, Gustav Eje},
  journal={ACM Transactions on Graphics (TOG)},
  volume={42},
  number={4},
  pages={1--20},
  year={2023},
  publisher={ACM New York, NY, USA}
}

@article{xiao2025motionstreamer,
  title={MotionStreamer: Streaming Motion Generation via Diffusion-based Autoregressive Model in Causal Latent Space},
  author={Xiao, Lixing and Lu, Shunlin and Pi, Huaijin and Fan, Ke and Pan, Liang and Zhou, Yueer and Feng, Ziyong and Zhou, Xiaowei and Peng, Sida and Wang, Jingbo},
  journal={arXiv preprint arXiv:2503.15451},
  year={2025}
}

@article{ji2025sport,
  title={Sport: From zero-shot prompts to real-time motion generation},
  author={Ji, Bin and Pan, Ye and Liu, Zhimeng and Tan, Shuai and Yang, Xiaokang},
  journal={IEEE Transactions on Visualization and Computer Graphics},
  year={2025},
  publisher={IEEE}
}

@inproceedings{zhong2024smoodi,
  title={Smoodi: Stylized motion diffusion model},
  author={Zhong, Lei and Xie, Yiming and Jampani, Varun and Sun, Deqing and Jiang, Huaizu},
  booktitle={European Conference on Computer Vision},
  pages={405--421},
  year={2024},
  organization={Springer}
}

@inproceedings{tao2022style,
  title={Style-ERD: Responsive and coherent online motion style transfer},
  author={Tao, Tianxin and Zhan, Xiaohang and Chen, Zhongquan and van de Panne, Michiel},
  booktitle={Proceedings of the IEEE/CVF Conference on Computer Vision and Pattern Recognition},
  pages={6593--6603},
  year={2022}
}

@article{smith2019efficient,
  title={Efficient neural networks for real-time motion style transfer},
  author={Smith, Harrison Jesse and Cao, Chen and Neff, Michael and Wang, Yingying},
  journal={Proceedings of the ACM on Computer Graphics and Interactive Techniques},
  volume={2},
  number={2},
  pages={1--17},
  year={2019},
  publisher={ACM New York, NY, USA}
}

@article{holden2017fast,
  title={Fast neural style transfer for motion data},
  author={Holden, Daniel and Habibie, Ikhsanul and Kusajima, Ikuo and Komura, Taku},
  journal={IEEE computer graphics and applications},
  volume={37},
  number={4},
  pages={42--49},
  year={2017},
  publisher={IEEE}
}

@article{aberman2020unpaired,
  title={Unpaired motion style transfer from video to animation},
  author={Aberman, Kfir and Weng, Yijia and Lischinski, Dani and Cohen-Or, Daniel and Chen, Baoquan},
  journal={ACM Transactions On Graphics (TOG)},
  volume={39},
  number={4},
  pages={64--1},
  year={2020},
  publisher={ACM New York, NY, USA}
}

@article{yumer2016spectral,
  title={Spectral style transfer for human motion between independent actions},
  author={Yumer, M Ersin and Mitra, Niloy J},
  journal={ACM Transactions on Graphics (TOG)},
  volume={35},
  number={4},
  pages={1--8},
  year={2016},
  publisher={ACM New York, NY, USA}
}

@inproceedings{du2019stylistic,
  title={Stylistic locomotion modeling and synthesis using variational generative models},
  author={Du, Han and Herrmann, Erik and Sprenger, Janis and Fischer, Klaus and Slusallek, Philipp},
  booktitle={Proceedings of the 12th ACM SIGGRAPH Conference on Motion, Interaction and Games},
  pages={1--10},
  year={2019}
}

@article{holden2016deep,
  title={A deep learning framework for character motion synthesis and editing},
  author={Holden, Daniel and Saito, Jun and Komura, Taku},
  journal={ACM Transactions on Graphics (ToG)},
  volume={35},
  number={4},
  pages={1--11},
  year={2016},
  publisher={ACM New York, NY, USA}
}

@article{xia2015realtime,
  title={Realtime style transfer for unlabeled heterogeneous human motion},
  author={Xia, Shihong and Wang, Congyi and Chai, Jinxiang and Hodgins, Jessica},
  journal={ACM Transactions on Graphics (TOG)},
  volume={34},
  number={4},
  pages={1--10},
  year={2015},
  publisher={ACM New York, NY, USA}
}

@inproceedings{guo2024generative,
  title     = {Generative Human Motion Stylization in Latent Space},
  author    = {Guo, Yuhan and Mu, Yuting and Zuo, Yiming and Dai, Jianrong and Yan, Xiaolong and Lu, Yiyang and Cheng, Xu},
  booktitle = {Proceedings of the International Conference on Learning Representations (ICLR)},
  year      = {2024},
  url       = {https://arxiv.org/abs/2401.13505}
}

@inproceedings{wen2021autoregressive,
  title     = {Autoregressive Stylized Motion Synthesis With Generative Flow},
  author    = {Wen, Yu-Hui and Chen, Ching-Chih and Hsu, Kun-Huang and Wang, Jun-Cheng},
  booktitle = {Proceedings of the IEEE/CVF Conference on Computer Vision and Pattern Recognition (CVPR)},
  year      = {2021},
  pages     = {14504--14513},
  url       = {https://arxiv.org/abs/2105.01963}
}

@inproceedings{nair2024deep,
  title={Deep motion masking for secure, usable, and scalable real-time anonymization of ecological virtual reality motion data},
  author={Nair, Vivek and Guo, Wenbo and O'Brien, James F and Rosenberg, Louis and Song, Dawn},
  booktitle={2024 IEEE Conference on Virtual Reality and 3D User Interfaces Abstracts and Workshops (VRW)},
  pages={493--500},
  year={2024},
  organization={IEEE}
}

@article{truong2025beyondmimic,
  title={BeyondMimic: From Motion Tracking to Versatile Humanoid Control via Guided Diffusion},
  author={Truong, Takara E and Liao, Qiayuan and Huang, Xiaoyu and Tevet, Guy and Liu, C Karen and Sreenath, Koushil},
  journal={arXiv preprint arXiv:2508.08241},
  year={2025}
}

@article{huang2025diffuse,
  title={Diffuse-cloc: Guided diffusion for physics-based character look-ahead control},
  author={Huang, Xiaoyu and Truong, Takara and Zhang, Yunbo and Yu, Fangzhou and Sleiman, Jean Pierre and Hodgins, Jessica and Sreenath, Koushil and Farshidian, Farbod},
  journal={ACM Transactions on Graphics (TOG)},
  volume={44},
  number={4},
  pages={1--12},
  year={2025},
  publisher={ACM New York, NY, USA}
}

@inproceedings{de2022automated,
  title={Automated play-testing through RL based human-like play-styles generation},
  author={de Woillemont, Pierre Le Pelletier and Labory, R{\'e}mi and Corruble, Vincent},
  booktitle={Proceedings of the AAAI Conference on Artificial Intelligence and Interactive Digital Entertainment},
  volume={18},
  number={1},
  pages={146--154},
  year={2022}
}

@article{ioannidis2025diffusion,
  title={Diffusion-based planning with learned viability filters},
  author={Ioannidis, Nicholas and Reda, Daniele and Cohan, Setareh and van de Panne, Michiel},
  journal={Proceedings of the ACM on Computer Graphics and Interactive Techniques},
  volume={8},
  number={4},
  pages={1--23},
  year={2025},
  publisher={ACM New York, NY}
}

@article{dosovitskiy2020image,
  title={An image is worth 16x16 words: Transformers for image recognition at scale},
  author={Dosovitskiy, Alexey and Beyer, Lucas and Kolesnikov, Alexander and Weissenborn, Dirk and Zhai, Xiaohua and Unterthiner, Thomas and Dehghani, Mostafa and Minderer, Matthias and Heigold, Georg and Gelly, Sylvain and others},
  journal={arXiv preprint arXiv:2010.11929},
  year={2020}
}

@inproceedings{devlin2019bert,
  title={Bert: Pre-training of deep bidirectional transformers for language understanding},
  author={Devlin, Jacob and Chang, Ming-Wei and Lee, Kenton and Toutanova, Kristina},
  booktitle={Proceedings of the 2019 conference of the North American chapter of the association for computational linguistics: human language technologies, volume 1 (long and short papers)},
  pages={4171--4186},
  year={2019}
}

@inproceedings{zhou2019continuity,
  title={On the continuity of rotation representations in neural networks},
  author={Zhou, Yi and Barnes, Connelly and Lu, Jingwan and Yang, Jimei and Li, Hao},
  booktitle={Proceedings of the IEEE/CVF conference on computer vision and pattern recognition},
  pages={5745--5753},
  year={2019}
}

@article{barquero2025sparse,
  title={From Sparse Signal to Smooth Motion: Real-Time Motion Generation with Rolling Prediction Models},
  author={Barquero, German and Bertsch, Nadine and Marramreddy, Manojkumar and Chac{\'o}n, Carlos and Arcadu, Filippo and Rigual, Ferran and He, Nicky Sijia and Palmero, Cristina and Escalera, Sergio and Ye, Yuting and others},
  journal={arXiv preprint arXiv:2504.05265},
  year={2025}
}

@inproceedings{shi2025simulating,
  title={Simulating Errors in Touchscreen Typing},
  author={Shi, Danqing and Zhu, Yujun and Fernandes Junior, Francisco Erivaldo and Zhai, Shumin and Oulasvirta, Antti},
  booktitle={Proceedings of the 2025 CHI Conference on Human Factors in Computing Systems},
  pages={1--13},
  year={2025}
}

@inproceedings{kristensen2020estimating,
  title={Estimating player completion rate in mobile puzzle games using reinforcement learning},
  author={Kristensen, Jeppe Theiss and Valdivia, Arturo and Burelli, Paolo},
  booktitle={2020 IEEE Conference on Games (CoG)},
  pages={636--639},
  year={2020},
  organization={IEEE}
}

@article{Nair:2023:ROT,
  doi = {10.48550/arXiv.2305.14320},
  title = {Results of the 2023 Census of Beat Saber Users: Virtual Reality Gaming Population Insights and Factors Affecting Virtual Reality E-Sports Performance},
  author = {Vivek Nair and Viktor Radulov and James F. O'Brien},
  month = may,
  year = 2023,
  pages = {1-19},
  url = {http://graphics.berkeley.edu/papers/Nair-ROT-2023-05/},
}

@phdthesis{Ruhf2020,
  author  = "Kaitlyn Danielle Ruhf",
  title   = "Physically Active Virtual Reality and Parkinson’s Disease: A Pilot Study",
  school  = "Wake Forest University",
  year    = "2020"
}

@article{Grospretre2023,
  doi = {10.1111/cogs.13278},
  title = {How Exergaming with Virtual Reality Enhances Specific Cognitive and Visuo-Motor Abilities: An Explorative Study},
  author = {Sidney Grosprêtre and Philémon Marcel-Millet and Pauline Eon and Bettina Wollesen},
  month = apr,
  year = 2023,
  vol = 24,
  num = 4,
  publisher = "Wiley",
  booktitle = "Cognitive Science",
  pages = {e13278},
  url = {https://doi.org/10.1111/cogs.13278},
}

@phdthesis{Tuong2019,
  author  = "Tuong Thai",
  title   = "The Influence Of Exergaming On Heart Rate, Perceived Exertion, Motivation To Exercise, And Time Spent Exercising",
  school  = "Salem University",
  year    = "2021"
}

@inproceedings{mahmood2019amass,
  title={AMASS: Archive of motion capture as surface shapes},
  author={Mahmood, Naureen and Ghorbani, Nima and Troje, Nikolaus F and Pons-Moll, Gerard and Black, Michael J},
  booktitle={Proceedings of the IEEE/CVF international conference on computer vision},
  pages={5442--5451},
  year={2019}
}

@article{makoviychuk2021isaac,
  title={Isaac gym: High performance gpu-based physics simulation for robot learning},
  author={Makoviychuk, Viktor and Wawrzyniak, Lukasz and Guo, Yunrong and Lu, Michelle and Storey, Kier and Macklin, Miles and Hoeller, David and Rudin, Nikita and Allshire, Arthur and Handa, Ankur and others},
  journal={arXiv preprint arXiv:2108.10470},
  year={2021}
}

@misc{BSMGWiki,
  author       = {{Beat Saber Modding Group}},
  shortauthor  = {MT},
  title        = "Beat Saber Modding Group Wiki",
  howpublished = "Web page",
  month        = "November",
  year         = "2019",
  url         = "https://bsmg.wiki/",
  key={BSMG}
}

@misc{BeatSaver,
  author       = "BeatSaver Team",
  title        = "BeatSaver",
  howpublished = "Web page",
  month        = "April",
  year         = "2021",
  url={https://beatsaver.com/},
  key={BeatSaver}
}

@inproceedings{jokinen2021touchscreen,
  title={Touchscreen typing as optimal supervisory control},
  author={Jokinen, Jussi and Acharya, Aditya and Uzair, Mohammad and Jiang, Xinhui and Oulasvirta, Antti},
  booktitle={Proceedings of the 2021 CHI conference on human factors in computing systems},
  pages={1--14},
  year={2021}
}

@inproceedings{oulasvirta2022computational,
  title={Computational rationality as a theory of interaction},
  author={Oulasvirta, Antti and Jokinen, Jussi PP and Howes, Andrew},
  booktitle={Proceedings of the 2022 CHI Conference on Human Factors in Computing Systems},
  pages={1--14},
  year={2022}
}

@inproceedings{roohi2020predicting,
  title={Predicting game difficulty and churn without players},
  author={Roohi, Shaghayegh and Relas, Asko and Takatalo, Jari and Heiskanen, Henri and H{\"a}m{\"a}l{\"a}inen, Perttu},
  booktitle={Proceedings of the Annual Symposium on Computer-Human Interaction in Play},
  pages={585--593},
  year={2020}
}

@inproceedings{gudmundsson2018human,
  title={Human-like playtesting with deep learning},
  author={Gudmundsson, Stefan Freyr and Eisen, Philipp and Poromaa, Erik and Nodet, Alex and Purmonen, Sami and Kozakowski, Bartlomiej and Meurling, Richard and Cao, Lele},
  booktitle={2018 IEEE Conference on Computational Intelligence and Games (CIG)},
  pages={1--8},
  year={2018},
  organization={IEEE}
}

@article{roohi2021predicting,
  title={Predicting game difficulty and engagement using AI players},
  author={Roohi, Shaghayegh and Guckelsberger, Christian and Relas, Asko and Heiskanen, Henri and Takatalo, Jari and H{\"a}m{\"a}l{\"a}inen, Perttu},
  journal={Proceedings of the ACM on Human-Computer Interaction},
  volume={5},
  number={CHI PLAY},
  pages={1--17},
  year={2021},
  publisher={ACM New York, NY, USA}
}

@inproceedings{du2023avatars,
  title={Avatars grow legs: Generating smooth human motion from sparse tracking inputs with diffusion model},
  author={Du, Yuming and Kips, Robin and Pumarola, Albert and Starke, Sebastian and Thabet, Ali and Sanakoyeu, Artsiom},
  booktitle={Proceedings of the IEEE/CVF Conference on Computer Vision and Pattern Recognition},
  pages={481--490},
  year={2023}
}

@inproceedings{ye2022neural3points,
  title={Neural3Points: Learning to Generate Physically Realistic Full-body Motion for Virtual Reality Users},
  author={Ye, Yongjing and Liu, Libin and Hu, Lei and Xia, Shihong},
  booktitle={Computer Graphics Forum},
  volume={41},
  number={8},
  pages={183--194},
  year={2022},
  organization={Wiley Online Library}
}

@inproceedings{kristensen2022personalized,
  title={Personalized game difficulty prediction using factorization machines},
  author={Kristensen, Jeppe Theiss and Guckelsberger, Christian and Burelli, Paolo and H{\"a}m{\"a}l{\"a}inen, Perttu},
  booktitle={Proceedings of the 35th Annual ACM Symposium on User Interface Software and Technology},
  pages={1--13},
  year={2022}
}

@inproceedings{rendle2010factorization,
  title={Factorization machines},
  author={Rendle, Steffen},
  booktitle={2010 IEEE International conference on data mining},
  pages={995--1000},
  year={2010},
  organization={IEEE}
}

@inproceedings{
tevet2023human,
title={Human Motion Diffusion Model},
author={Guy Tevet and Sigal Raab and Brian Gordon and Yoni Shafir and Daniel Cohen-or and Amit Haim Bermano},
booktitle={The Eleventh International Conference on Learning Representations },
year={2023},
url={https://openreview.net/forum?id=SJ1kSyO2jwu}
}

@article{holden2017phase,
  title={Phase-functioned neural networks for character control},
  author={Holden, Daniel and Komura, Taku and Saito, Jun},
  journal={ACM Transactions on Graphics (TOG)},
  volume={36},
  number={4},
  pages={1--13},
  year={2017},
  publisher={ACM New York, NY, USA}
}

@article{shi2024interactive,
  title={Interactive character control with auto-regressive motion diffusion models},
  author={Shi, Yi and Wang, Jingbo and Jiang, Xuekun and Lin, Bingkun and Dai, Bo and Peng, Xue Bin},
  journal={ACM Transactions on Graphics (TOG)},
  volume={43},
  number={4},
  pages={1--14},
  year={2024},
  publisher={ACM New York, NY, USA}
}

@article{ling2020character,
  title={Character controllers using motion vaes},
  author={Ling, Hung Yu and Zinno, Fabio and Cheng, George and Van De Panne, Michiel},
  journal={ACM Transactions on Graphics (TOG)},
  volume={39},
  number={4},
  pages={40--1},
  year={2020},
  publisher={ACM New York, NY, USA}
}

@article{zhang2018mode,
  title={Mode-adaptive neural networks for quadruped motion control},
  author={Zhang, He and Starke, Sebastian and Komura, Taku and Saito, Jun},
  journal={ACM Transactions on Graphics (TOG)},
  volume={37},
  number={4},
  pages={1--11},
  year={2018},
  publisher={ACM New York, NY, USA}
}

@inproceedings{cheema2023discovering,
  title={Discovering Fatigued Movements for Virtual Character Animation},
  author={Cheema, Noshaba and Xu, Rui and Kim, Nam Hee and H{\"a}m{\"a}l{\"a}inen, Perttu and Golyanik, Vladislav and Habermann, Marc and Theobalt, Christian and Slusallek, Philipp},
  booktitle={SIGGRAPH Asia 2023 Conference Papers},
  pages={1--12},
  year={2023}
}

@inproceedings{cheema2020predicting,
  title={Predicting mid-air interaction movements and fatigue using deep reinforcement learning},
  author={Cheema, Noshaba and Frey-Law, Laura A and Naderi, Kourosh and Lehtinen, Jaakko and Slusallek, Philipp and H{\"a}m{\"a}l{\"a}inen, Perttu},
  booktitle={Proceedings of the 2020 CHI Conference on Human Factors in Computing Systems},
  pages={1--13},
  year={2020}
}

@inproceedings{fischer2024sim2vr,
  title={SIM2VR: Towards Automated Biomechanical Testing in VR},
  author={Fischer, Florian and Ikkala, Aleksi and Klar, Markus and Fleig, Arthur and Bachinski, Miroslav and Murray-Smith, Roderick and H{\"a}m{\"a}l{\"a}inen, Perttu and Oulasvirta, Antti and M{\"u}ller, J{\"o}rg},
  booktitle={Proceedings of the 37th Annual ACM Symposium on User Interface Software and Technology},
  pages={1--15},
  year={2024}
}

@article{nair2023berkeley,
  title={Berkeley open extended reality recordings 2023 (boxrr-23): 4.7 million motion capture recordings from 105,852 extended reality device users},
  author={Nair, Vivek and Guo, Wenbo and Wang, Rui and O'Brien, James F and Rosenberg, Louis and Song, Dawn},
  journal={arXiv preprint arXiv:2310.00430},
  year={2023}
}

@inproceedings{ikkala2022breathing,
  title={Breathing life into biomechanical user models},
  author={Ikkala, Aleksi and Fischer, Florian and Klar, Markus and Bachinski, Miroslav and Fleig, Arthur and Howes, Andrew and H{\"a}m{\"a}l{\"a}inen, Perttu and M{\"u}ller, J{\"o}rg and Murray-Smith, Roderick and Oulasvirta, Antti},
  booktitle={Proceedings of the 35th Annual ACM Symposium on User Interface Software and Technology},
  pages={1--14},
  year={2022}
}

@inproceedings{luo2023perpetual,
  title={Perpetual humanoid control for real-time simulated avatars},
  author={Luo, Zhengyi and Cao, Jinkun and Kitani, Kris and Xu, Weipeng and others},
  booktitle={Proceedings of the IEEE/CVF International Conference on Computer Vision},
  pages={10895--10904},
  year={2023}
}

@article{starke2024categorical,
  title={Categorical Codebook Matching for Embodied Character Controllers},
  author={Starke, Sebastian and Starke, Paul and He, Nicky and Komura, Taku and Ye, Yuting},
  journal={ACM Transactions on Graphics (TOG)},
  volume={43},
  number={4},
  pages={1--14},
  year={2024},
  publisher={ACM New York, NY, USA}
}

@inproceedings{
jang2017categorical,
title={Categorical Reparameterization with Gumbel-Softmax},
author={Eric Jang and Shixiang Gu and Ben Poole},
booktitle={International Conference on Learning Representations},
year={2017},
url={https://openreview.net/forum?id=rkE3y85ee}
}

@article{vaswani2017attention,
  title={Attention is all you need},
  author={Vaswani, Ashish and Shazeer, Noam and Parmar, Niki and Uszkoreit, Jakob and Jones, Llion and Gomez, Aidan N and Kaiser, {\L}ukasz and Polosukhin, Illia},
  journal={Advances in neural information processing systems},
  volume={30},
  year={2017}
}

@String{Computing = "Computing" }

@String{Computer = "{IEEE} Computer" }

@String{Springer = "Springer-Verlag" }

@ArtifactSoftware{R,
    title = {R: A Language and Environment for Statistical Computing},
    author = {{R Core Team}},
    organization = {R Foundation for Statistical Computing},
    address = {Vienna, Austria},
    year = {2019},
    url = {https://www.R-project.org/},
}
\clearpage

\clearpage
\newpage
\appendix

\section{Beat Saber: A Benchmark VR Application}
\label{sec:beatsaber}
\emph{Beat Saber} is a skill-intensive VR rhythm game in which players swing virtual sabers to cut colored notes with the correct hand and direction, while avoiding bombs with the sabers and obstacles with the head.
The game has been established as a prominent benchmark VR application in various computational domains, including data anonymization (\textit{e.g.}, \citen{nair2024deep}), motion generation (\textit{e.g.}, \citen{starke2024categorical,barquero2025sparse}), and \emph{in silico} user modeling (\textit{e.g.}, \citen{fischer2024sim2vr}), as well as in non-computational fields such as health research \cite{Tuong2019,Grospretre2023} and neurological therapy \cite{Ruhf2020}.

\noindent \textit{Computational challenges for Beat Saber.}
Computationally, \emph{Beat Saber} challenges agents to map complex spatio-temporal input to body movements: players must precisely cut many dozens of notes, each varying in color, position, and timing, in rapid succession.
While the objective is simple, the game does not explicitly instruct on the whole-body coordination required, giving rise to highly diverse movement styles across the player population.
Moreover, \emph{Beat Saber}'s space of object configuration is huge: with 2 note colors, 9 note directions, and 12 grid positions, a sequence of 10 colored notes would already yield 1e23 possible combinations.

\noindent \textit{Beat Saber is difficult to curate.}
\emph{Beat Saber}'s popularity is partly owed to its culture of user-created content, comprising over 100,000 custom \emph{maps} authored by online creators referred to as \emph{mappers} \cite{Nair:2023:ROT}.
Mappers have creative freedom to choreograph interesting and complex patterns that encourage a variety of movements from flowing dance-like motions to rapid striking motions. 
While each map is manually assigned an overall difficulty rating among Easy, Normal, Hard, Expert, or Expert+, in practice, human play is required to assess a map's actual difficulty.
This is particularly so for maps intended for advanced players, as these maps often creatively break design guidelines.
Moreover, the experience of difficulty levels is often subjective: a player might find a particular map easier or harder than how an average player would experience the same map, based on the players' particular gameplay history and playstyle.
These pain points lead to difficulties in curating the vast game content, despite \emph{Beat Saber} being a skill-intensive game that could be served better with recommendations according to a more objective criterion.
In this work, we propose the use of a player model in conjunction with collaborative filtering to perform personalized score prediction, showing promise in this direction.

\section{Data Quality Control and Preprocessing} \label{appendix:QC}

We apply the following quality control (QC) measures to BOXRR-23, which consists of $3p$ movement sequences, each corresponding to a player's attempt on a map.
We first remove every replay with missing map information or score record.
As \emph{Beat Saber} features many \emph{mods} and \emph{game modes} that significantly alter the game's behavior, we simplify our task by removing records that include mods or non-standard game modes.
Additionally, records with excessive jitter or no movement are excluded, as well as records with incorrect left-handed indication (\emph{i.e.}, the record metadata shows left-handed but the gameplay is actually right-handed).
As the BeatSaver database is concurrently receiving updates without versioning, we simply remove maps that have received updates since the original $3p$ data were collected.

For the pose data QC, we standardize all players' height to 1.5044m at rest, matching that of the standard humanoid character for the physics-based tracking experiments in \Cref{sec:robotic_agent}.
As \emph{Beat Saber}'s game grid's height is adjusted based on the player's height, we apply a height offset to the headset and handhelds' coordinates.
As issues in hardware calibration can cause an incorrect global offset in the recorded $3p$ data, we center the $xyz$ coordinates by subtracting the median value from each dimension.
Finally, we correct for any coordinate-shifting anomalies during gameplay by detecting clusters of headset positions using kernel density estimation and removing any resulting modes.

\section{Network Architectures for Miscelleneous Models} \label{appendix:architecture}

Here, we outline the architectures used for the Gumbel-Softmax VAE, oracle classifier, and the Transformer baseline for PSP.
All of the models share the common trait of being sequence-based models, and for the interest of consistency, we elect to use Transformer as the backbone for all of the models.
The Transformer hyperparameters, such as number of layers and attention heads, remain the same across all models.
For the most thorough details on our implementation, we refer the reader to the official codebase linked to the project webpage: \url{https://robo-saber.github.io/}.

\noindent \textit{Gumbel-Softmax VAE.}
The length-$T$ pose sequence $\ThreePChunk$ is further processed into a length-$3T$ sequence of 9-dimensional vectors, as to model the dense relationship between body parts across time.
We use multi-headed self-attention with GELU activation, and then aggregate the final layer's latent sequence with an MLP head, flattening the whole sequence into a vector and finally encoding it into $\ThreePLogits$.

\noindent \textit{Oracle player classifier.}
The oracle receives 6 gameplay segments and identifies which held-out player produced them.
The oracle player classifier is very similar to the style encoder $\StyleEncoder$, which takes $\RefSegments$ and $\RefThreePChunk$ to produce $\StyleEmbeddings$.
Instead of an embedding, the classifier uses an MLP head to produce player classification logits.
We stay consistent with the use of all-zero sentinel token for predicting these logits.
To compute the logits for a generated trajectory, we take the mean across 6 predicted logits, produced from feeding randomly-chosen gameplay segments from the generated trajectory to the oracle.

\section{TorchSaber: Implementation Details} \label{appendix:torchsaber}

In \Cref{sec:selection}, we introduced TorchSaber (TS), the GPU-accelerated and simplified \emph{Beat Saber} simulator, which is incorporated into the sampling process and evaluation of our method. TS is based on a vectorized implementation of the slab method for collision detection, utilizing PyTorch for GPU-accelerated tensor operations.
Given two consecutive frames, we produce 1.2-meter-long line segments for the left and right sabers. We place 5 keypoints along each line segment at regular intervals. Representing the displacement between frames allows us to compute 5 rays, which describe the position changes of these keypoints.

We compute collision masks between displacement rays and note boxes using the slab method. A collision is registered if any displacement ray penetrates a box boundary. The head-to-obstacle collisions are computed simply as a point-box collision, using the normal vector of the obstacle box’s faces and the head's displacement with respect to the obstacle box’s vertices.

\noindent \textit{Geometries.} For computing collisions, TS utilizes the following geometries, based on available \emph{Beat Saber} documentation. First, the headset is modeled as a point in 3D.
The sabers are modeled as line segments, each 1.2 meters long.
Colored and bomb notes are modeled as cubes sized $(0.5, 0.5, 0.5)$. The boxes for the obstacles are scaled according to their durations, width, and height specified in the map data.
Each object is instantiated at the point determined by its appearance in the map data. For each object, the $x$-position (forward) is calculated from its relative timestamp, multiplied by the note jump distance specified by the map.
The $y$- and $z$-position (up) of each object is calculated based on a $4 \times 3$ grid.
The grid's $z$-offset is computed based on the player's height $H$.
We find $1.05 - \frac{H}{2}$ to be an appropriate offset.
Emulating the original game’s behavior, we set the objects' $x$-offset so they arrive just in front of the player at the targeted beat. We allow objects to persist in view up to 0.1 seconds after the beat, so that they exit when they are behind the player.
These parameters are optimized in the direction that maximizes the match between the computed TS and the recorded replay scores of real players.

\section{Reward Function Details} \label{appendix:reward_def}

Here, we define the reward terms that appear in \Cref{sec:selection}. The TorchSaber score $r_\text{TS}$ is as described in \Cref{sec:method}, based on the note-saber, bomb-saber, and head-obstacle collision flags.
For note-saber collisions, cut directions, cut velocities, and color correctness are taken into account to determine whether the cut was good.
The cut direction is declared as good if the dot product between the normalized saber tip velocity and the unit vector pointing to the note direction is greater than $0$.
For the ``omnidirectional'' notes, any cut is declared a good cut.
Simplifying \emph{Beat Saber}'s intricate scoring criteria, we compute the swing angle score for every ``good cut'', \textit{i.e.}, collision events with correct directions and colors.
The maximum score of $1$ per colored note is achieved if the correct saber's orientations $24$ frames previous to (pre-swing) and after (post-swing) the good cut are at least -100 and 60 degrees, respectively.

The total TS score for a map is computed by taking the average score across colored notes appearing.
For the sake of simplicity and ease of implementation, we excluded combos and cut angles from consideration, which still retained a strong positive correlation between TS scores and real \emph{Beat Saber} scores from BOXRR-23.

For candidate trajectory selection, we compute this score at each generator query step for each candidate.
For evaluating the full real and synthetic trajectories, we treat the entire play sequence as a segment to compute $r_\text{TS}$.

The bomb-saber collision penalty $r_\text{Bomb}$ is similarly calculated.
At each length-$T$ segment, given $n_\text{Bombs}$ that appear, we compute $n_\text{BombCollision}$, the total number of appeared bombs that have collided with any of the sabers. Then $r_\text{Bombs}$ is simply calculated as $\frac{n_\text{BombCollision}}{n_\text{Bombs}}$.

\section{Factorization Machines} \label{appendix:fm}
We follow \citen{kristensen2022personalized} and employ factorization machines (FMs, \citen{rendle2010factorization}) for CF.
FMs learn embeddings for each player and map, along with a simple linear model, such that the score for the player-map pair is predicted as:
\begin{align}
    \hat y^\text{player $\cdot$ map} &= \left(\mathbf{z}^\text{player} \cdot \mathbf{z}^\text{map}\right) +  w_0 + w^\text{player} + w^\text{map}
\end{align}
where $y^\text{player $\cdot$ map} \in \mathbb{R}$ is the predicted score, $\mathbf{z}^\text{player}$, $\mathbf{z}^\text{map}$ are the learned embeddings for the player and the map, and $w_0$, $w^\text{player}$, and $w^\text{map}$ are the global bias and player/map-specific weights of the linear model, respectively.

Simply, the embeddings and the linear model are optimized by a gradient-based optimizer on the squared error of the score:
\begin{align}
    \mathcal{L}_\text{FM} &= \left( y^\text{player $\cdot$ map} - \hat y^\text{player $\cdot$ map}\right)^2
\end{align}
For more details of FMs, we refer the reader to \citen{rendle2010factorization}.

\section{Physics-basd Tracking Implementation} \label{appendix:tracking}

Built within Isaac Gym \cite{makoviychuk2021isaac}, PHC \cite{luo2023perpetual} produces full-body joint actuations that align the robot's head and hands in both position and rotation with the input $3p$ pose at each timestep.
Using the usual DRL notation, the tracking controller $\pi$ outputs the actions $\mathbf{a}_t$ at each timestep $t$ as:
\begin{align}
    \mathbf{a}_t &= \pi\left(\cdot \mid \mathbf{s}_t, \mathbf{p}_t \right)
\end{align}
where $\mathbf{s}_t$ is the state of the simulated character and $\mathbf{p}_t$ is the target $3p$ pose.
The entire $3p$ trajectory may be produced in advance and subsequently provided to the tracking controller as a target sequence, thereby fully decoupling kinematic planning from physics-based control.
As depicted in \Cref{fig:teaser}, the system's overall behavior of synthesizing movements based on game state observations is unchanged. Accordingly, Robo-Saber operates as an open-loop, whole-body robotic agent for VR gameplay.

We first tested the pre-trained $3p$ tracking variant of the PHC controller without modification. We found that it could not maintain the character's balance with \emph{Beat Saber}-specific movements; therefore, we fine-tuned the tracking policy using custom mocap sequences.
An experienced \emph{Beat Saber} player performed 16 play sequences in total while wearing an inertial mocap suit, together with a VR headset and controllers, covering Normal, Hard, Expert, and Expert+ difficulty levels.
For the PHC finetuning, we produced a 50-50 mixture of the original AMASS \cite{mahmood2019amass} training data and new training data in order to prevent the pretrained controller from forgetting its tracking abilities. The sabers are attached to the PHC character's hands, aligned with the direction of the fingers.

\section{Table of Hyperparameters} \label{appendix:hyperparameters}

\Cref{table:hyperparameters} discloses every hyperparameter involved in producing this work.

\begin{table*}[]
\centering
\begin{tabular}{lll}
Symbol & Description & Value \\ \hline
$T$ & Number of frames per $3p$ motion chunk & 16 \\
- & Interval for interpolating the $T$ frames & 4 \\
$n$ & Length of latent sequence for each domain (colored notes, bomb notes, obstacles) & 20 \\
$s$ & The default lookahead, in seconds, used in training & 2.0 \\
$h$ & Number of frames used as $3p$ history input & $2$ \\
$\lambda_\text{Match}$ & Jensen-Shannon divergence-based matching loss weight & 1e-4 \\
$\lambda_\text{Bomb}$ & Reward weight for saber-bomb collision penalty & 10 \\
$\lambda_\text{Obstacle}$ & Reward weight for head-obstacle collision penalty & 10 \\
- & Batch size for kinematic $3p$ generator training & 128 \\
- & $3p$ generator model AdamW learning rate & 5e-5 \\
- & Number of Transformer encoder layers & 4 \\
- & Number of Transformer attention heads & 4 \\
$|\mathbf{z^\text{player}}|, |\mathbf{z^\text{map}}|$ & Embedding size for factorization machine (FM) & 16 \\
- & Batch size for FM training & 512 \\
- & Embedding \& linear weight dropout rate for FM training & 0.5 \\
- & FM optimizer AdamW weight decay rate & 1e-1 \\
- & FM optimizer AdamW learning rate & 3e-4
\end{tabular} 
\caption{Hyperparameters and their values used for our experiments.}
\label{table:hyperparameters}
\end{table*}

\end{document}